\documentclass[journal]{IEEEtran}
\IEEEoverridecommandlockouts
\usepackage[noadjust]{cite}
\usepackage{amsmath,amssymb,amsfonts}
\usepackage{amsthm}
\usepackage{algorithmic}
\usepackage{relsize}
\usepackage{algorithm}
\usepackage{graphicx}
\usepackage{subcaption}
\usepackage{comment}
\usepackage{textcomp}
\usepackage{xcolor}
 \usepackage{url}
 \usepackage{xcolor}
\usepackage{tikz}
\def\BibTeX{{\rm B\kern-.05em{\sc i\kern-.025em b}\kern-.08em
    T\kern-.1667em\lower.7ex\hbox{E}\kern-.125emX}}
\usepackage{enumerate}
\usepackage{paralist} 
\usepackage{afterpage}
\usepackage{amsbsy}

\usepackage{mathtools}
\mathtoolsset{showonlyrefs}
\usepackage{placeins}

\allowdisplaybreaks
\begin{document}
	
	\title{Graph Fourier Transform: A Stable Approximation}
	\author{Jo\~ao Domingos and Jos\'{e} M. F. Moura
		\thanks{\hspace{-.61cm} Jo\~ao Domingos and Jos\'{e} M. F. Moura {\{ph: (412)-268-6341, fax: (412)-268-3890\}}
			are with the Department of Electrical and Computer Engineering, Carnegie Mellon University, 5000 Forbes Av, Pittsburgh, PA 15217 USA, {\tt\small joaodomingos42@hotmail.com, moura@andrew.cmu.edu.} Jo\~ao Domingos is also with Instituto Sistemas e Rob\'otica, Instituto Superior T\'ecnico, Universidade de Lisboa, Portugal.}
		\thanks{\hspace{-.25cm}This work was partially supported by NSF grant CCF 1513936.
		}
	}
	%
	\newtheorem{theorem}{Theorem}
	\newtheorem{lemma}{Lemma}
	\newtheorem{remark}{Remark}
	\newtheorem{example}{Example}
	\newtheorem{result}{Result}
	\newtheorem{corollary}{Corollary}[theorem]
	\maketitle

\begin{abstract}
In Graph Signal Processing (GSP), data dependencies are represented by a graph whose nodes label the data and the edges capture dependencies among nodes. The graph is represented by a weighted adjacency matrix~$A$ that, in GSP, generalizes the Discrete Signal Processing (DSP) shift operator $z^{-1}$. The (right) eigenvectors of the shift~$A$ (graph spectral components) diagonalize~$A$ and lead to a graph Fourier basis~$F$ that provides a graph spectral representation of the graph signal. The inverse of the (matrix of the) graph Fourier basis~$F$ is the Graph Fourier transform (GFT), $F^{-1}$. Often, including in real world examples, this diagonalization is numerically unstable. This paper develops an approach to compute an accurate approximation to~$F$ and $F^{-1}$, while insuring their numerical stability, by means of solving a non convex optimization problem. To address the non-convexity, we propose an algorithm, the stable graph Fourier basis algorithm (SGFA) that we prove to exponentially increase the accuracy of the approximating~$F$ per iteration. Likewise, we can apply SGFA to $A^H$ and, hence, approximate the stable left eigenvectors for the graph shift~$A$ and directly compute the GFT.

We evaluate empirically the quality of SGFA by applying it to graph shifts~$A$ drawn from two real world problems, the 2004 US political blogs graph and the Manhattan road map, carrying out a comprehensive study on tradeoffs between different SGFA parameters. \textcolor{black}{We also confirm our conclusions by applying SGFA on very sparse and very dense directed Erd\H os-R\'enyi graphs.}
\end{abstract}


\section{Introduction}
\label{sec:introduction}
A truism of current technology is the explosive growth of data\textemdash \textit{Big Data}, characterized variously by\footnote{Volume, variety, velocity, veracity, value, variability, and visualization.} three, four, five, or seven V's. Of these, we focus on variety that may reflect different data formats, arising from diverse data sources and applications. Examples include tweets, networks of coauthors and citations \cite{Newman:04}, hyperlinked blogs \cite{political_blog_data}, phone users, customers of service providers, friends in a social network, individuals in populations, traffic flows in urban environments \cite{dataset_road_2,dataset_road}, or physical measurements like temperature or humidity in a network of meteorological stations spread over a whole country, among many other examples, e.g., \cite{Jackson:08,Newman-2010}, of current interest. Data from these applications contrast with traditional time series, audio or speech signals, images, video, or point clouds where the data samples are indexed by time instants, pixels, voxels, or pixels and time taking values on a regular one-, two-, three- or even four-dimensional grid. This variety of Big Data also often leads to data being referred to as \textit{un}structured, not fitting neatly on a table. This \textit{un}structured data is structured through a graph $\mathcal{G}=(\mathcal{V},\mathcal{E})$ where nodes or vertices in the set~$\mathcal{V}$ represent the sources or labels of the data and (directed or undirected) edges in the set~$\mathcal{E}$ capture dependencies or relations among the data of different nodes. For example, traffic counts in a metropolitan area are directly dependent if corresponding to close by locations of the same street. In this paper, the graph~$\mathcal{G}$ is general and characterized by an adjacency matrix~$A$ \cite{Chung:96}. As explained in Section~\ref{sec:GSP}, in Graph Signal Processing~(GSP) \cite{GSP_1}, the adjacency matrix~$A$ plays the role of the shift $z^{-1}$ in discrete signal processing~(DSP) \cite{siebert-1986,Oppenheim:99,mitra-1998}. We refer to~$A$ as the graph shift. In GSP, the eigenvalues $\left\{\lambda_i\right\}$ and (right) eigenvectors $\left\{f_i\right\}$, $1\leq i\leq n$, of~$A$ are the graph frequencies and graph spectral components \cite{GSP_1,ShumanNFOV:13}, respectively, extending to signals defined on a graph~$\mathcal{G}$ the common concepts of frequency and harmonic components for time signals. The matrix~$F$ of the eigenvectors or graph spectral components of~$A$ will be referred to as the graph Fourier basis. \textcolor{black}{If~$A$ is diagonalizable, the inverse of~$F$ is the graph Fourier transform\footnote{\label{ftn:Finv1}\textcolor{black}{If~$A$ is not diagonalizable, see \cite{derimoura-2017} on how to define the GFT.}} (GFT), $F^{-1}=W^H$. The matrix~$F$ and its inverse~$F^{-1}=W^H$ are unique up to choice of basis in the graph signal space \cite{Pueschel:08a,Pueschel:08b}. In DSP, the GFT is the discrete Fourier transform~(DFT).}

For generic directed graphs (digraphs), $A$ is not symmetric, the eigenvalues $\lambda_i$ may be complex valued, its \textcolor{black}{eigenvectors or} graph Fourier components~$f_i$ are not orthogonal and the graph Fourier basis~$F$ is not unitary $\left(F^{-1}\neq F^H\right)$ or fail to be a complete basis \cite{GSP_1,deri-moura}, and the columns~$f_i$ of~$F$ are in general not of unit norm. Authors have taken different approaches to avoid these issues and preserve~$F$ to be unitary. For example,
\begin{inparaenum}[1)]
\item \cite{giraultgoncalvesfleury-2015,gavilizhang-2017} redefine the shift matrix to make it a norm preserving operator;
     \item \cite{marquessegarraleusribeiro-2016,marques2017stationary} consider the case of a normal shift $\left(AA^H=A^HA\right.$ and~$A$ is unitarily diagonalizable, but with eigenvalues not necessarily real valued$\left.\vphantom{AA^H}\right)$;
          \item more often, the literature considers the graph to be undirected, so, the shift is symmetric, or take the shift to be the graph Laplacian~$L$ \cite{ShumanNFOV:13}, either of which is diagonalizable by a unitary operator\footnote{\label{ftn:Laplacian}The Laplacian~$L$ is a second order operator like a second order derivative or difference, while~$A$ is first order like first order derivative or difference.  \textcolor{black}{Adopting~$L$ as shift is then equivalent to working with a time shift $z^{-2}$ in traditional DSP or linear systems, or to restrict signals to have even symmetry like autocorrelations and filters to be polynomials (or rational functions) in powers of $z^{-2}$. In contrast, working with~$A$ as shift corresponds in DSP to the time shift $z^{-1}$ and filters to be polynomials (or rational functions) in $z^{-1}$}.}; and
              \item \cite{SardellittiBarbarossaLorenzo-2017} considers directed graphs but redefines the GFT to keep it orthogonal and avoid Jordan decompositions.
              \textcolor{black}{However, it may be hard to develop a graph filtering theory for \cite{SardellittiBarbarossaLorenzo-2017}.}
                  \end{inparaenum}

\textcolor{black}{Our goal is to consider spectral analysis for data supported by generic graphs\textemdash graphs not restricted by any special structure\textemdash to address the difficulties associated with finding a numerically stable graph Fourier transform and Fourier basis~$F$. These problems may stem from
\begin{inparaenum}[1)]
\item the lack of orthogonality between the $f_i$ (eigenvectors of~$A$ and columns of~$F$), some of which may be close to being parallel,
\item possible large scale differences between the $f_i$'s, and,
 for defective~$A$, 
 \item the set of eigenvectors $\left\{f_i\right\}$ not being a complete basis and $A$ not being diagonalizable.
\end{inparaenum} 
We develop an optimization approach to approximately diagonalize the matrix~$A$ by a numerically stable, non unitary, Fourier basis~$F$. The algorithm provides a tunable tradeoff between the accuracy of the diagonalization $\left(\vphantom{\left\|AF-F\Lambda\right\|_\mathcal{F}}\right.\!$measured for example by the error\footnote{\textcolor{black}{The subindex~$\mathcal{F}$ stands for Frobenius norm, $\|A\|_{\mathcal{F}}=\sqrt{\sum_{i,j} A_{i,j}^2}$.}} $\left\|AF-F\Lambda\right\|_\mathcal{F}$,  where $\Lambda$  is the diagonal matrix of eigenvalues of~$A$$\left.\vphantom{\left\|AF-F\Lambda\right\|_\mathcal{F}}\!\right)$ and the stability of~$F$ $\left(\vphantom{\sigma_{\min}}\right.$\!measured by the minimum singular value $\sigma_{\min}(F)$$\left.\vphantom{\sigma_{\min}}\!\right)$}.

 \textcolor{black}{In fact, attempting to diagonalize shift matrices (or adjacency matrices) of generic graphs by standard methods like the routine {\tt{eig}} in MATLAB$^\circledR$ leads often to highly numerically unstable Fourier basis~$F$, i.e., the matrix~$F$ has a very large condition number\footnote{\textcolor{black}{Empirically, we observed that the maximum singular value $\sigma_{\max}(F)$ is not very large. It is the $\sigma_{\min}(F)$ that is very small. We focus this discussion on $\sigma_{\min}(F)$.}} $\kappa$ and a very small minimum singular value $\sigma_{\min}(F)\ll10^{-12}$. Numerically unstable Fourier basis~$F$ is highly undesirable since then the GFT, $F^{-1}$, is badly scaled, becoming difficult to carry out graph spectral analysis, graph spectral decompositions, graph Fourier transforms, and related concepts in GSP. To confirm that such matrices occur frequently, we study the class of directed Erd\H os-R\'enyi random graphs generated with probability of connection~$p\in [0\,\,\,1]$. \textcolor{black}{For not very large values of~$p$, Erd\H os-R\'enyi random graphs are highly sparse, a characteristic found in many real world graphs. For completeness, we include numerical studies for two types of Erd\H os-R\'enyi random graphs: 
 \begin{inparaenum}[a)]
 \item models without self-loops, obtained when the probability of connection between a node and itself is zero, and 
 \item models with possible self-loops where a node may connect with itself with probability $p$. 
 \end{inparaenum}
 We consider $100$ values of~$p$ discretizing $[0\,\,\,1]$. For each of these 100 values of~$p$, we generated $10^3$ directed Erd\H os-R\'enyi graphs for both models (without and with possible self-loops). We then used MATLAB$^\circledR$ {\tt{eig}} to find the Fourier basis~$F$ diagonalizing their adjacency matrix~$A$ and computed for each such~$F$ the corresponding $\sigma_{\min}(F)$. Figure~\ref{fig:tails_erdos_renyi} plots the empirical probability that $\sigma_{\min}(F)\leq 10^{-12}$ for forty values of~$p$ tested in the ranges\footnote{Results for other values of~$p$ are not shown since the empirical probability that $\sigma_{\min}(F)\leq 10^{-12}$ is zero.} $[0\,\,\,.09]$ and $[.93\,\,\,1]$, corresponding to very low and very high probability of connection~$p$. The value of $10^{-12}$ upper bounding $\sigma_{\min}(F)$ reflects the very bad conditioning of the Fourier basis~$F$ computed by MATLAB$^\circledR$ {\tt{eig}}. In both models, the extreme points $p=0$ and $p=1$ yield a numerically stable Fourier basis~$F$ with probability one since the generated (isolated nodes and complete) graphs are trivially undirected.
As can be seen, for both with (blue plot) and without (red plot) self-loops
  \begin{figure}[htbp]
 	\centering
 		\includegraphics[width=8.4cm]{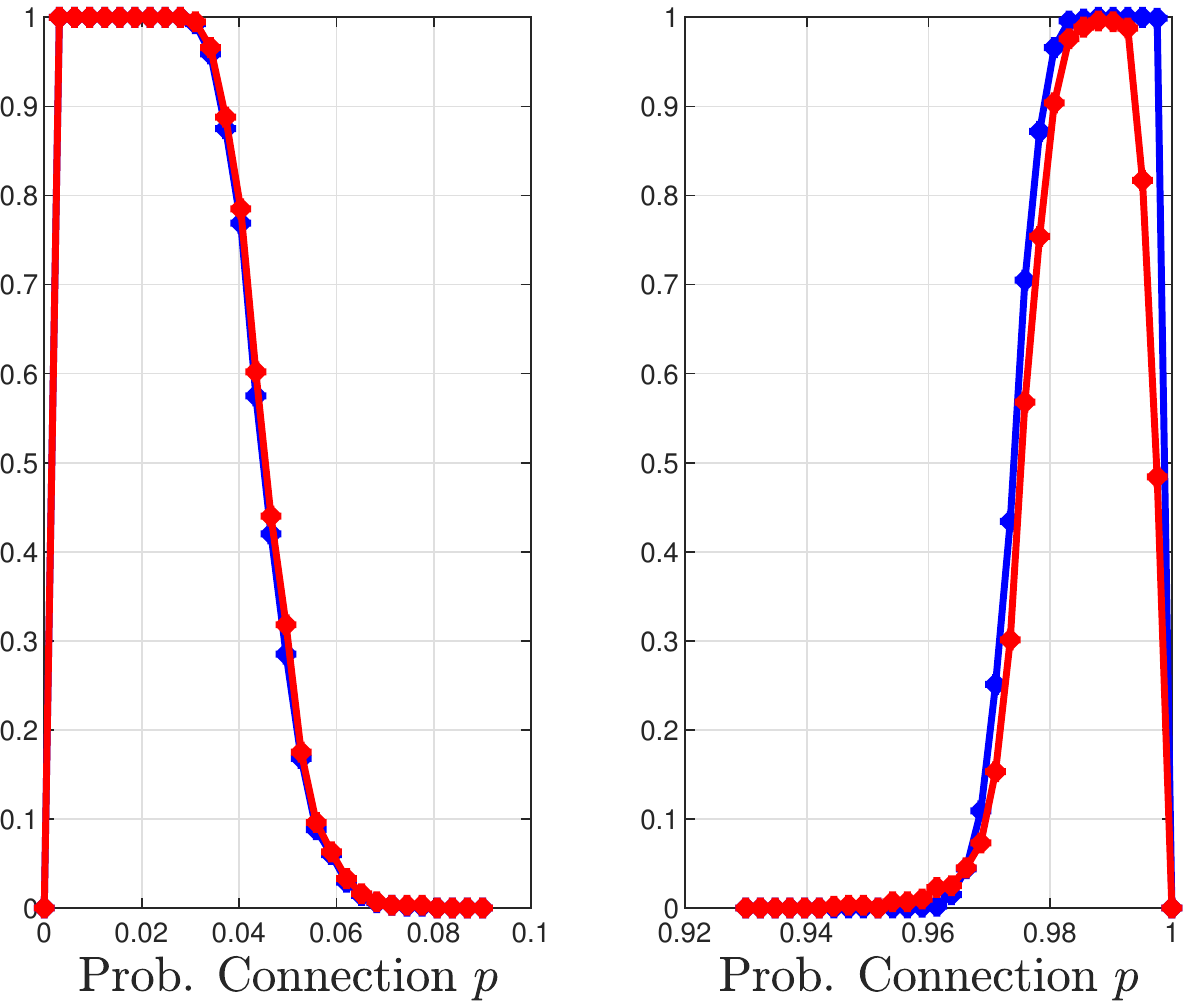}
 	\caption{\textcolor{black}{Empirical probability that $\sigma_{\min}(F)\leq 10^{-12}$ vs.~probability of connection~$p$ for two types of Erd\H os-R\'enyi models with $n=100$ nodes: in red, models without self-loops and, in blue, models with possible self-loops.  Left and right tails for $1000$ Monte Carlo trials.}}
 	\label{fig:tails_erdos_renyi}
 \end{figure}
 and for both very sparse and very dense random \textit{directed} Erd\H os-R\'enyi graph models, the Fourier bases~$F$ computed by MATLAB$^\circledR$ {\tt{eig}} tend to be numerically unstable with high probability. When we allow for self-loops (red plot), we note for high values of~$p$ that the (red) plot has a slightly smoother transition to the extreme (completely connected) case $p=1$.}}
 \textcolor{black}{Beyond random networks, this is also the case with real world networks, like we observed in other work, when we analyzed  \cite{GSP_1} the political blogs network \cite{political_blog_data} shown in figure~\ref{fig_political} and when we studied \cite{dataset_road_2,dataset_road} the Manhattan road map shown in figure~\ref{fig_road_net}. Using MATLAB$^\circledR$ {\tt{eig}} leads to~$F$ with $\sigma_{\min}(F)\approx10^{-33}$ for the political blogs and $\sigma_{\min}(F)\approx10^{-19}$ for the Manhattan road map. We expand on these in section~\ref{sec_political} and section~\ref{sec_road}, respectively.
 }
 \textcolor{black}{\begin{remark}\label{rem:jordan1} As we mentioned, the paper gives a method to find an approximate stable approximation to the Fourier basis~$F$ when the eigenvector matrix is unstable either because
  \begin{inparaenum}[1)]
  \item \label{inpar:Fdoesnotexist} no full rank~$F$ exists; or
  \item \label{inpar:Funstable} full rank $F$ may exist, but the computed~$F$ is numerically unstable.
  \end{inparaenum}
 A Fourier basis~$F$ may not exist because the shift~$A$ is defective. This may occur when the graph shift has a repeated eigenvalue~$\lambda$ with (numerically computed) algebraic multiplicity greater than one\footnote{\label{ftn:alggeomultiplicities}\textcolor{black}{A matrix is diagonalizable iff for every eigenvalue the algebraic and geometric multiplicities are equal.}}. As studied in \cite{derimoura-2017}, this is the case with the political blogs network and the Manhattan road map,  
   where, in both cases, as computed with MATLAB$^\circledR$ {\tt{eig}}, $\lambda=0$ has algebraic multiplicity of several hundred (see below in figure~\ref{fig_spectrum_angles} a plot of the eigenvalues for the Manhattan road map). Numerically computing an eigenvalue with large algebraic multiplicity is a well known difficult problem. It may give rise to a cluster of values with the true eigenvalue possibly at its center \cite{kahan1972conserving,zeng_jordanformbkch} or the true eigenvalue may be best approximated by a pseudoeigenvalue~\cite{zeng_defeigenval,trefethen2005spectra}. Rounding errors may make it not possible to determine if an eigenvalue is actually a repeated eigenvalue or a cluster of (not necessarily close together) eigenvalues. In other words, determining true versus spurious eigenvalues in the numerically computed spectrum of the shift matrix may be ambiguous. Further, as observed by \cite{GolubWilkinson-1976}, with repeated eigenvalues, the direct computation of the eigendecomposition of the shift~$A$ is in general numerically unstable (even more so, if it involves the Jordan decomposition \cite{BeelenVanDooren-1990,derimoura-2017} where a small perturbation of the input matrix may drastically change the computed Jordan form). The paper will not dwell further on this; we simply state that our algorithm can be applied to compute a stable~$F$ and stable~$F^{-1}$ that accurately approximate the diagonalization of~$A$ in either the no full rank~$F$ case~\ref{inpar:Fdoesnotexist}) or the full rank~$F$ case~\ref{inpar:Funstable}) above. \hfill$\small\blacksquare$
   \end{remark}
   }
   The examples above of random graphs and real world applications \textcolor{black}{point to the difficulty encountered often} of computing the Fourier basis~$F$ and its inverse, the GFT, $F^{-1}$. \textcolor{black}{They confirm the interest and the need to develop a numerically efficient method to find an approximate stable Fourier basis~$F$, our goal with this work.}
%
 This paper develops SGFA, the stable graph Fourier basis algorithm, to compute an \textit{accurate} \textit{stable} approximation to~$F$ for generic \textcolor{black}{digraph} shifts~$A$ (not necessarily symmetric). SGFA is a two stage procedure that, iteratively, attempts
\begin{inparaenum}[1)]
\item to diagonalize a triangular decomposition of~$A$, and
\item to optimize the numerical stability of the resulting~$F$.
\end{inparaenum}
There are trade-offs between accuracy (degree of diagonalizability of the decomposition as measured by how close the numerically stable approximation of~$F$ is to the true~$F$) and stability (well conditioning of the resulting approximation of~$F$).

We can compute the GFT, $F^{-1}$, by numerically inverting the graph Fourier basis~$F$ obtained with SGFA. But this may affect its numerical accuracy, since we first compute the (right) eigenvectors and then invert the Fourier basis~$F$. To avoid this, we compute directly the GFT by applying SGFA to the $A^H=A^T$, since the right eigenvectors of $A^H$ are the left eigenvectors\footnote{\textcolor{black}{Let $A=F\,\Lambda F^{-1}=W^{-H}\,\Lambda\,W^H$, then $W^H\,A=\Lambda\, W^H$, $A^H=W\,\Lambda^\star W^{-1}$, and the columns of~$W$ are the left eigenvectors of~$A$ and~$W$ is the Fourier basis of $A^H$.}} of~$A$. In other words, diagonalizing~$A^H$ with SGFA computes directly a stable approximation to the GFT, $F^{-1}$, without inverting the SGFA computed~$F$. \textcolor{black}{Since we do not know the true~$F$ and true GFT, $F^{-1}$,} applying SGFA to both~$A$ and $A^H$ provides the opportunity to study the quality of the numerically computed graph Fourier basis~$F$ by comparing its numerically computed inverse with the SGFA directly computed GFT, and, vice-versa, comparing the directly computed graph Fourier basis with the (numerically computed) inverse of the matrix of (left) eigenvectors.

To analyze tradeoffs between accuracy and stability and to confirm the quality of the diagonalization and of the approximations to~$F$, we carry out a number of studies:
\begin{inparaenum}[1)]
\item We evaluate how close $F \cdot F^{-1}$ and $F^{-1}\cdot F$ are to the identity matrix~$I$, where $F^{-1}$ is the numerically computed inverse of~$F$. If SGFA computes accurately a stable~$F$, then $F^{-1}$ is accurately computed, and $F \cdot F^{-1}$ and $F^{-1}\cdot F$ should be close to~$I$.
        \item We test how close the columns of~$F$ are to eigenvectors of~$A$ by comparing $A\cdot F$ to $F\cdot\Lambda$, where $\Lambda$ is the diagonal matrix of the graph frequencies (eigenvalues of~$A$). This is carried out both in terms of how close the magnitudes of $A\cdot f_i$ and $\lambda_i f_i$ are, where $f_i$ is the eigenvector of~$A$ corresponding to the eigenvalue $\lambda_i$, and in terms of how aligned $A\cdot f_i$ and $f_i$ are (how small the angle between the two vectors is).
    \item We compare how close the inverse of the SGFA computed graph Fourier basis~$F$  is to the (Hermitian of the) matrix of directly computed left eigenvectors of~$A$ and vice-versa.
            \end{inparaenum}
\textcolor{black}{
We carry out extensive empirical studies with two real world graphs, the political blogs network~\cite{political_blog_data} and the Manhattan Road Map \cite{dataset_road_2,dataset_road}. The random models of figure~\ref{fig:tails_erdos_renyi} are also considered in section~\ref{section:random_graph_experiments}.}


We summarize the remaining of the paper. In section~\ref{sec:GSP}, we briefly review GSP concepts and present two interesting graphs arising in applications. Section~\ref{problem_formulation} introduces the problem of computing numerically the graph Fourier basis~$F$ for generic graphs~$A$. Section~\ref{sec_SGFA} introduces, motivates, and analyzes the Stable Graph Fourier basis Approximation~(SGFA) algorithm. We prove that the accuracy of our approximation to~$F$ improves (at least) exponentially, per iteration. In sections~\ref{sec_political} and~\ref{sec_road}, we apply SGFA to compute the graph Fourier basis~$F$ and the GFT for the $2004$ US political blogs graph \cite{political_blog_data} and for the Manhattan Road Map \cite{dataset_road_2,dataset_road} and carry out a comprehensive empirical study to illustrate accuracy, stability, and tradeoffs of the SGFA. \textcolor{black}{The random models of figure~\ref{fig:tails_erdos_renyi} are studied in section~\ref{section:random_graph_experiments}.} Section~\ref{sec:freq_order} displays examples of low, medium, and high frequency of the graph spectral components of the Manhattan road map as computed by SGFA presented in the paper. Finally, section~\ref{sec:conclusion} concludes the paper.
%
%
\section{Graph Signal Processing}
\label{sec:GSP}
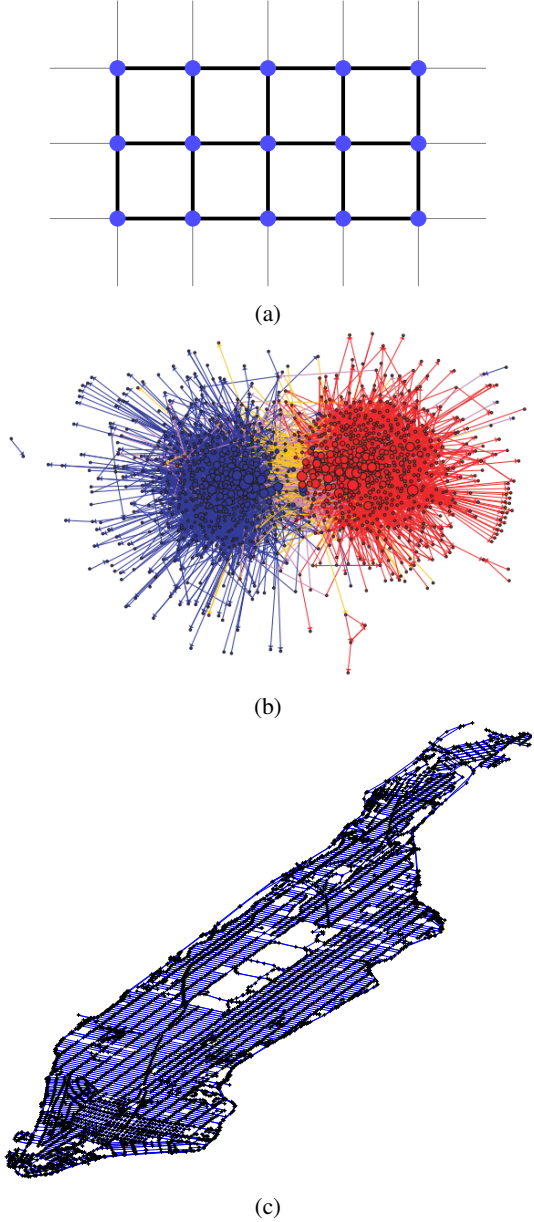
\begin{figure}[htbp]
	\begin{subfigure}{0.5\textwidth}
		\centering
\begin{tikzpicture}
\draw[step=1cm,gray,very thin]
 (-0.9,-0.9) grid (4.9,2.9);
  \draw [line width=0.5mm, -] (0,0) -- (1,0);
  \draw [line width=0.5mm, -] (1,0) -- (2,0);
  \draw [line width=0.5mm, -] (2,0) -- (3,0);
  \draw[line width=0.5mm, -](3,0) -- (4,0);

    \draw [line width=0.5mm, -] (0,1) -- (1,1);
    \draw [line width=0.5mm, -] (1,1) -- (2,1);
    \draw[line width=0.5mm, -] (2,1) -- (3,1);
    \draw[line width=0.5mm, -] (3,1) -- (4,1);

    \draw [line width=0.5mm, -] (0,2) -- (1,2);
    \draw [line width=0.5mm, -] (1,2) -- (2,2);
    \draw [line width=0.5mm, -] (2,2) -- (3,2);
    \draw [line width=0.5mm, -] (3,2) -- (4,2);

    \draw[line width=0.5mm, -](0,0) -- (0,1);
    \draw [line width=0.5mm, -] (1,0) -- (1,1);
    \draw [line width=0.5mm, -] (2,0) -- (2,1);
    \draw [line width=0.5mm, -] (3,0) -- (3,1);
    \draw [line width=0.5mm, -] (4,0) -- (4,1);

    \draw [line width=0.5mm, -] (0,1) -- (0,2);
    \draw[line width=0.5mm, -] (1,1) -- (1,2);
    \draw [line width=0.5mm, -] (2,1) -- (2,2);
    \draw [line width=0.5mm, -] (3,1) -- (3,2);
    \draw [line width=0.5mm, -] (4,1) -- (4,2);

\foreach \x in {0,1,2,3,4}{
	\foreach \y in {0,1,2}{
		\draw [fill=blue!70, color=blue!70] (\x,\y) circle (0.1cm);
	}
}
\end{tikzpicture}
	\caption{}
\label{fig_rect}
\end{subfigure}
	\begin{subfigure}{0.5\textwidth}
	\centering
	\includegraphics[width=7cm]{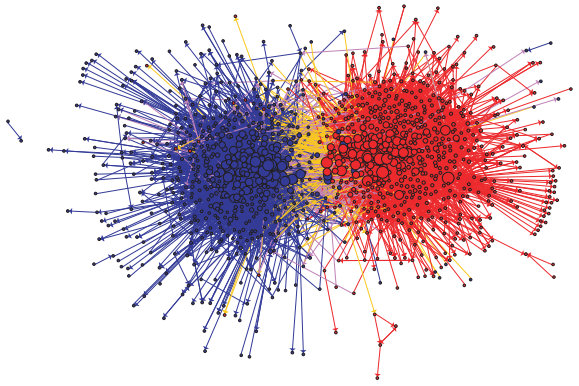}
	\caption{}
	\label{fig_political}
\end{subfigure}
\begin{subfigure}{0.5\textwidth}
\centering
\includegraphics[width=7cm]{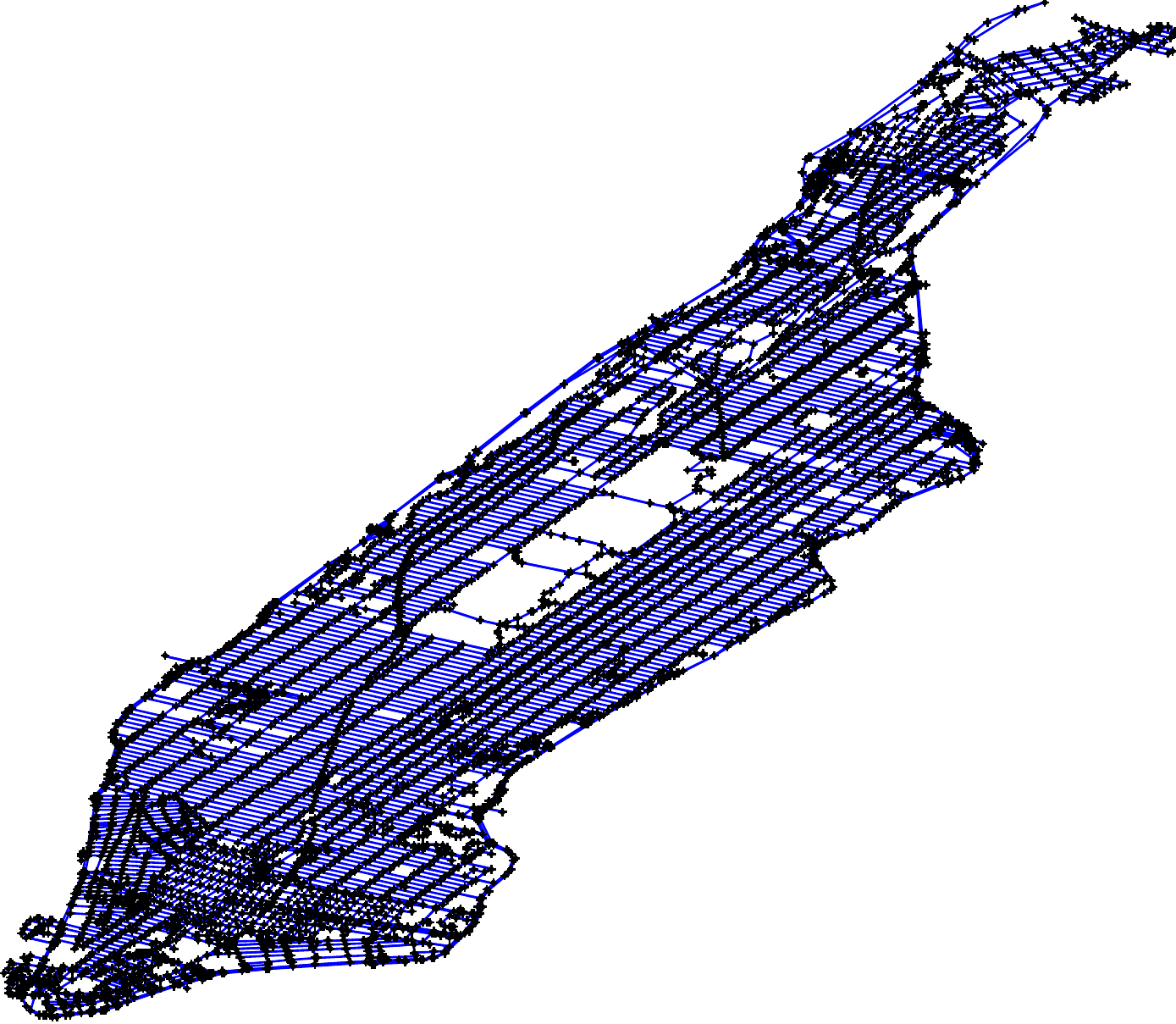}
\caption{}
\label{fig_road_net}
\end{subfigure}
\caption{Different graphs in the context of GSP.~\textbf{(a)} A 2D rectangular lattice indexing pixels in a digital image;~\textbf{(b)} graph of political blogs; and~\textbf{(c)} Manhattan road map.}
\label{fig:graphs}
\end{figure}

Graph Signal Processing (GSP)  \cite{GSP_1,ShumanNFOV:13,GSP_frequencyanalysis,GSP_Big_Data}, see also overview \cite{ortegafrossardkovacevicmouravandergheynst-2018}, considers the problem of analyzing and processing signals indexed by the nodes of a graph $\mathcal{G}=(\mathcal{V},A)$. We will follow \cite{GSP_1,GSP_frequencyanalysis,GSP_Big_Data} that applies to general graphs~$\mathcal{G}$, directed or undirected, while \cite{ShumanNFOV:13} considers undirected graphs. The set of nodes is $\mathcal{V}=\{v_0,\ldots,v_{n-1}\}$ and the $n\times n$ graph weighted adjacency matrix is~$A$. Entry $(i,j)$ of~$A$ represents a directed edge between nodes $v_i$ and $v_j$ with weight $A_{i,j}$. The graph signal~$s$ is a complex valued vector $s=\left[s_0\cdots s_{n-1}\right]^T\in \mathcal{C}^n$ such that the $i$th-component $s_i$ of~$s$ corresponds to the element of $s$ indexed by node $i$ in graph $\mathcal{G}$, i.e., $s$ maps nodes in graph $\mathcal{G}$ to complex numbers in $\mathcal{C}$. The application at hand dictates the structure of  $\mathcal{G}$, as highlighted in the three graphs of Figure~\ref{fig:graphs}. In figure~\ref{fig_rect}, $\mathcal{G}$ represents the graph associated with an image where each node corresponds to a pixel, and, for a common image model, each pixel value (pixel color or intensity) is related to the values of its four adjacent pixels. This relation for the graph in figure~\ref{fig_rect} is symmetric, hence all edges are undirected and have the same weight, with possible exceptions of boundary nodes that may have directed edges and/or different edge weights, depending on boundary conditions  \cite{Pueschel:08b}. Figure~\ref{fig_political} \cite{political_blog_data} represents a directed network, where each node corresponds to a political blog and the edges represent hyperlinks between blogs. The colors of the nodes in figure~\ref{fig_political} are the graph signal and represent the political orientation of the blog (red for conservative, blue for liberal). Orange edges go from liberal to conservative, and purple ones from conservative to liberal; blue and red edges hyperlink blogs of the same color. The size of each blog reflects the number of other blogs that link to it, as according to~\cite{political_blog_data}. Figure~\ref{fig_road_net} models the Manhattan Road Map where each node (in black) corresponds to a two dimensional coordinate location (latitude, longitude) \cite{dataset_road} and the directed edges (in blue) represent one- or two-way streets between locations, as verified by Google Maps~\cite{dataset_road_2}.
 In GSP \cite{GSP_1,GSP_frequencyanalysis,GSP_Big_Data}, the adjacency matrix~$A$ plays the role of the shift operator $z^{-1}$ in DSP. Its eigendecomposition is
\begin{align}
\label{eqn:shiftA-1}
A{}&=F\Lambda F^{-1},
\end{align}
assuming~$A$ is not defective. We let
\begin{align}
\label{eqn:shiftA-2}
F{}&=\left[f_0\cdots f_{n-1}\right]\\
\label{eqn:shiftA-3}
\Lambda{}&=\textrm{diag}\left[\lambda_0,\cdots,\lambda_{n-1}\right]\\
\label{eqn:first-b}
F^{-1}{}&=W^H=\left[\begin{array}{c}
w_0^H\\
\vdots\\
w_{n-1}^H
\end{array}
\right].
\end{align}
From~\eqref{eqn:shiftA-1}, \eqref{eqn:shiftA-2}, and~\eqref{eqn:shiftA-3}, we obtain
\begin{align}
\label{eqn:first-1a}
AF{}&=F\Lambda\\
\label{eqn:first}
Af_i{}&=\lambda_i f_i, \:\: i=0,\cdots,n-1.
\end{align}
Likewise, from~\eqref{eqn:shiftA-3}, \eqref{eqn:shiftA-1}, and~\eqref{eqn:first-b}, we get
\begin{align}
\label{eqn:second-1a}
F^{-1}A{}&=\Lambda F^{-1} \:\left(\textrm{or   } W^HA{}=\Lambda W^H\right)\\
\label{eqn:second}
w_j^H A{}&=\lambda_j w_j^H, \:\: j=0,\cdots,n-1.
\end{align}
Matrix~$\Lambda$ is the diagonal matrix of the eigenvalues $\lambda_0, \cdots, \lambda_{n-1}$ of~$A$, $F$ is the matrix of (right) eigenvectors $f_0,\cdots, f_{n-1}$ of~$A$, and $F^{-1}=W^H$ is the matrix of (left) eigenvectors\footnote{By convention, the left eigenvectors of~$A$ are the columns of the matrix~$W$ and not the rows of~$W^H$.} $w_0,\cdots, w_{n-1}$ of~$A$.

The eigenvectors $f_0,\cdots, f_{n-1}$  (columns of~$F$) are the graph frequency or graph spectral components (corresponding to the harmonic components of time signals), and we will refer to them as the Fourier basis, and the eigenvalues $\lambda_0, \cdots, \lambda_{n-1}$ are the graph frequencies\footnote{With time signals, say continuous time, it is common to call frequencies $f=\frac{1}{j2\pi}\ln \lambda=\frac{1}{j2\pi}\ln e^{j2\pi f}$. Here, the graph frequencies are the eigenvalues~$\lambda$ themselves.}. The matrix $F^{-1}=W^H$ is the graph Fourier transform~(GFT).

For a graph signal~$s$, its graph Fourier transform $\widehat{s}$  is
\begin{align}
\label{eqn:shiftA-4}
\widehat{s}{}&=F^{-1} \cdot s=W^H \cdot s\\
\label{eqn:shiftA-4a}
{}&=\left[w_0^H s\cdots w_{n-1}^H s\right]^T\\
\label{eqn:shiftA-4b}
{}&=\left[\left\langle w_0,s\right\rangle \cdots \left\langle w_{n-1},s\right\rangle\right]^T,
\end{align}
where $\left\langle w_j,s\right\rangle$ is the complex inner product of the left eigenvectors~$w_j$ and~$s$. Equation~\eqref{eqn:shiftA-4} analyzes the graph signal~$s$ in terms of its graph Fourier coefficients $\widehat{s}_0, \cdots, \widehat{s}_{n-1}$, the entries of~$\widehat{s}$.

Similarly, the graph signal~$s$ is obtained from its graph Fourier transform $\widehat{s}$ through the inverse GFT, $F$,
\begin{align}
\label{eqn:shiftA-5}
s{}&=F \cdot \widehat{s}\\
\label{eqn:shiftA-5a}
{}&=\widehat{s}_0 f_0+\cdots+\widehat{s}_{n-1} f_{n-1}.
\end{align}
Equation~\eqref{eqn:shiftA-5} shows that $\widehat{s}$ is the representation of the graph signal~$s$ on the graph Fourier basis~$F$, while~\eqref{eqn:shiftA-5a} synthesizes the graph signal~$s$ as a linear combination of the graph Fourier components $f_0,\cdots,f_{n-1}$, with the coefficients of this linear combination being the graph spectral coefficients $\widehat{s}_j$ of~$s$.

If we take the Hermitian of both sides of Equation~\eqref{eqn:shiftA-1}, we obtain the eigendecomposition of the Hermitian of~$A$
\begin{align}
\label{eqn:shiftAH-1}
A^H{}&=F^{-H}\Lambda^\star F^H\\
\label{eqn:shiftAH-2}
{}&=W\Lambda^\star F^H,
\end{align}
where $\Lambda^*$ is the conjugate of the matrix~$\Lambda$ of the eigenvalues of~$A$. From~\eqref{eqn:shiftAH-2}, it follows
\begin{align}
\label{eqn:shiftAH-3}
A^HW{}&=W\Lambda^\star.
\end{align}
 This shows that the right eigenvectors of $A^H$ are the (left) eigenvectors of~$A$ (written as column vectors). If~$A$ is real valued (which we will assume in the sequel, unless otherwise stated), $A^H=A^T$.

 Graph Signal Processing has experienced significant research activity in the last few years, a very incomplete list of references or topics include on sampling \cite{chenvarmasandryhailakovacevic-2015,annisgaddeortega-2016,marquessegarraleusribeiro-2016,tsitsverobarbarossadilorenzo-2016,tanaka-2018,shimoura-2019graph}, on extending concepts of randomness and stationarity \cite{perraudinvandergheynst-2017,marques2017stationary,pasdeloupgriponmercierpastorrabbat-2018}, on recovering the underlying graphs from data \cite{dongthanoufrossardvandergheynst-2016,segarra2016network,meimoura-2017,meimoura-2018,egilmezpavezortega-2019}, on extensions to multirate graph signal processing and wavelets \cite{tekevaidyanathan-2017,zengcheungortega-2017}, on denoising \cite{chen2014signal,onukionoyamagishitanaka-2016} and other signal reconstruction problems \cite{chen2015signal-2}, as well as many others. Not much work however has been pursued on computing the graph Fourier transform and related topics for actual real world networks, the focus of this paper.
\section{Problem Formulation}
\label{problem_formulation}
The paper considers an algorithm to compute an accurate and stable  diagonalization~\eqref{eqn:shiftA-1} of~$A$. This is important when either the matrix~$F$ of eigenvectors is theoretically not full rank ($A$ is defective), see footnote~\ref{ftn:alggeomultiplicities}, or it is numerically unstable (very small minimum singular value). Although the diagonal structure of~\eqref{eqn:shiftA-1} might not be achieved, it can be approximated to arbitrary precision as stated next.
\begin{theorem}[\cite{Horn}, Theorem 2.4.7.2]\label{thm:theorem_1}
	Any $n$-dimensional square matrix $A$ is similar to an upper triangular matrix
	\begin{equation}
\label{eqn:theorem_1}
	\begin{aligned}  \forall A,\enspace \exists \enspace  F(\epsilon),T(\epsilon):\enspace
	A=F(\epsilon) T(\epsilon) F(\epsilon)^{-1},
	\end{aligned}
	\end{equation}
	where $\epsilon>0$ denotes arbitrary precision, the eigenvalues of $A$ are on the main diagonal of $T(\epsilon)$, and $T(\epsilon)$ is upper triangular
	\begin{equation}
\label{eqn:theorem_1a}
	T_{i,i}(\epsilon)=\lambda_i(A),\enspace \enspace \left|T_{i,j}(\epsilon)\right|\leq \epsilon \text{ for } j>i.
	\end{equation}
	\label{th:theorem_triangular_approx}
\end{theorem}
By~\eqref{eqn:theorem_1a}, the triangular matrix $T(\epsilon)$ has infinitesimal small energy on the upper diagonal entries.

The proof of Theorem~\ref{th:theorem_triangular_approx} is constructive\textemdash \cite{Horn} provides closed form expression for both $F(\epsilon)$ and $T(\epsilon)$ through the complex Schur decomposition of~$A$. Although their matrix $F(\epsilon)$ is provably invertible, a simple argument shows that the minimum singular value $\sigma_{\min}$ of $F(\epsilon)$ is upper bounded by $\epsilon^{n-1}$; so, in practice, it will not be numerically stable for a sufficiently low $\epsilon$ or large enough~$n$. In words, they trade the diagonalization degree of~$T$ with the numerical stability of~$F$. To prove this, we import their construction of $T(\epsilon)$ and $F(\epsilon)$ for an $\epsilon<1$. It suffices to analyse $\sigma_{\min}\left(F(\epsilon)\right)$. Without loss of generality, we need only consider the particular instance $\epsilon<1$, since, for any scalar $\phi\geq\epsilon$, we get that if $\left|T_{i,j}\right|\leq \epsilon$ then $\left|T_{i,j}\right|\leq \epsilon\leq\phi$. Matrices $T(\epsilon)$ and $F(\epsilon)$ are \cite{Horn}
\begin{align}
\label{eqn:const_F_epsilon}
&(U,T):\enspace UU^H=I,\enspace AU=TU,\enspace T_{i,j}=0,\enspace j\leq i  \\
&t:=\underset{j>i}{\max}\enspace  \left|T_{i,j}\right|\\
& D_\theta:=\begin{bmatrix}
1 & & & \\
 & \theta &  \\
& & \ddots &  \\
& & & \theta^{n-1}\end{bmatrix} \nonumber \\
&\!\!\!(F, T)(\epsilon)\!=\!\begin{cases} \!\!\left(UD_\epsilon,\enspace D_\epsilon^{-1} T D_\epsilon\right) &\!\!\!\!\mbox{if }t\leq 1 \\
&\\
\label{eqn:const_F_epsilonb}
\!\!\left(UD_{1/t}D_\epsilon ,\enspace D_\epsilon^{-1} D_{1/t}^{-1} T  D_{1/t} D_\epsilon\right) & \!\!\!\!\mbox{if } t> 1, \end{cases}
\end{align}
\textcolor{black}{
	where $U$ is the orthogonal matrix of Schur vectors for~$A$}.
As stated in Theorem~\ref{thm:theorem_1}, the matrix $\Lambda$ of eigenvalues of~$A$ is the matrix of diagonal entries of the triangular matrix $T(\epsilon)$.

Let $t>1$ in~\eqref{eqn:const_F_epsilon}; the alternative is similarly addressed.
\begin{align}
\sigma^2_{\min}\left(F(\epsilon)\right)&= \sigma^2_{\min}\left(UD_{1/t}D_\epsilon\right) \\
&= \underset{\|x\|=1}{\min} \left\|UD_{1/t}D_\epsilon x\right\|_2^2 \nonumber \\
&\leq \|U\|_2^2 \underset{\|x\|=1}{\min}  \left\|D_{1/t}D_\epsilon x\right\|_2^2 \nonumber \\
&\leq \left\|D_{1/t}\right\|_2^2 \underset{\left\|x\right\|=1}{\min}  \left\|D_\epsilon x\right\|^2 \nonumber \\
&\leq  \sigma^2_{\min}\left(D_\epsilon\right) \nonumber \\
&= \epsilon^{2(n-1)}.
\label{eqn:bound_sigma_min}
\end{align}
The last equation shows as desired that $\sigma_{\min}\left(F(\epsilon)\right)\leq \epsilon^{(n-1)}$.  Inspired by this result, we propose to approximate the diagonal structure of~$A$ in~\eqref{eqn:shiftA-1}, while maintaining the numerical stability of the Fourier Basis as measured by the minimum singular value $\sigma_{\min}(F)$, by solving the following problem
\begin{equation}
\begin{aligned}
& \underset{F,T \in \mathbb{C}^{n\times n}}{\text{minimize}}
& &\left\|AF-F\Lambda\right\|_\mathcal{F} \\
& \text{subject to}
& & AF=FT,\enspace \enspace  \sigma_{\min}\left(F\right)\geq \alpha, \\
& & & T_{i,j}=0\enspace \enspace i<j,\enspace \enspace  T_{i,i}=\lambda_i,\\
\end{aligned}
\label{eqn:opt}
\end{equation}
where  $\|.\|_\mathcal{F}$  denotes the Frobenius norm,  $\alpha\leq 1$ is an arbitrary strictly positive constant, and  $\Lambda$ is obtained from the diagonal entries of~$T$, see~\eqref{eqn:theorem_1a}. To impose the  diagonal structure of~\eqref{eqn:shiftA-1}, we can choose other measures of accuracy like
$\sqrt{\sum_{j>i}  \left|T_{i,j}\right|^2}=\left\|T-\Lambda\right\|_\mathcal{F}$ instead of $\left\|F(T-\Lambda)\right\|_\mathcal{F}$. The algorithm of Section~\ref{sec_SGFA} is invariant to both formulations, and we can show that the same type of theoretical reasoning applies (Result~\ref{theor_2} and Theorem~\ref{theorem_exponential}).

 For generic stability margin~$\alpha$, problem~\eqref{eqn:opt} is challenging since
 \begin{inparaenum}[1)]
 \item we have a bilinear interaction between matrices $(F,T)$; and
 \item  $\sigma_{\min}$ is a non concave function.
 \end{inparaenum}
\begin{remark}\label{rmk:rescaling1}
Even though by Theorem~\ref{thm:theorem_1} any matrix is arbitrarily close to being similar to a diagonal matrix, the solution of the problem~\eqref{eqn:opt} is not simply to rescale the triangular~$T$ in Theorem~\ref{thm:theorem_1} till $\sigma_{\min}\left(F\right)\geq \alpha$. In fact, the objective in problem~\eqref{eqn:opt} is not invariant to scaling. If we scale~$F$, we obtain a feasible solution, but the objective function will become worst. In fact, assume we find a pair $(F,T)$ such that  $\sigma_{\min}\left(F\right)>0$ and  all constraints of problem~\eqref{eqn:opt} hold, but $\sigma_{\min}\left(F\right)\ngeq \alpha$. In this case, if, to solve problem~\eqref{eqn:opt}, we simply scaled~$F$ by $\gamma /\sigma_{\min}\left(F\right)$, even if $\gamma F/\sigma_{\min}\left(F\right)$ is still feasible for any $\gamma \geq \alpha$, the objective value is also scaled by $\gamma/\sigma_{\min}\left(F\right)> 1$ and so it gets worse. In summary, the solution is not simply to find an~$F$ that satisfies all constraints except $\sigma_{\min}\left(F\right)\geq \alpha$ and then rescale this~$F$.
\end{remark}
\section{Stable Accurate Approximation to the GFT}
\label{sec_SGFA}
In this section, we present a local solution to problem~\eqref{eqn:opt}, the Stable Graph Fourier Algorithm (SGFA).
\subsection{Stable Graph Fourier Algorithm (SGFA)}
\label{subsec:sgfa}
Although non-convex, problem~\eqref{eqn:opt} can be globally solved for a small $\alpha$, as we describe in the next result.
\begin{result}
	For any $\epsilon>0$, if $\alpha>0$ is sufficiently small, then problem~\eqref{eqn:opt} can be optimally solved up to $\epsilon$, by using in~\eqref{eqn:const_F_epsilonb}
\begin{align}
\label{eqn:theorem2a}
T\left(\frac{\epsilon}{\left\|F\right\|_\mathcal{F}\sqrt{ 0.5n(n-1)}} \right) \textrm {  and } F\left(\frac{\epsilon}{\left\|F\right\|_\mathcal{F}\sqrt{ 0.5n(n-1)}} \right).
\end{align}
	\label{theor_2}
\end{result}
\vspace{-1cm}
\begin{proof}
	Result~\ref{theor_2} follows since the objective in problem~\eqref{eqn:opt} is upper bounded by the off diagonal energy of matrix $T$. To be concrete, we show that the objective in problem~\eqref{eqn:opt} is upper bounded by $\epsilon$. To see this, take $T(\phi )$ and $F(\phi )$ from~\eqref{eqn:const_F_epsilonb}, and for $\phi=\frac{\epsilon}{\left\|F\right\|_\mathcal{F}\sqrt{ 0.5n(n-1)}}$, we get successively
	\begin{align}
	\left\|AF-F\Lambda\right\|_\mathcal{F}&=\left\|F(T-\Lambda)\right\|_\mathcal{F} \\
	&\leq \left\|F\right\|_\mathcal{F} \left\|T-\Lambda\right\|_\mathcal{F}  \nonumber\\
	&\leq  \left\|F\right\|_\mathcal{F} \sqrt{0.5n(n-1) \phi^2} \nonumber\\
	&=\epsilon. \nonumber
	\end{align}
The third inequality follows from the second because the $.5n(n-1)$ upper elements of the triangular matrix~$T$ are bounded by~$\phi$ and from the assumed expression for~$\phi$. \textcolor{black}{The proof would be done if $\epsilon$ was a free variable. Note however that, by~\eqref{eqn:bound_sigma_min}, there exists a lower bound on $\epsilon$ namely $\epsilon\geq \alpha^{1/(n-1)} ||F||_\mathcal{F} \sqrt{0.5n(n-1)}$. Since $\alpha$ is assumed arbitrarily small, the bound $ \alpha^{1/(n-1)} ||F||_\mathcal{F} \sqrt{0.5n(n-1)}$ can be made arbitrarily small and the optimality result follows. }
\end{proof}
Our heuristic for problem~\eqref{eqn:opt} is based on Result~\ref{theor_2}: if we find a point $(F,T)$ such that the off diagonal energy of $T$ is arbitrarily small and matrix $F$ is numerically stable, then $(F,T)$ is (close to being) globally optimal for problem~\eqref{eqn:opt}. We then propose the next simple iterative scheme. Start with a feasible initial point $\left(F_0,T_0\right)$ and proceed by updating both variables as follows:
\begin{inparaenum}[1)]
\item contract the upper diagonal energy of $T$; and
\item compute a Fourier Basis~$F$, compliant with~$T$, that has maximal $\sigma_{\min}$.
\end{inparaenum}
Contraction is obtained by simply multiplying the off diagonals elements of $T_k$ by a factor $\beta<1$.

Maximizing the minimum singular value of an arbitrary matrix is challenging, since its $\sigma_{\min}$ is a non-concave function. Instead, we consider the following general concave bound \cite{Horn}
\begin{align}
\label{eqn:sigma_min_inequality}
\hspace{-.5cm}\forall \left(F_k,F_{k+1}\right)\!\!:\!  \sigma_{\min}\left(F_{k+1}\right) \!\!\geq & \sigma_{\min}\!\left(F_k\right)\!-\!\left\|F_{k+1}-F_k\right\|_\mathcal{F}\!,
\end{align}
where $F_k$ is the (constant matrix of) Fourier Basis from iteration~$k$. Hence, the Fourier Basis $F_k$ is updated as follows
\begin{equation}
\begin{aligned}
F_{k+1}\in \enspace
& \underset{F}{\arg \max}
& &\sigma_{\min}\left(F_k\right)-\left\|F_k-F\right\|_\mathcal{F}\\
& \text{subject to}
& & AF=FT_{k+1},
\end{aligned}
\label{eqn:opt_update_F}
\end{equation}
which is equivalent to
\begin{equation}
\begin{aligned}
F_{k+1}\in \enspace
& \underset{F}{\arg \min}
& &\left\|F_k-F\right\|_\mathcal{F}\\
& \text{subject to}
& & AF=FT_{k+1}.
\end{aligned}
\label{eqn:opt_update_F_2}
\end{equation}
 Remark that problem~\eqref{eqn:opt_update_F_2} is always feasible and, hence, the iterative scheme is well posed for any iteration number $k$. Regardless of matrix $A$, one possible starting point $\left(F_0,T_0\right)$ comes from the complex Schur decomposition of $A$:
		\begin{equation}
A=F_0 T_0 F_0^H,\enspace F_0^HF_0=I_n,\enspace \left\{T_0\right\}_{ij}=0\enspace i>j.
\label{eqn:triang}
\end{equation}
The pair $\left(F_0,T_0\right)$ is feasible for problem~\eqref{eqn:opt}, since $\sigma_{\min}\left(F_0\right)=1\geq \alpha$ and the diagonal elements of $T_0$ correspond to eigenvalues of~$A$. SGFA is in algorithm~\ref{alg:SGFA_alg} where the numerical tolerance $\alpha$ defines the stopping criteria.
  \begin{algorithm}[htbp]
	\caption{Stable Graph Fourier Algorithm-SGFA}\label{alg:SGFA_alg}
	\begin{enumerate}
		\item Input Parameters: contraction factor $0<\beta<1$ and stability tolerance $0<\alpha\leq 1$.
		\vspace{0.2cm}
		\item Compute the complex triangular decomposition of matrix $A$, i.e., compute $F_0,T_0$ such that~\eqref{eqn:triang} holds.
			\vspace{0.2cm}
		\item \label{alg:step3} Update $T_k$ by contracting the upper diagonal elements:
		\begin{equation}
		\left\{T_{k+1}\right\}_{i,j}=\begin{cases}
		\beta \left\{T_{k}\right\}_{i,j}, & \text{for } j>i\\
		\phantom{\beta}\left\{T_{k}\right\}_{i,j} & \text{otherwise}
		\end{cases}.
		\end{equation}
		\item \label{alg:step4} Update $F_k$ by solving optimization problem:
		\begin{equation}
		\begin{aligned}
		F_{k+1} \in \enspace
		& \underset{F\in \mathbb{C}^{n\times n}}{\text{arg min}}
		& & \left\|F-F_k\right\|_\mathcal{F} \\
		& \text{subject to}
		& & AF=FT_{k+1}, \label{eqn:opt_inner_step}
		\end{aligned}
		\end{equation}
        through an iterative solver.
        \vspace{.25cm}
		\item \label{alg:step5} Update $\left(T_k,F_k\right)$ until $\sigma_{\min}\left(F_k\right)<\alpha$.
	\end{enumerate}
\end{algorithm}
The next theorem proves that SGFA exponentially decreases the objective of problem~\eqref{eqn:opt}, by exploring the orthogonal projection nature of the update $F_{k+1}$.
\begin{theorem}
	If SGFA runs for~$N$ iterations, the objective of problem~\eqref{eqn:opt} decays, at least, exponentially fast in~$N$,
	\begin{align}
	\label{eqn:bound_objective}
	\left\|AF_N-F_N\Lambda\right\|_\mathcal{F}&\leq \beta^{N} \left\|T_0-\Lambda\right\|_\mathcal{F} \left\|F_0\right\|_\mathcal{F}.
	\end{align}
	\label{theorem_exponential}
\end{theorem}
\vspace*{-1cm}
\begin{proof}
	Bound~\eqref{eqn:bound_objective} follows from  $\left\|F_{k+1}\right\|_\mathcal{F}\leq \left\|F_{k}\right\|_\mathcal{F}$. To show this, note that $F_{k+1}$ corresponds to the orthogonal projection of $F_k$ on the linear subspace $\Phi_{k+1}$ defined by
	\begin{equation}
\label{eqn:subspacePhi}
	\Phi_{k+1}:=\left\{F:AF=FT_{k+1}\right\}.
	\end{equation}
	 Hence, $F_k$ can be uniquely decomposed as
\begin{align}
\label{eqn:orthogonaldecompFk}
F_k=F_{k+1}+W_{k+1}
\end{align}
 where $F_{k+1}\in\Phi_{k+1}$ and $W_{k+1}$ belongs to the orthogonal complement of $\Phi_{k+1}$, i.e., $	W_{k+1}\in\Phi_{k+1}^\bot $. By the orthogonality between $W_{k+1}$ and $F_{k+1}$, we can conclude that
	\begin{align}
	\label{eqn:proof_aux_1}
	0&\leq \left\|F_{k+1}-F_k\right\|_\mathcal{F}^2 \\
	  &= \left\|F_{k+1}\right\|_\mathcal{F}^2+\left\|F_k\right\|_\mathcal{F}^2 -\text{trace}\left(F_{k+1}^HF_k\right)-\text{trace}\left(F_{k}^HF_{k+1}\right)  \nonumber \\
	  &= \left\|F_{k+1}\right\|_\mathcal{F}^2+\left\|F_k\right\|_\mathcal{F}^2-2\,\text{trace}\left(F_{k+1}^HF_{k+1}\right) \nonumber \\
	  &= \left\|F_k\right\|_\mathcal{F}^2-\left\|F_{k+1}\right\|_\mathcal{F}^2.
	\label{eqn:proof_aux_2b}
	\end{align}
The first equality follows by expanding the right-hand-side of the first inequality and recalling that the inner product of two matrices is given by the trace, and the third equality follows by recognizing that by orthogonality $\textrm{trace}\left(F_{k+1}^HW_{k+1}\right)=0$. 	We now show that the exponential bound~\eqref{eqn:bound_objective} is a consequence of~\eqref{eqn:proof_aux_1}. By step~\ref{alg:step4} of the SGFA, algorithm~\ref{alg:SGFA_alg}, $AF_{k}=F_kT_{k}$. Then
	\begin{align} \left\|AF_{k}-F_{k}\Lambda\right\|_\mathcal{F}&=\left\|F_k\left(T_{k}-\Lambda\right)\right\|_\mathcal{F} \\
	&\leq \left\|T_{k}-\Lambda\right\|_\mathcal{F}\left\|F_k\right\|_\mathcal{F} \nonumber \\
	&=   \beta^{k} \left\|T_0-\Lambda\right\|_\mathcal{F}\left\|F_k\right\|_\mathcal{F} \nonumber \\
	&\leq \beta^{k} \left\|T_0-\Lambda\right\|_\mathcal{F}\left\|F_0\right\|_\mathcal{F}. \nonumber
	\end{align}
The last inequality follows from step~\ref{alg:step3} of the SGFA, algorithm~\ref{alg:SGFA_alg}, and~\eqref{eqn:proof_aux_2b}.
\end{proof}
\vspace*{-1cm}
\textcolor{black}{\begin{example}[Algorithm~\ref{alg:SGFA_alg}: Tradeoffs]\label{exp:algorithm1} We illustrate with an analytical example the tradeoffs between the algorithm parameters~$\beta$ and~$\alpha$, and the error $\left\|AF_k-F_k\Lambda\right\|_\mathcal{F}$ as a function of the iteration number~$k$. Consider the nilpotent graph shift
		\begin{align}
	A=\begin{bmatrix}
	0 & 1 &  &  \\
	 & 0 & \ddots &  \\
	  &   & \ddots & 1  \\
 &  &  & 0  
	\end{bmatrix}\in \mathbf{R}^{n\times n}
		\end{align} 
		that has eigenvalue zero with algebraic multiplicity~$n$ and geometric multiplicity~$1$. This graph shift is a $n\times n$ Jordan block associated with $\lambda=0$ and, so, it is not diagonalizable. We apply algorithm~\ref{alg:SGFA_alg} and derive a closed form expression for the approximate Fourier basis $F_k$ of~$A$. Since~$A$ is upper triangular, the initialization step of algorithm~\ref{alg:SGFA_alg} is
$	A\,F_0= T_0\,F_0
$ with	$T_0=A$, $F_0=I_n$. The iterates $\left\{T_k,F_k \right\}_{k=1}^{+\infty}$ are
	\begin{align}
	T_k&=\begin{bmatrix}
	0 & \beta^k &  &  \\
	& 0 & \ddots &  \\
	&   & \ddots & \beta^k  \\
	&  &  & 0
	\end{bmatrix}
		\label{eqn:simple_example_T}
		\\
	F_k&=\underset{F}{\arg \min}\left\{\,\left\|F-F_{k-1}\right\|_\mathcal{F},\enspace A\,F=F\,T_k\right\}
		\label{eqn:simple_example_F}
		\\
	&=	\gamma_k\, \textrm{diag}\left[1,\beta^k,\cdots,\beta^{k(n-1)}\right]\
	\label{eqn:simple_example_F_result}\\
\gamma_k:&=\displaystyle \prod_{i=1}^{k} \frac{1+\beta^{2i-1}+\dots+\beta^{2i\,(n-1)-(n-1)}}{1+\beta^{2i}+\dots+\beta^{2i\,(n-1)}}
\label{eqn:simple_example-gammak}
	\end{align}
	Result~\eqref{eqn:simple_example_F_result} follows from the structure of~$A$ and the quadratic nature of~\eqref{eqn:simple_example_F},
\begin{align}
\begin{aligned}
& \underset{F}{\text{minimize}}
& & \left\|F-F_{k-1}\right\|_\mathcal{F} \\
& \text{subject to}
& & A\,F=F\,T_k\\
&&\Leftrightarrow\\
& \underset{F}{\text{minimize}}
& & \sum_{i,j=1}^n \Big(F_{i,j}-\{F_{k-1}\}_{i,j}\Big)^2  \\
& \text{subject to}
& & \Big\{F_{i,j}\Big\}_{i,j=2,\dots,n}
=\beta^k\,\Big\{F_{i,j}\Big\}_{i,j=1,\dots,n-1}
\label{eqn:update_F_trivial_A_derivation}
\end{aligned}
\end{align}
~\\
Since $F_{0}$ is diagonal, it follows that $F_k$ will remain diagonal because in problem~\eqref{eqn:update_F_trivial_A_derivation} the objective is minimized by selecting $F_{i,j}=0$ for $i\neq j$ (which is feasible). Given that $F_{i,i}=\beta^{k\,(i-1)}\,F_{1,1}$ for $i=2,\dots,n$, the convex objective~\eqref{eqn:update_F_trivial_A_derivation} only needs to be optimized with respect to the scalar variable $F_{1,1}$. The optimal $F_{1,1}$ 
 given by
\begin{align}
F_{1,1}= \frac{ \{F_{k-1}\}_{1,1}+\beta^k\, \{F_{k-1}\}_{2,2}+\dots+\beta^{k(n-1)}\, \{F_{k-1}\}_{n,n} } {1+\beta^{2k}+\dots+\beta^{2k(n-1)}}.
\label{eqn:recursion_F_1_1}
\end{align}
Expression~\eqref{eqn:simple_example_F_result} follows by setting $F=F_k$ in~\eqref{eqn:recursion_F_1_1} and unrolling the resulting recursion. The factor $\gamma_k$ in~\eqref{eqn:simple_example-gammak} will converge to an $\ell\geq 1$ since $\beta<1$. To show this, first note that
\begin{align}
\gamma_k\leq \prod_{i=1}^{k}\{ {1+\beta^{2i-1}+\dots+\beta^{2i\,(n-1)-(n-1)}}\}:=\hat{\gamma}_k
\end{align}
since the denominator of~\eqref{eqn:simple_example-gammak} is upper bounded by one. To show that ${\gamma}_k$ converges, we show instaed that $\hat{\gamma}_k$ converges. Note
\begin{align}
\log\{\hat{\gamma}_k\}&=\sum_{i=1}^{k}\log \{1+ h_i  \}
\end{align}
with $h_i:=\beta^{2i-1}+\dots+\beta^{2i\,(n-1)-(n-1)}$. Let us compare the two series $\sum_{i=1}^{k}\log \{1+ h_i  \}$ and $\sum_{i=1}^{k} h_i$ by the limit comparison test,
\begin{align}
\lim_{i\rightarrow +\infty} \frac{\log \{1+h_i\}}{h_i}=1,
\end{align}
since $h_i\rightarrow 0$. Hence $\log\{\widehat{\gamma}_k\}$ converges if and only if $\sum_{i=1}^{k} h_i$ converges. This last series trivially converges since we are summing $n-1$ geometric series with ratio lower than unit. So $\log\{\widehat{\gamma}_k\}$ converges and $\gamma_k$ converges to a limit~$\ell$. The limit~$\ell$ is greater than or equal to one, since the~$k$ terms being multiplied in~\eqref{eqn:simple_example-gammak} are greater than or equal to one. Using~\eqref{eqn:simple_example_F_result}, one can directly relate the approximation error $\left\|AF_k-F_k\Lambda\right\|_\mathcal{F}$ and the minimum singular value $\sigma_{\min}\left(F_k\right)$,
\begin{align}
\left\|AF_k-F_k\Lambda\right\|^2_\mathcal{F}&= \gamma_k^2\{ \beta^{2k}+\dots+\beta^{2k(n-1)}  \}
\label{eqn:approx_error_equal_min_singular_value_trivial_A}
\\
\label{eqn:approx_error_equal_min_singular_value_trivial_A-2}
\sigma_{\min}^2\left(F_k\right) &=\gamma^2_k\,\beta^{2k(n-1)}.
\end{align}
Equations~\eqref{eqn:approx_error_equal_min_singular_value_trivial_A} and~\eqref{eqn:approx_error_equal_min_singular_value_trivial_A-2}
 show for this simple example important facts:\begin{inparaenum}[1)]
	\item they give explicit expressions for both accuracy  $\left\|AF_k-F_k\Lambda\right\|_\mathcal{F}$ and stability $\sigma_{\min}^2\left(F_k\right)$;
	\item they show that the singular value $\sigma_{\min}\left(F_k\right)$ also decays exponentially fast with parameter~$\beta$, but at much faster rate than the approximation error $\left\|AF_k-F_k\Lambda\right\|_\mathcal{F}$ ($\beta^{k(n-1)}$ vs $\beta^{k}$); \item they show that $\sigma_{\max}(F_k)=1$, i.e., the condition number $\sigma_{\max}(F_k)/\sigma_{\min}(F_k)$ behaves as $1/\sigma_{\min}(F_k)$. The last two comments will be verified empirically for much broader examples.
	\item Because $\gamma_k\rightarrow \ell\geq 1$, $\lim_{k\rightarrow\infty}\sigma_{\min}\left(F_k\right) =\lim_{k\rightarrow\infty}\beta^{k(n-1)}\,\gamma_{k}=0$, showing that $F_k$ asymptotically becomes rank deficient.
	\item they impose a fundamental limit on the quality of our approximation, i.e., for any fixed stopping criteria~$\alpha$ the approximation error $\left\|AF_k-F_k\Lambda\right\|_\mathcal{F}$ cannot be made arbitrarily small. 
	\end{inparaenum}
	Indeed, for $\left\|AF_k-F_k\Lambda\right\|_\mathcal{F}$ to be arbitrarily small one would need to
	\begin{inparaenum}[a)]
	\item take infinitely many iterations~$k$, or 
	\item choose an infinitesimal small contraction factor~$\beta$.
	\end{inparaenum}
	In either case, the minimum singular value $\sigma_{\text{min}}(F_k)$ would also decay to zero, and we would get $\sigma_{\text{min}}(F_k)< \alpha$ after 
	\begin{inparaenum}[a)]
	\item sufficiently many iterations $k$, or 
	\item in the first iteration $k=1$ when $\beta$ is sufficiently small. 
	\end{inparaenum}
	So, this simple example shows that SGFA comes with a trade-off: in general, it is impossible to find a Fourier basis that is both arbitrarily stable and arbitrarily accurate. One must trade the two metrics to get an appropriate approximation that meets a stability criteria defined by $\alpha$.  
	\hfill$\small\square$        
\end{example}
}

\textcolor{black}{Sections~\ref{sec_political}, \ref{sec_road}, and ~\ref{section:random_graph_experiments} will confirm empirically the observation in example~\ref{exp:algorithm1} that there exists a tradeoff between contraction factor~$\beta$ and the minimum singular value $\sigma_{\min}\left(F_k\right)$. The sections verify empirically that, when~$\beta$ is small, a large contraction is performed in $T_k$ and $\sigma_{\min}\left(F_k\right)$ tends to decrease rapidly. This decay in $\sigma_{\min}\left(F_k\right)$ is approximately exponential for low values of $\beta$, replicating the behavior observed in the previous example. 
}
\vspace*{-.2cm}
\subsection{Updating the Fourier Basis}
\label{subsec:updatingFbasis}
To update the Fourier Basis~$F_k$ and solve problem~\eqref{eqn:opt_inner_step}, we consider the (\textcolor{black}{iterative}) LSQR solver implemented in MATLAB \cite{solver}. The method is based on the Golub-Kahan bidiagonalization process. It is algebraically equivalent to the standard method of conjugate gradient, but it has better numerical properties.

 To use LSQR, we first reformulate problem~\eqref{eqn:opt_inner_step} \textcolor{black}{and step~4 of algorithm~\ref{alg:SGFA_alg}} as follows,
\begin{equation}
\begin{aligned}
& \underset{f\in \mathbb{C}^{n^2}}{\text{minimize}}
& &\left\|f\right\|_2 \\
& \text{subject to}
& & \mathcal{A}f=b,
\end{aligned}
\label{eqn_reformulation_kron}
\end{equation}
where
\begin{align}
\mathcal{A}=I_n \otimes A - T_{k+1}^T \otimes I_{n}, \enspace b=\text{vec}\left(F_kT_{k+1}-AF_k\right).\nonumber
\end{align}
The symbol $\otimes$ denotes the Kronecker product. Formulation~\eqref{eqn_reformulation_kron} simply interprets matrix $F$ as a vector $f=\text{vec}\left(F\right)$ in $\mathbb{C}^{n^2}$, rewrites the underling matrix equality in vector form and shifts the solution by $\text{vec}(F_k)$ to get the desired format for the MATLAB solver~\cite{solver}.

LSQR has no need to store in memory matrix $\mathcal{A}\in \mathbb{C}^{n^2\times n^2}$, by defining a proxy to efficiently compute $\mathcal{A}f$ and $\mathcal{A}^Hf$. In our context, this is possible since
\begin{align}
\mathcal{A}f&=\text{vec}\left(AF-FT_{k+1}\right) \\
\mathcal{A}^Hf&=\text{vec}\left(A^TF-FT_{k+1}^H\right).
\label{eqn:A_f_macro}
\end{align}

 \textcolor{black}{We now consider the computational effort of SGFA and algorithm~\ref{alg:step3}. We start by estimating this effort for step~\ref{alg:step4} of algorithm~\ref{alg:SGFA_alg} using solver LSQR and as refomulated by~\eqref{eqn_reformulation_kron} and~\eqref{eqn:A_f_macro}. To compute $\mathcal{A}f$ and $\mathcal{A}^Hf$, we just store, multiply, and sum $n\times n $ matrices. To solve~\eqref{eqn_reformulation_kron}, by~\cite{solver}, LSQR will perform $10\,n^2$ multiplications plus computing $\mathcal{A}f$ and $\mathcal{A}^Hf$ per iteration. By~\eqref{eqn:A_f_macro}, computing $\mathcal{A}f$ and $\mathcal{A}^Hf$ involves computing $AF$ and $FT_{k+1}$  $\left(\vphantom{FT_{k+1}^H}\right.$\hspace{-.05cm}or $A^TF$ and $FT_{k+1}^H$$\left.\vphantom{FT_{k+1}^H}\right)$. We recall that in practice~$A$ is highly sparse, $F$ is dense, and $T_{k+1}$ will vary from an upper triangular to essentially a diagonal matrix (whose diagonal entries are fixed being equal to the eigenvalues of~$A$). So, computing $FT_{k+1}$ and $FT_{k+1}^H$ will require from $O\left(n^3\right)$   $\left(\vphantom{T_{k+1}}\right.$\hspace{-.05cm}when $T_{k+1}$ is triangular$\left.\vphantom{T_{k+1}}\right)$ to $O\left(n^2\right)$ $\left(\vphantom{T_{k+1}}\right.$\hspace{-.05cm}when $T_{k+1}$ is diagonal$\left.\vphantom{T_{k+1}}\right)$.
 We now consider the computational effort of calculating $AF$ and $A^TF$. In practice, graphs are highly sparse with node average degree\footnote{\textcolor{black}{For example, for \textcolor{black}{undirected} Erd\H os-R\'enyi graphs with probability of connection~$p$, the average degree is $pn$. With sharp threshold of connectedness $\frac{\ln n}{n}$, the average degree of a sparse connected Erd\H os-R\'enyi graph is of the order $\ln n$. \textcolor{black}{Recent work~\cite{directed_erdos_renyi_paper} shows that $\frac{\ln n}{n}$ is also a threshold for directed Erd\H os-R\'enyi graphs, now with probability~$p$ of being strongly connected.}}} $d\ll n$. Assuming that the average number of nonzero entries of each row of~$A$ is~$d\approx\ln n\ll n$, then the product $\mathcal{A}\,f$ $\left(\vphantom{\mathcal{A}^H\,f}\right.$\hspace{-.05cm}or $\mathcal{A}^H\,f$$\left.\vphantom{\mathcal{A}^H\,f}\right)$ has computational complexity $\mathcal{O}(dn^2)\sim \mathcal{O}\left(n^2\ln n\right)$, since matrix~$F$ is typically dense. Therefore, the overall computational cost of one iteration of LSQR varies from ${O}\left(n^3\right)$ to $\mathcal{O}\left(n^2\ln n\right)$.
 \\~\\
 \phantom{lin}To have a better estimate of the computational requirements of LSQR, we plot in figure~\ref{fig:running_times} the running time of a single iteration of LSQR (single iteration of step~\ref{alg:step4} of algorithm~\ref{alg:SGFA_alg}) when graph shift~$A$ is a directed Erd\H os-R\'enyi graph with probability of connection\footnote{\textcolor{black}{By using MATLAB$^\circledR$  {\tt{eig}} function, we always verify that $\sigma_{\min}(F)<10^{-12}$, where~$F$ denotes the eigenvector matrix of the graph shift~$A$.}} $p=\frac{1}{n}$, and matrices $\left(F_k\right)$ are complex matrices with real and complex parts sampled from an i.i.d.~standard Gaussian distribution and matrices $\left(T_k\right)$ are either triangular i.i.d.~standard Gaussian (red or top curve of figure~\ref{fig:running_times}) or diagonal (blue or bottom curve of figure~\ref{fig:running_times}). We plot these two curves since, as explained  previously, matrix $T_{k+1}$ will vary from an upper triangular to an essentially diagonal matrix as the iteration number $k$ increases. Hence, the curves of figure~\ref{fig:running_times} represent bounds on the expected computational time of a single iteration of LSQR (single iteration of step~\ref{alg:step4} of algorithm~\ref{alg:SGFA_alg}): at the beginning (initial iterations) the red (top) curve tracks the time it takes for an LSQR iteration, while for later iterations, as LSQR becomes faster, it is the blue (bottom) curve that tracks this time.
  \\~\\
  \phantom{lin}We now consider the time it takes one iteration of SGFA\textemdash step~\ref{alg:step3} (contracting $T_{k+1}$), multiple iterations of step~\ref{alg:step4} (LSQR), say~$K$, and step~\ref{alg:step5} of algorithm~\ref{alg:SGFA_alg}.
  We empirically verified that the complexity of an iteration of SGFA is essentially dominated by the LSQR steps.
  Then, the overall computational complexity of a single iteration of SGFA\footnote{\textcolor{black}{ In algorithm~\ref{alg:SGFA_alg} the complexity of contracting upper triangular matrix $T_k$ by $\beta$ -- step 3 -- is not considered since it is dominated by the other updates -- steps 4 and 5.}}, algorithm~\ref{alg:SGFA_alg}, varies from $\mathcal{O}(Kn^2\ln n+n^3)$ to $\mathcal{O}((K+1)\,n^3)$. The extra $n^3$ term comes from the computation of the stopping criterion of SGFA, step~\ref{alg:step5} of algorithm~\ref{alg:SGFA_alg}, that requires computation of the singular value of $F_k$ and will be ignored in the sequel\footnote{\textcolor{black}{Singular values are usually computed by LAPACK~\cite{anderson1999lapack} that
\begin{inparaenum}[1)]
\item \label{inpar:lapack1} reduces the square matrix $A$ to a bi-diagonal form $B$, and
    \item \label{inpar:lapack2} finds the singular values of $B$, which are equal to those of $A$.
         \end{inparaenum}
         Step~\ref{inpar:lapack1} uses Householder reductions with complexity $\mathcal{O}(n^3)$, while several alternatives exist for~\ref{inpar:lapack2} with cost $\mathcal{O}(n^2)$. We ignore this time in the discussion.}}.
\\~\\
\phantom{lin}We use figure~\ref{fig:running_times} to estimate the running time of SGFA. For example, for a graph with $n=1,500$ nodes, figure~\ref{fig:running_times} shows that a single iteration of LSQR takes from .0055~min to .14~min, while for a much larger graph, $n=10,000$, it takes from .27~min to 41.56~min. If we assume that we run $K=100$ iterations of LSQR (inner loop) to solve optimization~\eqref{eqn:opt_inner_step}, we conclude that one iteration of SGFA (with $K=100$ iterations of LSQR) takes for the graph with $n=1,500$ nodes on average from .5~min to 14~min, while the larger graph with $n=10,000$ nodes it takes from 27~min to roughly 70~h.    All numerical experiments were carried out on a personal computer with processor Intel Core i7-$2600$, CPU $3.4$~GHz, and $16$~GB of RAM\footnote{Such large graphs will be handled by more powerful computational resources. Also, the number of iterations of LSQR can be adjusted to improve the scalability of algorithm~\ref{alg:SGFA_alg}.}.
	\begin{figure}[htbp]
		\centering \includegraphics[width=8cm]{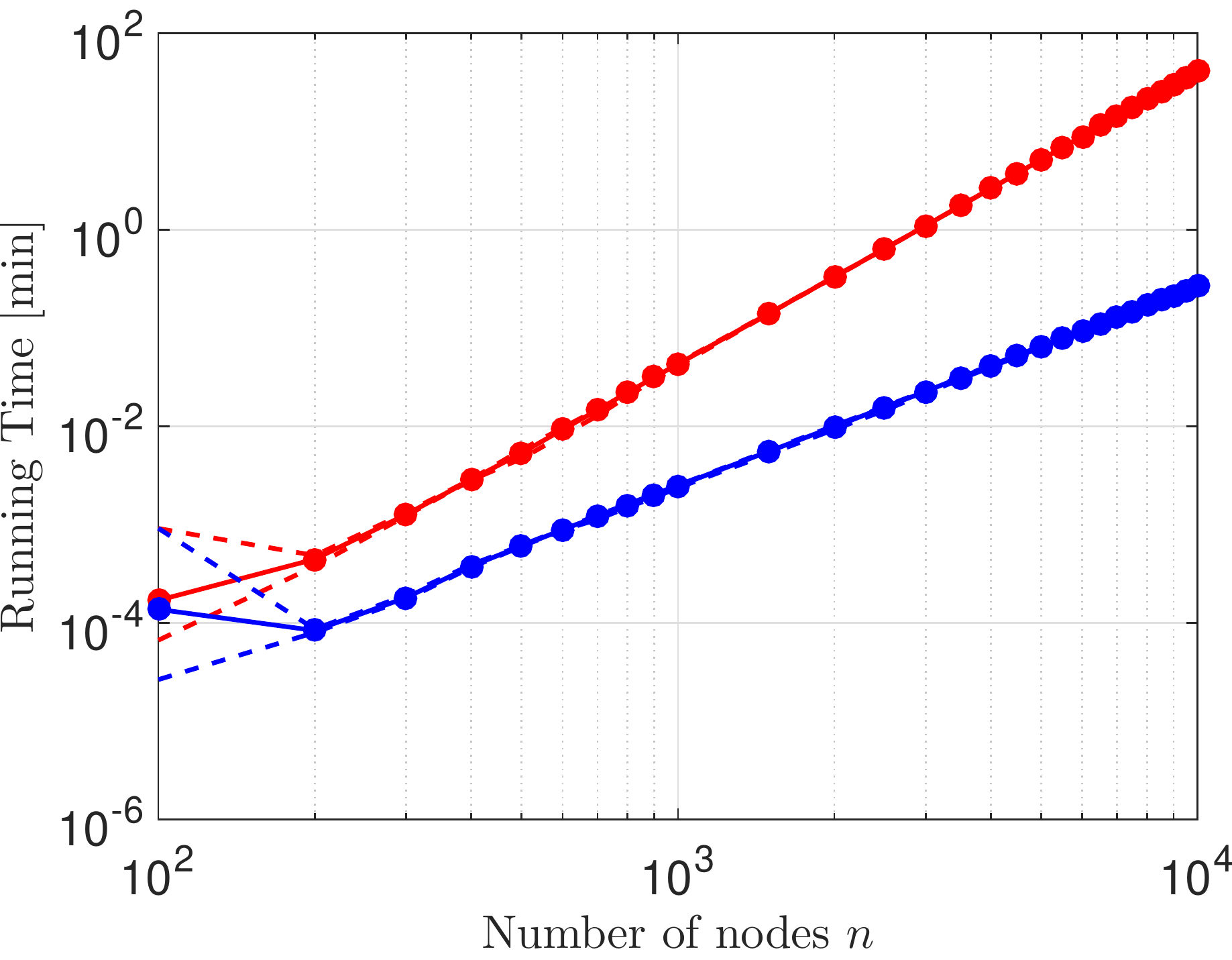}
		\caption{Running time of a single iteration of LSQR for~\eqref{eqn_reformulation_kron}: mean (dots) and minimum/maximum (dotted lines) times for $10$ random initializations. Matrix $T_{k+1}$ is upper triangular for the red plot and diagonal for the blue plot.}
		\label{fig:running_times}
	\end{figure}
	%
}
\section{Political Blogs}
\label{sec_political}
We now study empirically the SGFA, algorithm~\ref{alg:SGFA_alg}, by applying it to the shift matrix for the graph of the political blogs network.
\subsection{Dataset}\label{subsec:dataset}
 We consider the network of Figure~\ref{fig_political}\textemdash a graph of hyperlinks between $n=1,490$ weblogs, over the period of two months preceding the U.S. Presidential Election of 2004 \cite{political_blog_data}. The underlying adjacency matrix $A$ is directed and weighted: if node $v_i$ has $k$ URL references to node $v_j$ then $A_{i,j}=k$. The maximum number of URL references between two nodes is two. Matrix~$A$  is  made publicly available by Mark Newman\footnote{Visited March 2019: \def\UrlFont{\bfseries}\url{https://www.cise.ufl.edu/research/sparse/matrices/Newman/polblogs} }, together with some additional general information: $\sigma_{\min}(A)=0$ (default precision), there are $19,025$ non zero entries, and $\text{rank}(A)=784<1,490$; in this case $\lambda=0$ is a repeated eigenvalue. \textcolor{black}{By computing~$F$ with MATLAB$^\circledR$ \texttt{eig}, we find that $\sigma_{\min}(F)\approx1.1\times 10^{-33}$}. We now apply SGFA, algorithm~\ref{alg:SGFA_alg}, to find an approximation to the diagonalization of~$A$.
\subsection{Approximating a Stable Graph Fourier Transform}
\label{sec:approximating_political_blog}
In this section, we evaluate the performance of SGFA, algorithm~\ref{alg:SGFA_alg}, for different values of the threshold~$\alpha$ for $\sigma_{\min}(F_k)$ and contracting factor~$\beta$, namely, values of $(\alpha,\beta)$ on a two dimensional $6\times 14$ grid. As explained in Section~\ref{problem_formulation}, the objective is to get an accurate approximation of the Fourier Basis~$F$ (inverse GFT), while maintaining the numerical stability of~$F$. We discuss accuracy and stability.
\subsubsection{Accuracy}
\label{subsec:politicalblogaccuracy}
\begin{figure}[htbp]
	\centering
		\includegraphics[width=7.5cm]{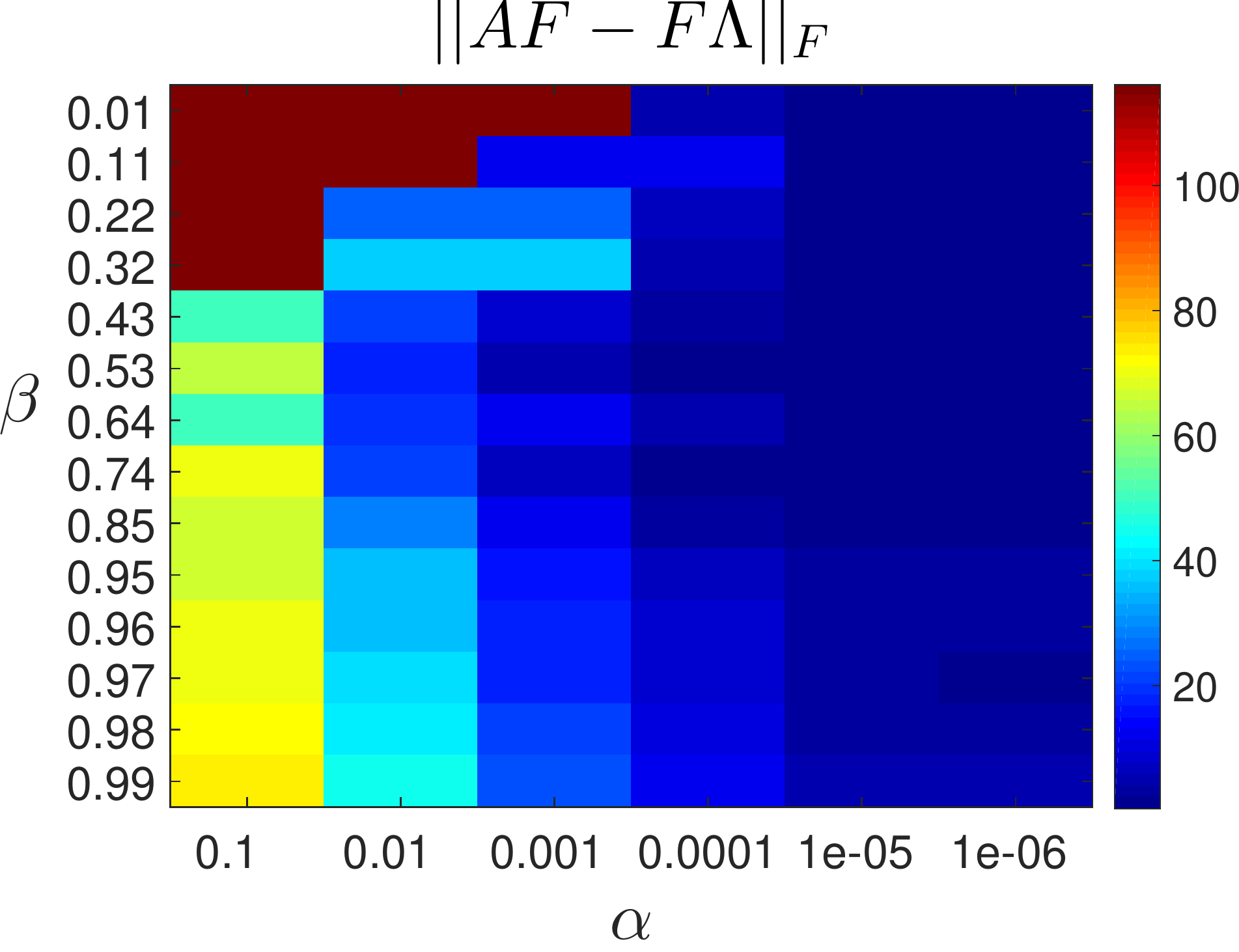}
	\caption{Political blogs network: Accuracy of SGFA, algorithm~\ref{alg:SGFA_alg}, $2$-D map as a function of algorithm parameters $(\alpha,\beta)$.
	}
	\label{fig:accuracy}
\end{figure}

Figure~\ref{fig:accuracy} plots the objective of problem~\eqref{eqn:opt}, $\left\|AF-F\Lambda\right\|_\mathcal{F}$, for the range of values of $(\alpha,\beta)$ indicated on the horizontal and vertical axes of the figure, respectively. The vertical color code bar on the right of the figure gives the values of $\left\|AF-F\Lambda\right\|_\mathcal{F}$. For the brown-red blocks on the upper left corner of the image ($.01\leq \beta\leq .32$ and $\alpha\leq 5\times 10^{-4}$), SGFA,   algorithm~\ref{alg:SGFA_alg}, only ran for a single iteration, since after the first iteration the stopping condition $\sigma_{\min}(F_1)<\alpha$ is already met; hence, the best estimate of the Fourier Basis is the starting~$F_0$ obtained by performing the complex Schur decomposition of~$A$ in~\eqref{eqn:triang}. For any value of the contracting factor~$\beta$, when the threshold~$\alpha$ decreases, the accuracy of the approximation improves as can be observed from the gradient of the colors in each row of Figure~\ref{fig:accuracy}. This is expected since, if the threshold~$\alpha$ decreases, our stability criterion is relaxed and SGFA, algorithm~\ref{alg:SGFA_alg}, runs for more iterations. \textcolor{black}{SGFA may stop before noticeable reduction of the objective. For example, for $\beta =.01$ and $\alpha\geq.001$ the algorithm stops with the approximation error still significant, see top left corner of the figure that gives $\left\|FA-\Lambda F\right\|_\mathcal{F}\geq 100$. To decrease the error objective to for example $\left\|FA-\Lambda F\right\|_\mathcal{F}\leq 20$, we can increase $\beta\geq .43$ with $\alpha\geq .01$. To reduce even more significantly the objective $\left\|FA-\Lambda F\right\|_\mathcal{F}\leq 10$, we can keep $\beta\geq .43$ but reduce $\alpha\approx 10^{-5}-10^{-6}$.}

\begin{figure}[htbp]
	\centering
	\includegraphics[width=6.5cm]{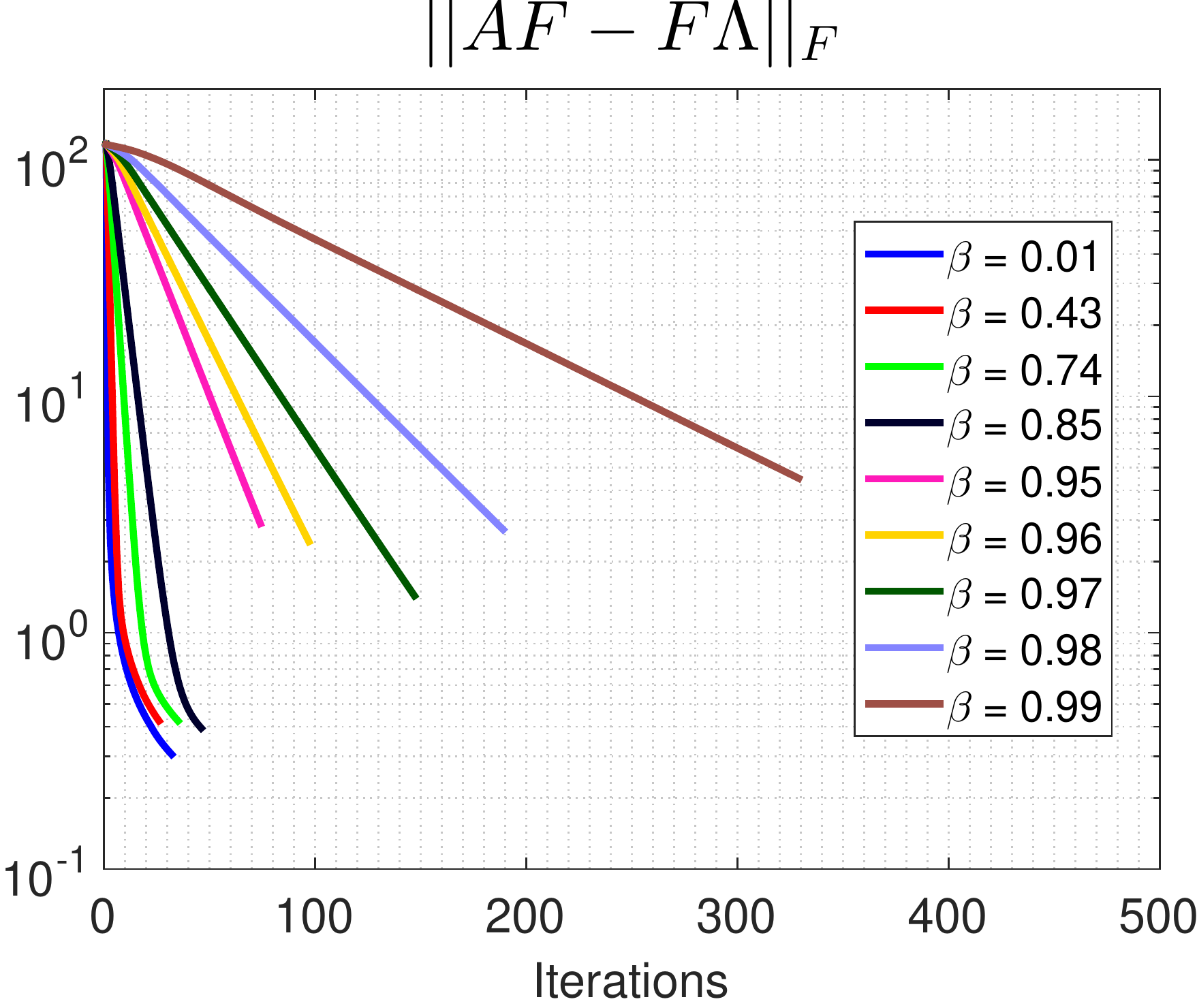}
	\caption{Political blogs network: Log scale accuracy of SGFA, algorithm~\ref{alg:SGFA_alg},  with threshold kept fixed, $\alpha=10^{-6}$.}
	\label{fig:accuracy_log}
\end{figure}

Theorem~\ref{theorem_exponential} proves that there is an exponential dependency between accuracy and the number of iterations of SGFA, algorithm~\ref{alg:SGFA_alg}. We consider this in figure~\ref{fig:accuracy_log} that plots in log scale $\left\|AF-F\Lambda\right\|_\mathcal{F}$ versus the number of iterations for the indicated values of the contracting factor~$\beta$, while maintaining fixed the threshold at $\alpha=10^{-6}$. Note the linear decay tendency of the several plots in figure~\ref{fig:accuracy_log}. It is interesting to note that for $\alpha\leq 10^{-5}$ the approximation error tends to be less sensitive to moderate values of $\beta\leq 0.85$, as seen from the corresponding plots of Figure~\ref{fig:accuracy_log} (the four left lines close to the vertical axis) that all terminate at a similar endpoint $\left\|AF-F\Lambda\right\|_\mathcal{F}\approx0.4$. This can also be concluded by observing that there are no visible significant color differences on the right of Figure~\ref{fig:accuracy}.

\begin{figure}[htbp]
	\centering
	\includegraphics[width=7cm]{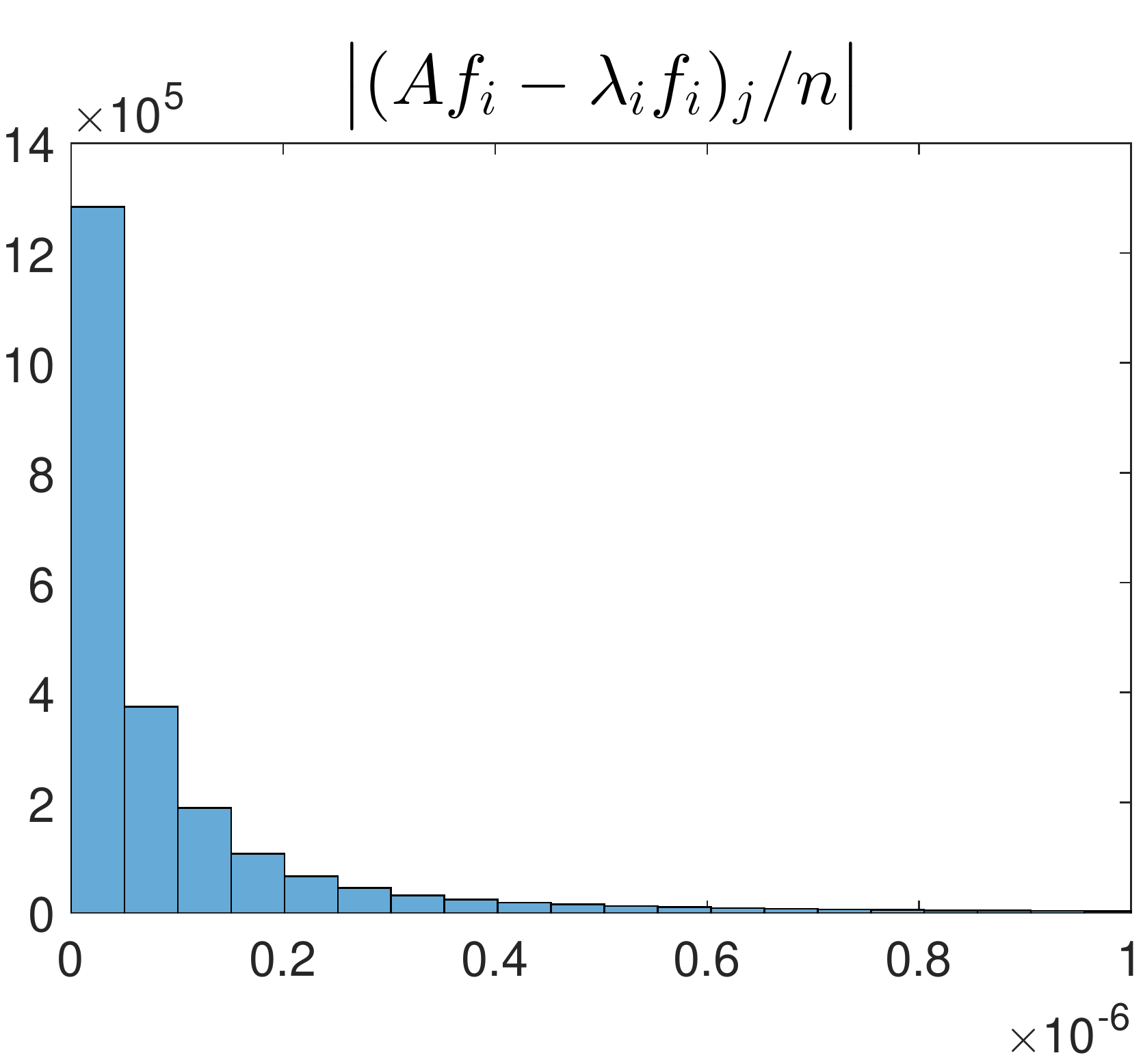}
	\caption{Political blogs network\textemdash Histogram of component wise errors: Other measure of the accuracy of SGFA, algorithm~\ref{alg:SGFA_alg}, with $(\alpha,\beta)=\left(10^{-6},0.74\right)$. The tail of the histogram was removed for visualization purposes. From the $1490^2$ data points, only $0.006\%$ had an error larger then $10^{-6}$, with maximum error $5.03\times 10^{-6}$.}
	\label{fig:accuracy_component}
\end{figure}

While Figure~\ref{fig:accuracy_log} gives the sum of the errors over all $1,490^2\approx 2.25\times 10^6$ entries, we consider now the accuracy over each normalized component of the eigenvector error given by
\begin{equation}
\left|\frac{\left(Af_i-\lambda_if_i\right)_j}{n}\right|,\enspace (i,j)\in \{1,\ldots,n\}^2.
\label{eqn:component_error}
\end{equation}
Figure~\ref{fig:accuracy_component} plots the histogram of the error~\eqref{eqn:component_error} after computing~$F$ with SGFA, algorithm~\ref{alg:SGFA_alg}, with $(\alpha,\beta)=(10^{-6},0.74)$. As can be seen, the component wise error given by~\eqref{eqn:component_error} and shown on the horizontal axis of the figure is on the order of $10^{-6}$ and, in fact, $75\%$ of all  $\left(Af_i-\lambda_if_i\right)_j$ have an absolute value lower than $10^{-7}$ (bars on the left of the figure).
\subsubsection{Stability}
\label{subsec:politicalblogstability}
\begin{figure}[htbp]
	\centering
	\includegraphics[width=6.5cm]{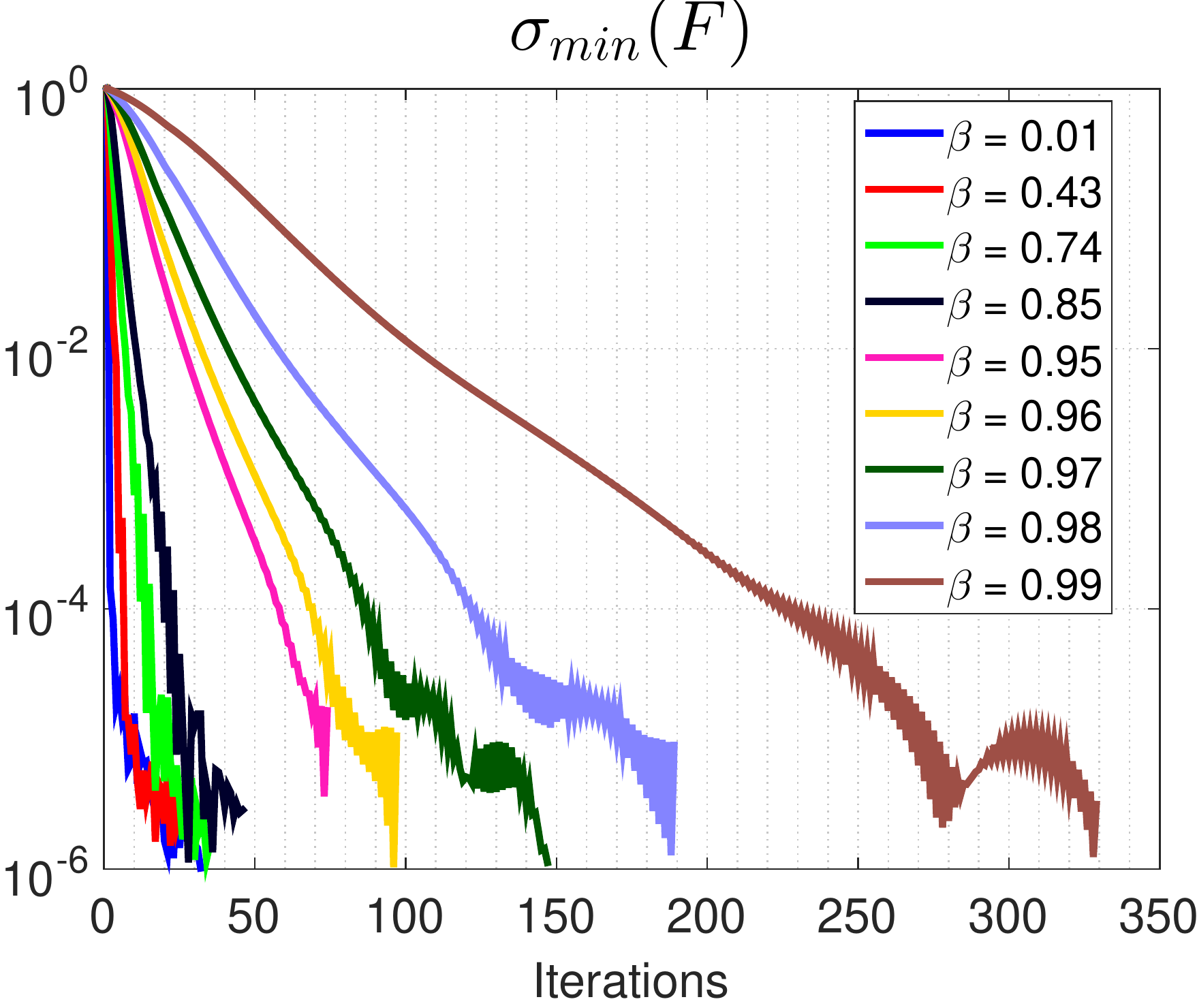}
	\caption{Political blogs network: Stability of SGFA, algorithm~\ref{alg:SGFA_alg}, as a function of number of iterations~$K$ with stopping criteria $\alpha=10^{-6}$: $\sigma_{\min}\left(F\right)$.}
	\label{fig:stability}
\end{figure}

\begin{figure}[htbp]
	\centering
	\begin{subfigure}{0.5\textwidth}
		\centering
		\includegraphics[width=6.5cm]{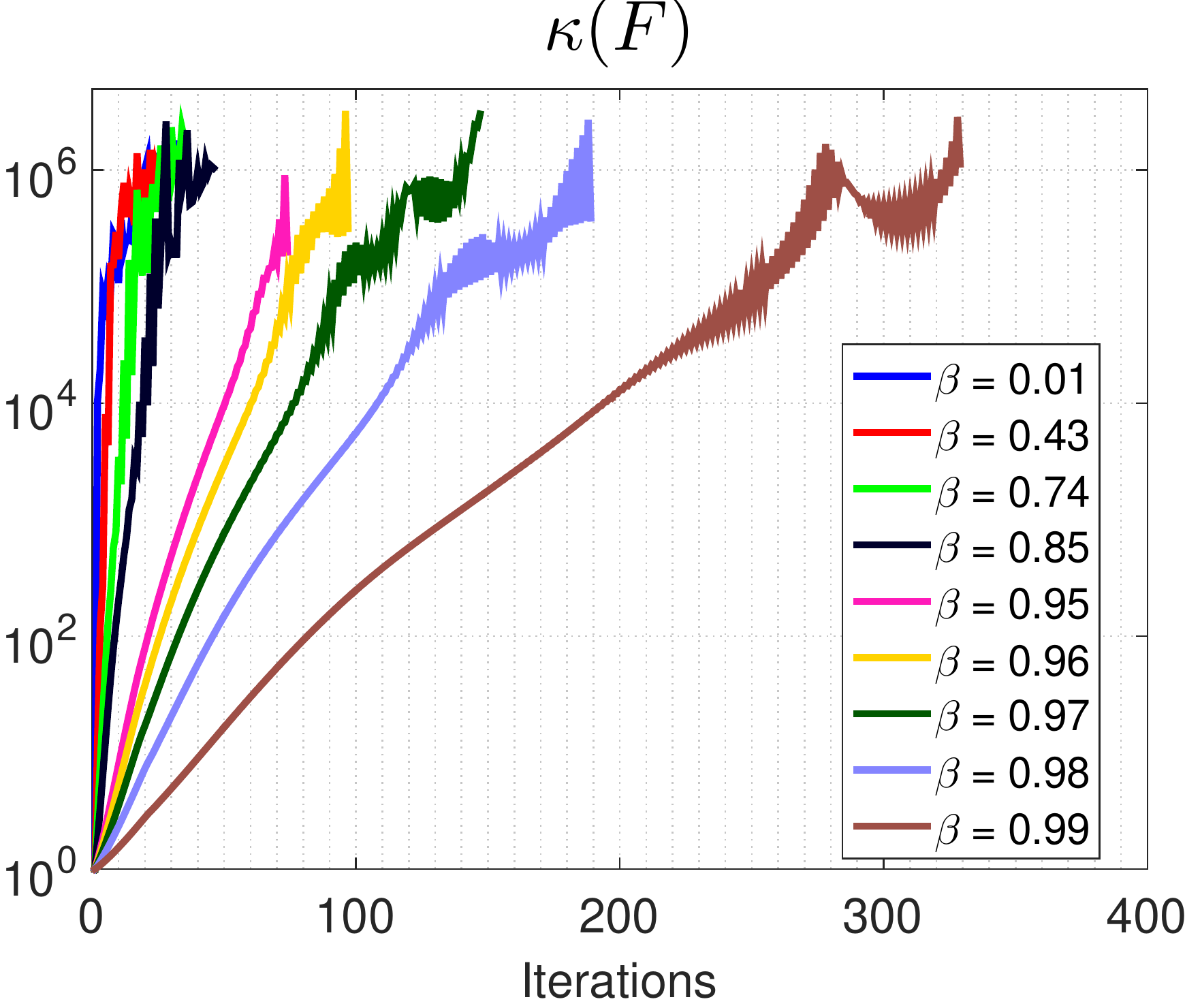}
		\caption{}
		\label{fig_a}
	\end{subfigure}
	\begin{subfigure}{0.5\textwidth}
		\centering
		\includegraphics[width=6.5cm]{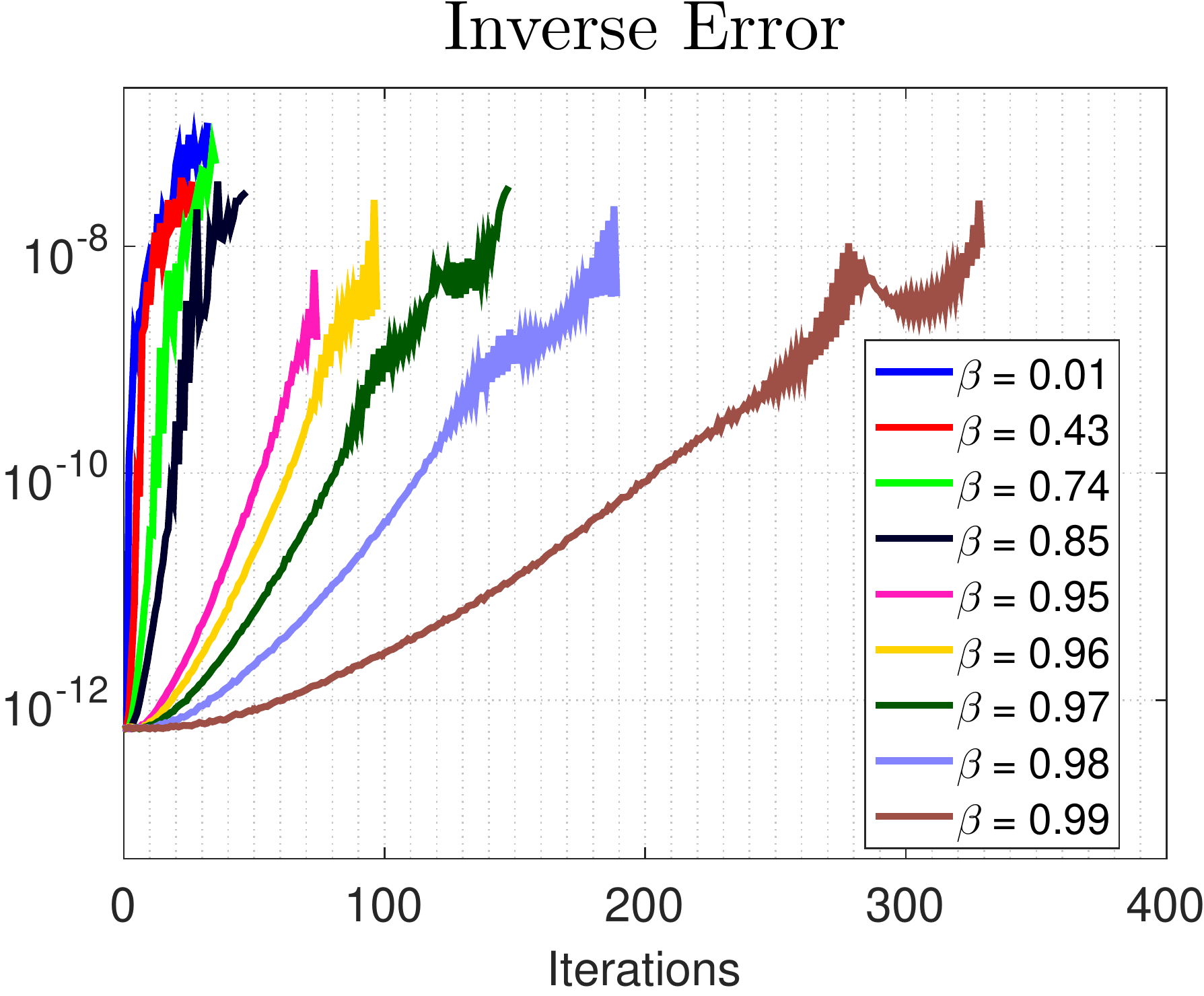}
		\caption{}
		\label{fig_b}
	\end{subfigure}
	\label{other_stability}
		\caption{Political blogs network: Other measures of stability of~$F$ as a function of the number of iterations~$K$ with the same stopping criteria $\alpha=10^{-6}$: \textbf{(a)} condition number $\kappa(F)$; \textbf{(b)} inverse error $\epsilon_{\textrm{inv}}\left(F\right)$  that quantifies the error of inverting matrix~$F$.
	}
\label{fig:other_measures_stability}
\end{figure}

Figures~\ref{fig:stability} and~\ref{fig:other_measures_stability} plot measures of numerical stability for the approximated Fourier Basis~$F$\textemdash the minimum singular value $\sigma_{\min}\left(F\right)$, the condition number $\kappa(F)$, and the inverse error $\epsilon_{\textrm{inv}}\left(F\right)$. The condition number $\kappa(.)$ in figure~\ref{fig_a} is
\begin{align}
\kappa(F)=\frac{\left\|F\right\|_2}{\sigma_{\min}\left(F\right)}.
\label{eqn:cond}
\end{align}
If the condition number $\kappa(F)$ is large the underlying matrix is close to  being singular, and, hence, \eqref{eqn:cond} provides another classical measure of numerical stability for~$F$.

The inverse error $\epsilon_{\textrm{inv}}\left(F\right)$ shown in figure~\ref{fig_b} is defined by
\begin{align}
 \epsilon_{\textrm{inv}}\left(F\right){}&=\text{max}\left\{\left\|FF^{-1}-I\right\|_\mathcal{F},\left\|F^{-1}F-I\right\|_\mathcal{F}\right\}.
 \label{eqn:inv_error}
 \end{align}
  It measures the quality of the numerically computed inverse $F^{-1}$. This is important in graph signal processing since, as it is well known \cite{GSP_1} and shown by~\eqref{eqn:shiftA-4}, $F^{-1}$ is the GFT.

  Contrasting the five lines on the right with the four left lines graphed in the three plots shown in figure~\ref{fig:stability} and  Figure~\ref{fig:other_measures_stability}, we see that the three stability measures of~$F$, $\sigma_{\min}\left(F\right)$, $\kappa(F)$, and $\epsilon_{\textrm{inv}}\left(F\right)$ all tend to converge at a much slower rate for $\beta\geq 0.85$ than with low-medium values of $\beta$, while still achieving, after a sufficiently large number of iterations, the same overall numerical stability. In practice, this suggests that one should avoid a high value of $\beta$: SGFA, algorithm~\ref{alg:SGFA_alg}, will take longer to achieve an equally stable  but less accurate Fourier basis (as shown in figure~\ref{fig:accuracy_log}).

Although the stopping criteria of SGFA, algorithm~\ref{alg:SGFA_alg}, is completely defined by $\sigma_{\min}\left(F_k\right)<\alpha$, we observe that in practice all three stability measures tend to scale linearly with the number of iterations~$K$ just as the accuracy measure $\|AF-FA\|_\mathcal{F}$ scales linearly with~$K$, see Figure~\ref{fig:accuracy}. This is more true with small values of~$\beta$. Further, from Figure~\ref{fig_a}, we observe that the limiting values of the minimum singular value and of the condition number of~$F$ tend to scale linearly with $\alpha$ or $1/\alpha$, respectively, i.e., $\sigma_{\min}\left(F\right)\approx\alpha\approx10^{-6}$ (Figure~\ref{fig:stability} or  $\kappa(F)\approx 1/\alpha\approx 10^6$ (Figure~\ref{fig_a})). This suggests that, empirically,  one can interpret the constant stopping criteria~$\alpha$ either in terms of $\sigma_{\min}$ or the condition number~$\kappa$, i.e., although SGFA, algorithm~\ref{alg:SGFA_alg}, was motivated by the convex upper bound of $\sigma_{\min}$ given in Equation~\eqref{eqn:sigma_min_inequality}, Figure~\ref{fig_a}  suggests that the termination criteria of algorithm~\ref{alg:SGFA_alg} could compare~$\kappa$ against $1/\alpha$ instead of $\sigma_{\min}$ against~$\alpha$.

  As a final comment, by numerically computing the inverse $F^{-1}$ of the approximate matrix~$F$ obtained with SGFA, algorithm~\ref{alg:SGFA_alg}, Figure~\ref{fig_b} shows that the inverse error $\epsilon_{\textrm{inv}}\left(F\right)$ measuring how far $FF^{-1}$ is away from the identity~$I$ as given by~\eqref{eqn:inv_error} is in the range $\left[10^{-7},10^{-8}\right]$ for a tolerance of $\alpha=10^{-6}$. Since the total number of entries of~$F$ is $n^2=1,490^2\approx 2.2\times 10^{6}$ this means that the error in any individual entry of $FF^{-1}$ is, on average, on the order of $10^{-13}$ to $10^{-14}$.
\subsubsection{Approximating Left vs Right Eigenvectors\textemdash Political Blogs Network}
\label{subsubsec:leftvsrighteigenvectors}
\begin{figure}[htbp]
	\centering
		\includegraphics[width=7cm]{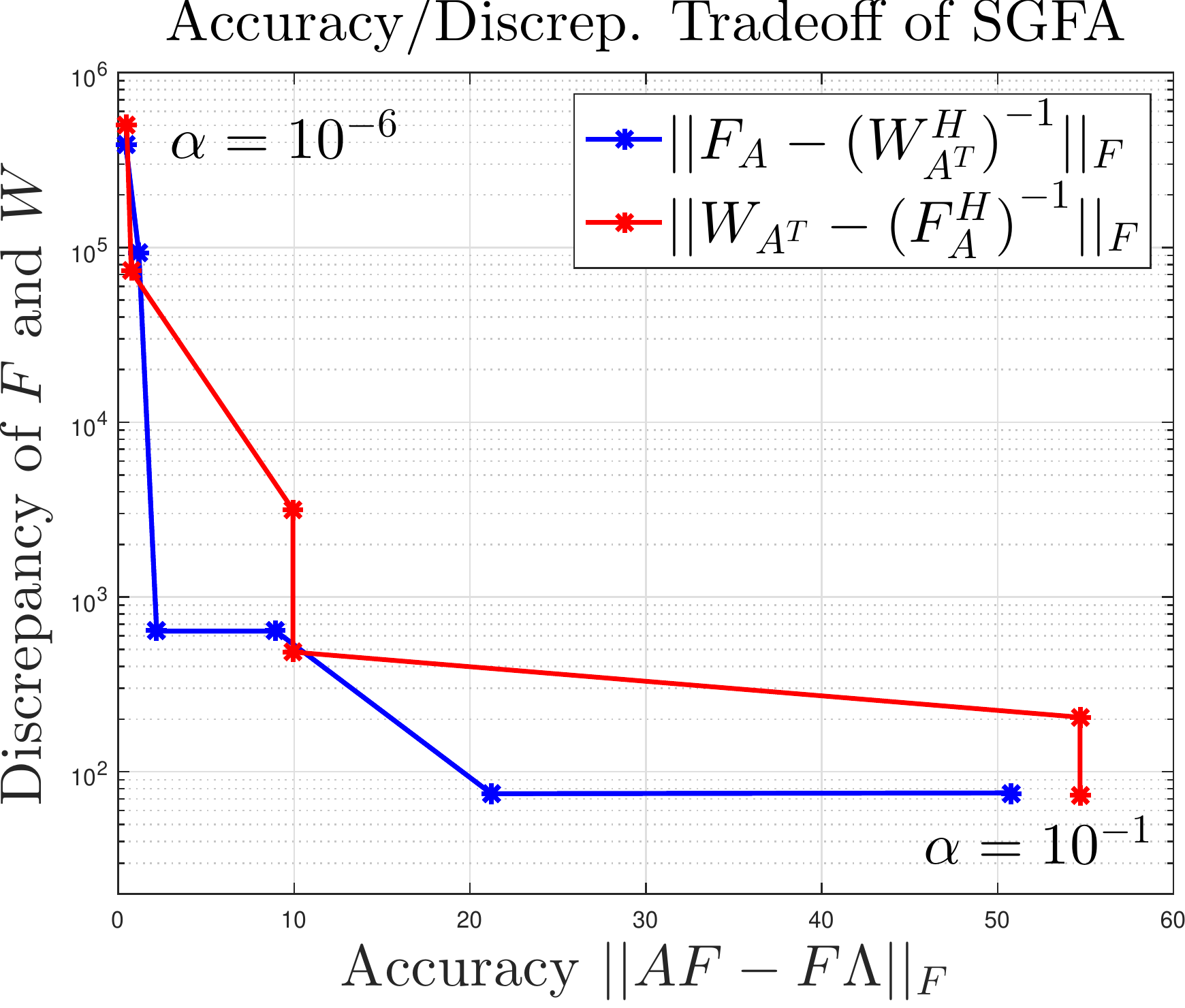}
	\caption{Political blogs network: Approximating left  and right eigenvectors, $F_{A}$ and $W_{A^T}$ (respectively), using SGFA, algorithm~\ref{alg:SGFA_alg},
	 with $\beta=0.43$ and considering different $\alpha$ levels.
}
	\label{fig:Left_Right_approx_political_blogs}
\end{figure}

All previous sections considered the problem of computing accurate and stable approximation of right eigenvectors~$F$ for an arbitrary graph shift~$A$. Note, however, that the same exact reasoning applies for (the conjugate transposed of the) left eigenvectors~$W^H=F^{-1}$, as defined in Equation~\eqref{eqn:first-b}, since the left eigenvectors of~$A$ correspond to the right eigenvectors of~$A^H$, i.e.,
\begin{align}
AF=F\Lambda  \Leftrightarrow  F^H A^H= \Lambda^* F^H  \Leftrightarrow  A^H W  = W \Lambda^*.
\label{eqn:left_right_eig}
\end{align}
So, one could, equivalently, apply SGFA, algorithm~\ref{alg:SGFA_alg}, with input~$A$ to obtain~$F$ and then invert it to compute~$F^{-1}$, or apply SGFA, algorithm~\ref{alg:SGFA_alg}, directly to~$A^H$ and obtain~$W$ and then invert $W^H$ to get~$F$, and choose the better approximation, in terms of accuracy and stability. If SGFA, algorithm~\ref{alg:SGFA_alg}, finds the global solution of~\eqref{eqn:opt}, Equation~\eqref{eqn:left_right_eig} implies that we will exactly have~$F^{-1}=W^H$, computed either way, given that the matrices are stable enough for numerical invertibility, i.e., parameter~$\alpha$ in SGFA, algorithm~\ref{alg:SGFA_alg}, is sufficiently high. Figure~\ref{fig:Left_Right_approx_political_blogs} plots accuracy and discrepancy results between~$F$ and~$W$, considering SGFA, algorithm~\ref{alg:SGFA_alg}, with input~$A$, ~$A^H=A^T$ and $\Lambda=\Lambda^*$  (A is real). The subscript~$F_A$ $\left(W_{A^T}\right)$ indicates that the right (left) eigenvectors of~$A$ $\left(A^T\right)$ were computed with SGFA, algorithm~\ref{alg:SGFA_alg}, with input~$A/A^T$, respectively. Each point of Figure~\ref{fig:Left_Right_approx_political_blogs} corresponds to a run of SGFA, algorithm~\ref{alg:SGFA_alg}, for~$A$ (blue curve) and~$A^T$ (red curve), for a specific stability parameter~$\alpha\in\{10^{-1},10^{-2},\dots,10^{-6}\}$ ($\alpha$ increases from left to right).
For very small~$\alpha$, both approximations, $F_A$ and~$W_{A^T}$  are accurate (low objective of problem~\eqref{eqn:opt} corresponds to points on the left of the horizontal axis) but the discrepancy between~$F_A$ and~$W_{A^T}$ is higher (higher values on the vertical scale). Increasing~$\alpha$ (moving from left to right), discrepancy decreases (lower values on the vertical axis), but the accuracy worsens (moving to the right on the horizontal axis). For a value of $\alpha=10^{-3}$, we find a good trade-off between these two factors: accurate approximation (error $\approx 9.5$) and small discrepancy (error~$\approx 0.37 n $).
\section{Manhattan Road Network}
\label{sec_road}
  We now carry out a similar study by applying SGFA, algorithm~\ref{alg:SGFA_alg}, to the Manhattan road map.
\subsection{Dataset}
\label{subsec:roaddataset}
 We start by briefly describing the Manhattan road network of Figure~\ref{fig_road_net}. This Network consists of 5,464 nodes that represent latitude and longitude coordinates \cite{dataset_road}, connected by 11,568 directed or undirected edges that represent one or two way streets as verified by Google Maps~\cite{dataset_road_2}. The underlying graph $\mathcal{G}=(\mathcal{V},A)$ is directed and unweighted: $A_{i,j}=1$ if and only if there is a directed edge from node~i to node~j. Otherwise $A_{i,j}=0$. As it is natural (no dead-end streets in a city), graph~$\mathcal{G}$ is strongly connected (there exists, at least, one path from each node to any other node). Matrix~$A$ has an eigenvalue $\lambda=0$ with high multiplicity as seen in figure~\ref{fig_spectrum_angles}. \textcolor{black}{Computing~$F$ with MATLAB$^\circledR$ \texttt{eig}, we find that $\sigma_{\min}(F)\approx6.2\times 10^{-19}$}.
\subsection{Approximating a Stable Fourier Basis}
\label{sec:approx_road_network}
Taking into account the numerical study of  Section~\ref{sec:approximating_political_blog}, we applied SGFA, algorithm~\ref{alg:SGFA_alg}, and choose a (fixed) stopping criterion $\alpha=10^{-6}$ and a grid of six values for the contraction factor~$.2\leq\beta\leq .9$.  We did not consider values $\beta>.9$ since as observed before this takes SGFA, algorithm~\ref{alg:SGFA_alg}, longer to converge with no noticeable performance improvement.
\subsubsection{Accuracy}
\label{subsec:roadaccuracy}
\begin{figure}[htbp]
	\centering
	\centering
	\includegraphics[width=6.5cm]{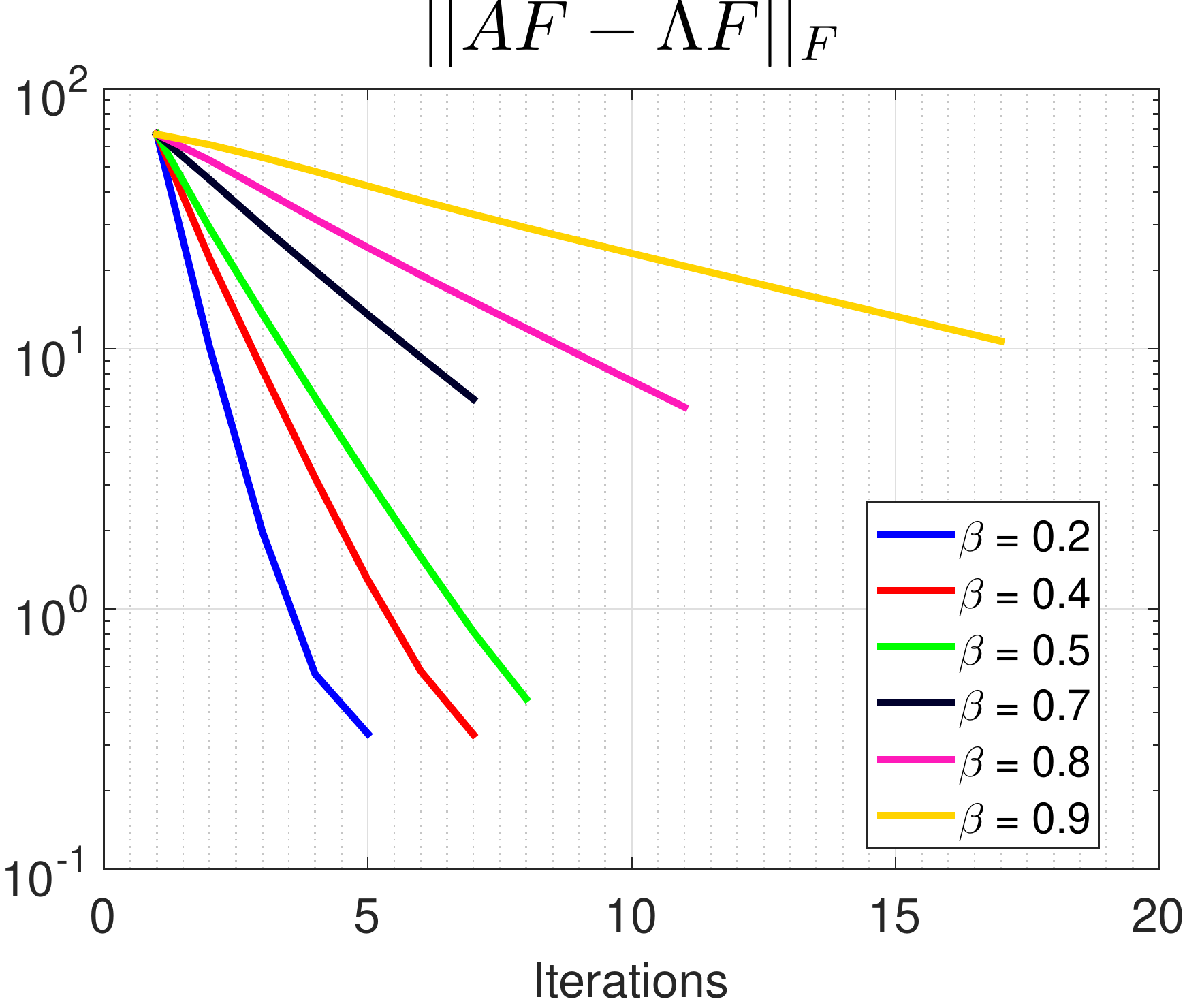}
	\caption{Manhattan road network: Accuracy of SGFA, algorithm~\ref{alg:SGFA_alg}.}
	\label{fig:road_numerics_accuracy}
\end{figure}

Figure \ref{fig:road_numerics_accuracy} displays the accuracy $\|AF-F\Lambda\|_\mathcal{F}$ (log scale) versus number of iterations (linear scale). It shows that accuracy increases exponentially with~$\beta$ (theorem \ref{theorem_exponential}); again, there is no reason to consider a high value for the contraction factor~$\beta$ since, for $\beta \geq 0.7$, SGFA, algorithm~\ref{alg:SGFA_alg}, will run for more iterations while achieving a worst overall approximation.
\subsubsection{Stability}
\label{subsec:roadstability}
\begin{figure}[htbp]
	\centering
	\begin{subfigure}{0.4\textwidth}
		\centering
		\includegraphics[width=6.5cm]{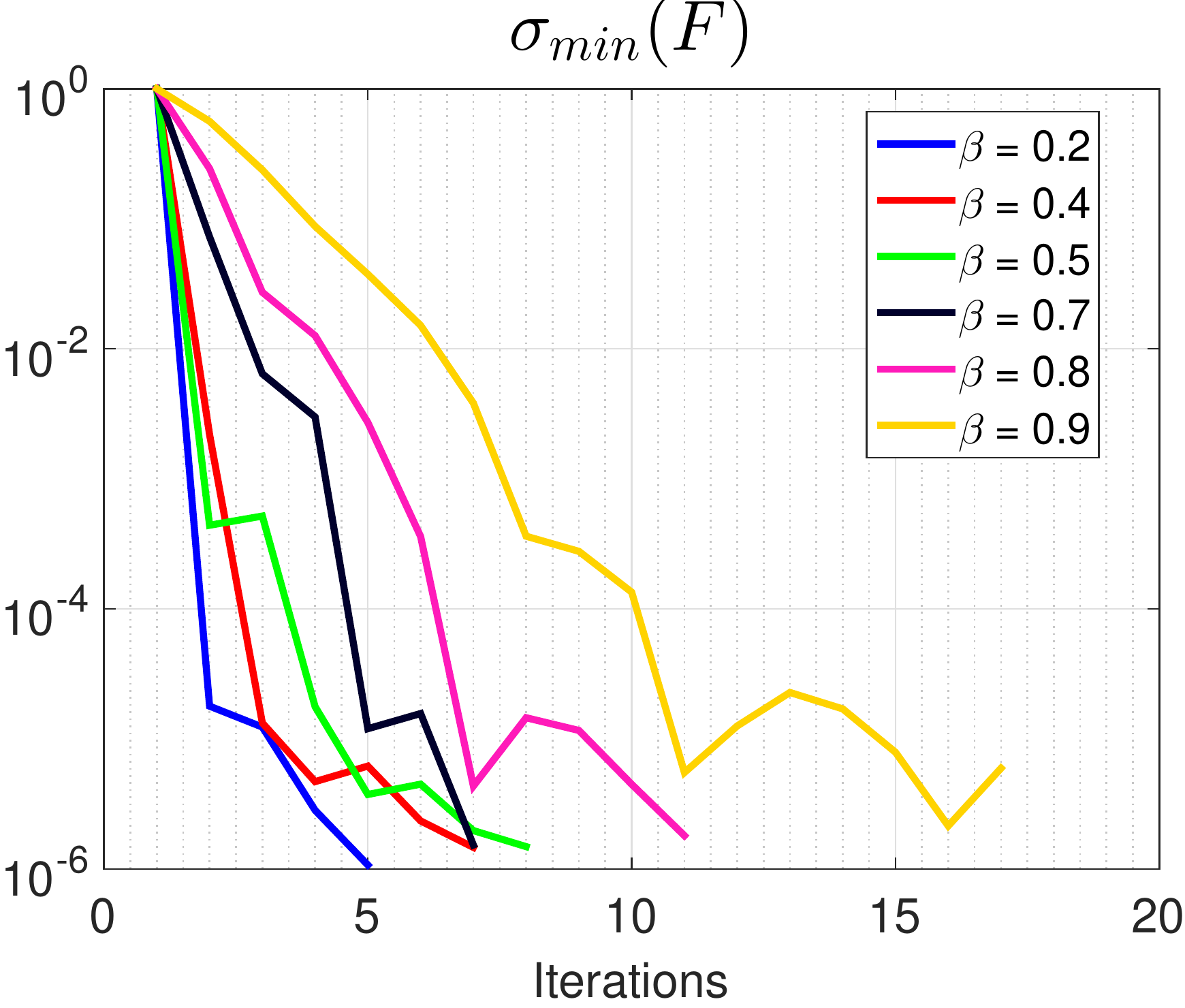}
		\caption{}
		\label{fig:road_numerics_min_svd}
	\end{subfigure}
	\begin{subfigure}{0.4\textwidth}
		\centering
		\includegraphics[width=6.5cm]{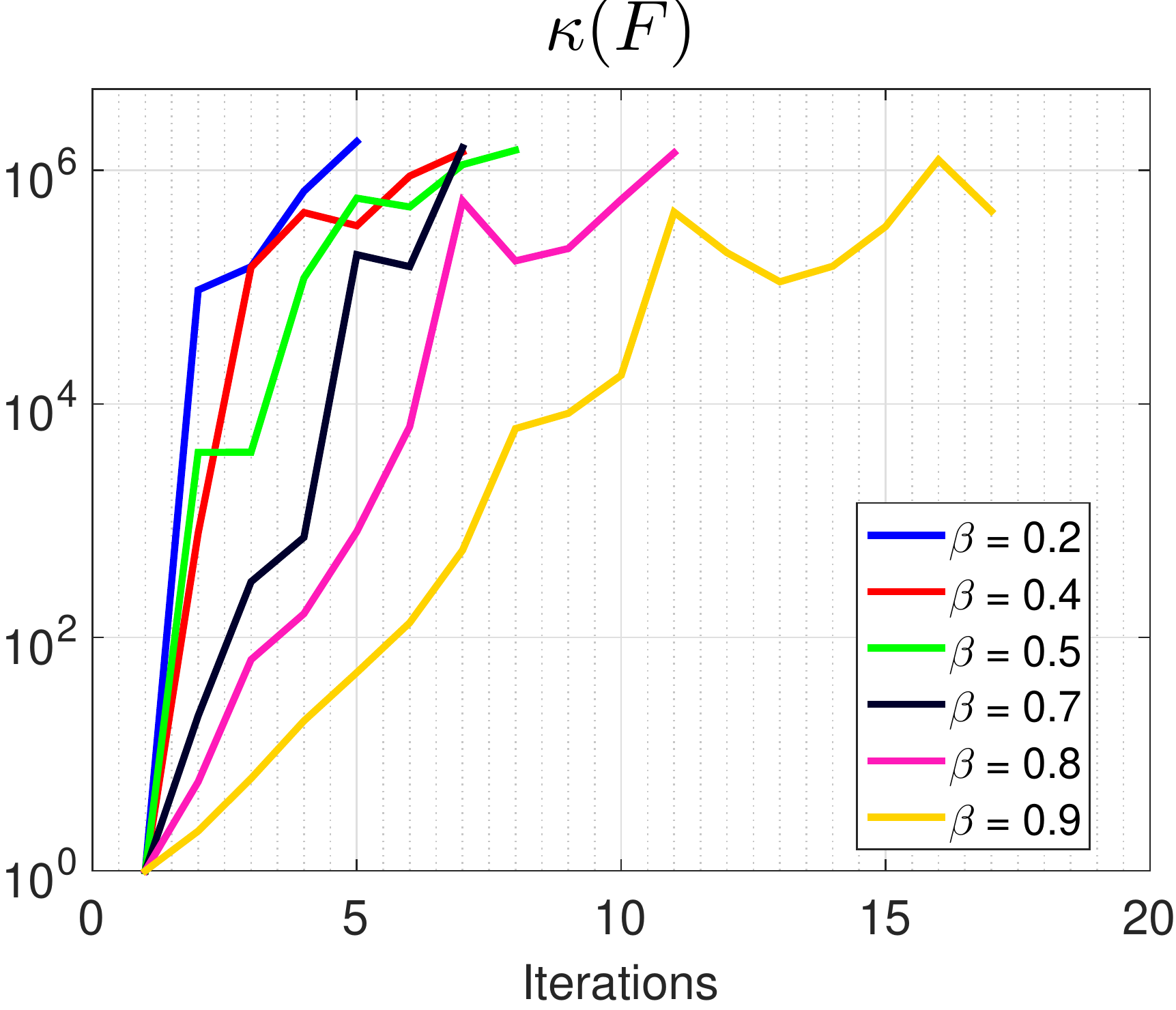}
		\caption{}
		\label{fig:road_numerics_cond}
	\end{subfigure}
	\begin{subfigure}{0.4\textwidth}
		\centering
		\centering
		\includegraphics[width=6.5cm]{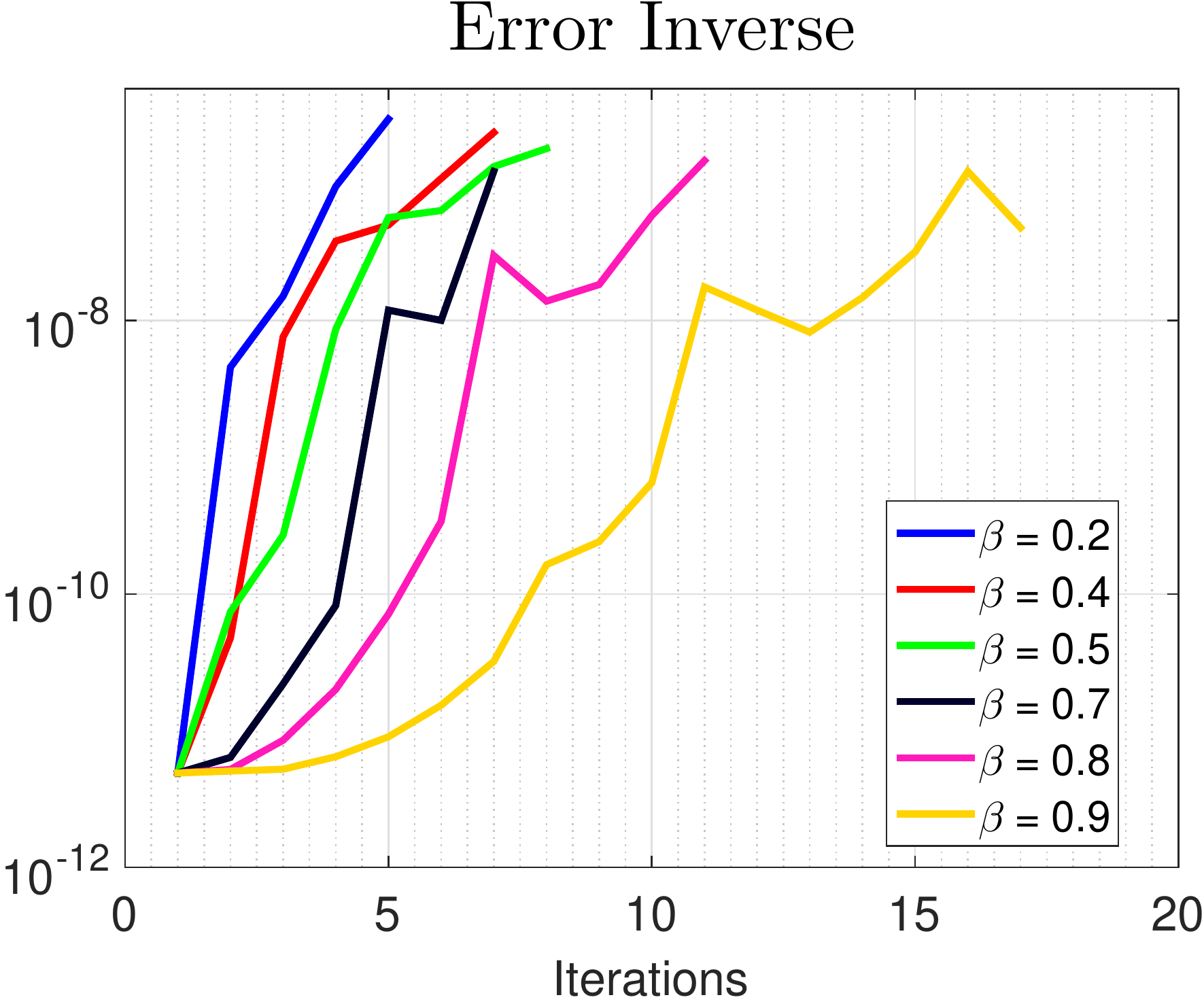}
		\caption{}
		\label{fig:road_numerics_inverse_error}
	\end{subfigure}
	\caption{Manhattan road network: Stability of SGFA, algorithm~\ref{alg:SGFA_alg}.}
\end{figure}

Figures~\ref{fig:road_numerics_min_svd}, \ref{fig:road_numerics_cond}, and~\ref{fig:road_numerics_inverse_error} plot the  stability measures of Section~\ref{sec_political}, now for the Manhattan road network. We observe a phase transition behavior with~$\beta$ where the stability measures converge at a much slower rate for $\beta > 0.7$. Again, we verify that the condition number $\kappa(F)$ tends to scale linearly with $1/\alpha$ , i.e., $\kappa(F) \approx 10^6$ as seen in figure~\ref{fig:road_numerics_cond}. As observed with the political blogs network, the same conclusion applies here: one should avoid very high values of~$\beta$ since there is no significant payoff in terms of numerical stability.

\begin{figure}[htbp]
  	\centering
  	\begin{subfigure}{0.5\textwidth}
  		\centering
  		\includegraphics[width=5.7cm]{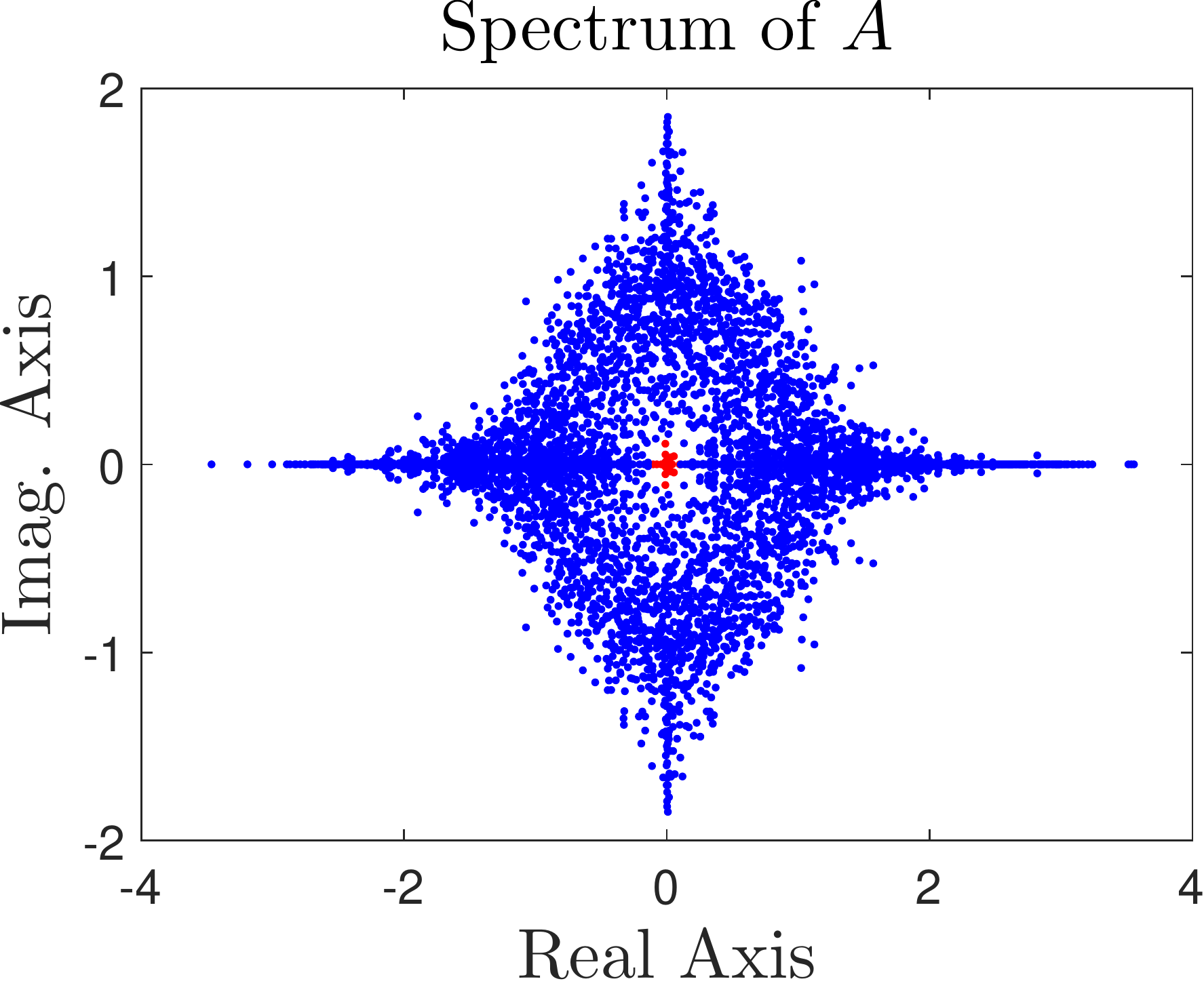}
  		\caption{}
  		\label{fig_spectrum_angles}
  	\end{subfigure}
  	\begin{subfigure}{0.5\textwidth}
  		\centering
  		\includegraphics[width=5.7cm]{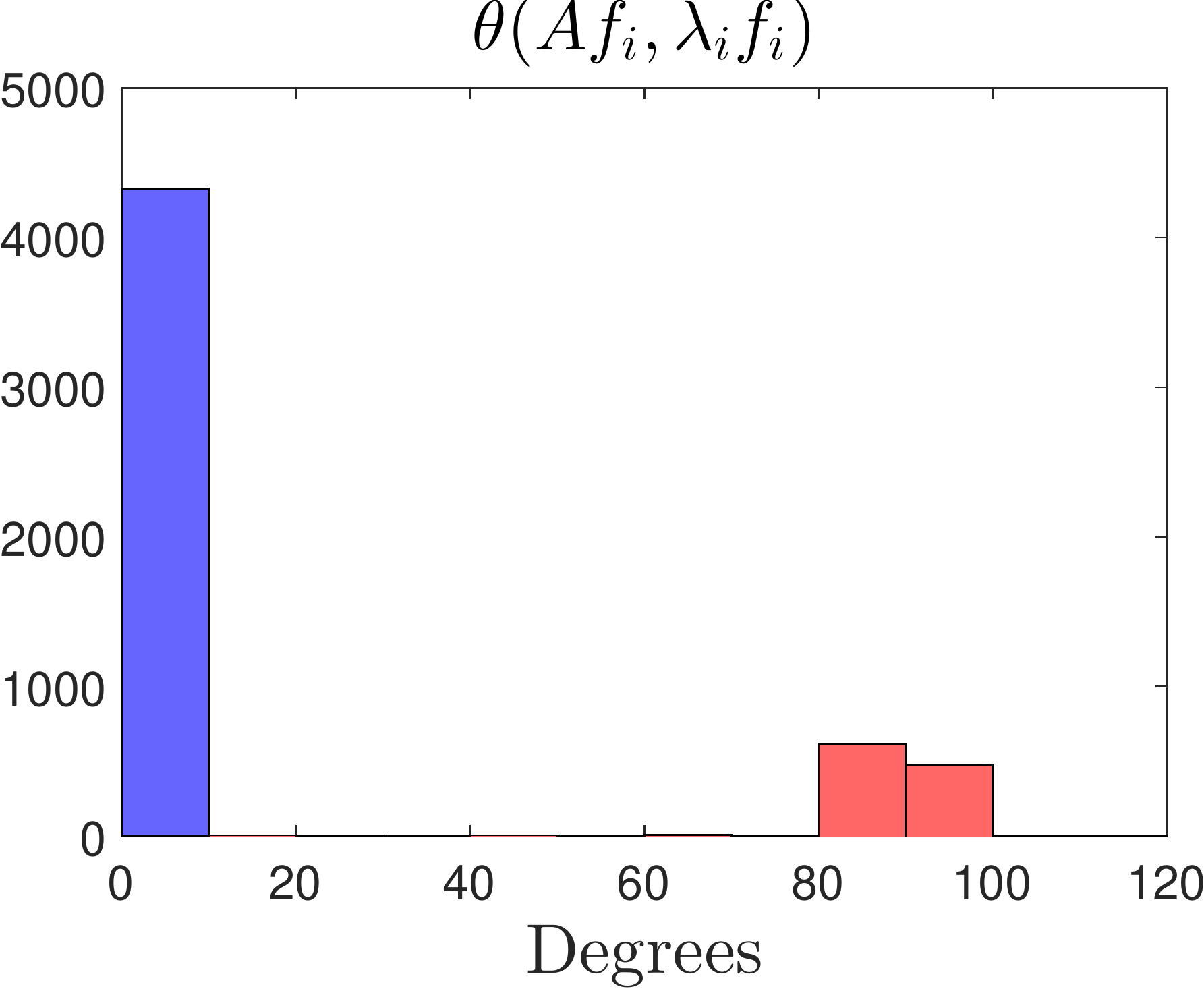}
  		\caption{}
  		\label{fig_hist_angles}
  	\end{subfigure}
  	\label{angles_plot}
  	\caption{Manhattan road network: Angle \eqref{eqn:angle_formula} stability measure. In~\textbf{(a)} the spectrum of~$A$ and in~\textbf{(b)} the histogram of the angles $\theta\left(Af_i,\lambda_i f_i\right)$. Blue indicates that $\theta\left(Af_i,\lambda_i f_i\right)<10^\circ$ and red represents the complement $\theta\left(Af_i,\lambda_i f_i\right)\geq 10^\circ$.
  	}
  	\label{fig:angles_plot}
  \end{figure}

 The accuracy measure $\left\|AF-F\Lambda\right\|_\mathcal{F}$ studies how close the magnitude of $Af_i$ is to $\lambda_if_i$. But both the magnitude and orientation of $f_i$ are important for achieving an accurate and stable approximated Fourier Basis~$F$.
 To further test how close the columns  $f_i$ of~$F$ are to being eigenvectors\footnote{Note that we do not normalize $f_i$ to norm~1 because normalizing $f_i$ will affect the numerical stability of~$F$, as shown in Section~\ref{problem_formulation} for $F(\epsilon)$. This was precisely the problem with Theorem~\ref{th:theorem_triangular_approx} that motivated formulation~\eqref{eqn:opt}.}, we compute the angle $\theta\left(Af_i,\lambda_i f_i\right)$ between the two vectors $Af_i$ and $\lambda_i f_i$. If the angle is small, the two vectors are aligned as they should be. This represents yet another important stability measure for SGFA, algorithm~\ref{alg:SGFA_alg}. Generally, $Af_i$ and $\lambda_i f_i$ are complex vectors, hence we compute $\theta(\cdot,\cdot)$ by relying on the real vector space $\mathbf{R}^{2n}$ that is isometric to $\mathbb{C}^{n}$ \cite{complex_angles}
  \begin{equation}
  \forall x,y\in \mathbb{C}^n\setminus\{0\}:\enspace   \theta(x,y)=\arccos\left(\frac{\text{Re}(x^Hy)}{\left\|x\right\|_2 \, \, \left\|y\right\|_2}\right).
  \label{eqn:angle_formula}
  \end{equation}
 Figure~\ref{fig_spectrum_angles} shows a scatter plot of the eigenvalues (spectrum) of~$A$. A majority of the eigenvalues form a (blue) cloud away from the origin, while the remaining form a dense (red) cluster very close to the origin. Figure~\ref{fig_hist_angles} is the histogram of the angles $\theta\left(Af_i,\lambda_i f_i\right)$ between $Af_i$ and $\lambda_if_i$ computed by~\eqref{eqn:angle_formula}. The histogram has three main bars, a large (blue) bar on the left collecting about $80\%$ of the angles $\theta\left(Af_i,\lambda_i f_i\right)<10^\circ$, and two small (red) bars on the right with the remaining $20\%$ of the angles $\theta\left(Af_i,\lambda_i f_i\right)\approx 90^\circ$. The large (blue) bar close to the origin corresponds to the $80\%$ of the eigenvectors $f_i$ associated with the eigenvalues away from zero (blue cloud) in figure~\ref{fig_hist_angles}, demonstrating that the vectors $Af_i$ and $\lambda_if_i$ are (almost) colinear as they should be. The two small (red) bars on the right of the histogram close to $90^\circ$ correspond to the $20\%$ of the eigenvectors $\lambda_if_i$ associated with the eigenvalues close to zero (dense red cluster) in figure~\ref{fig_hist_angles}, apparently showing that the vectors $Af_i$ and $\lambda_if_i$ are practically orthogonal. This is simply an artifact of how the angle is computed through~\eqref{eqn:angle_formula}. Because for these eigenvectors $f_i$ the associated eigenvalues $\lambda_i\approx 0$, then $\lambda_if_i$ is approximately the zero vector, and once a vector is the zero vector the inner product computed in~\eqref{eqn:angle_formula} is also zero leading to interpreting the angles as $\approx 90^\circ$.
\subsubsection{Approximating Left vs Right Eigenvectors\textemdash Manhattan Road Network}
\label{sec:roadapproximatingeigenvectors}
\begin{figure}[htbp]
  	\centering
  	\includegraphics[width=7cm]{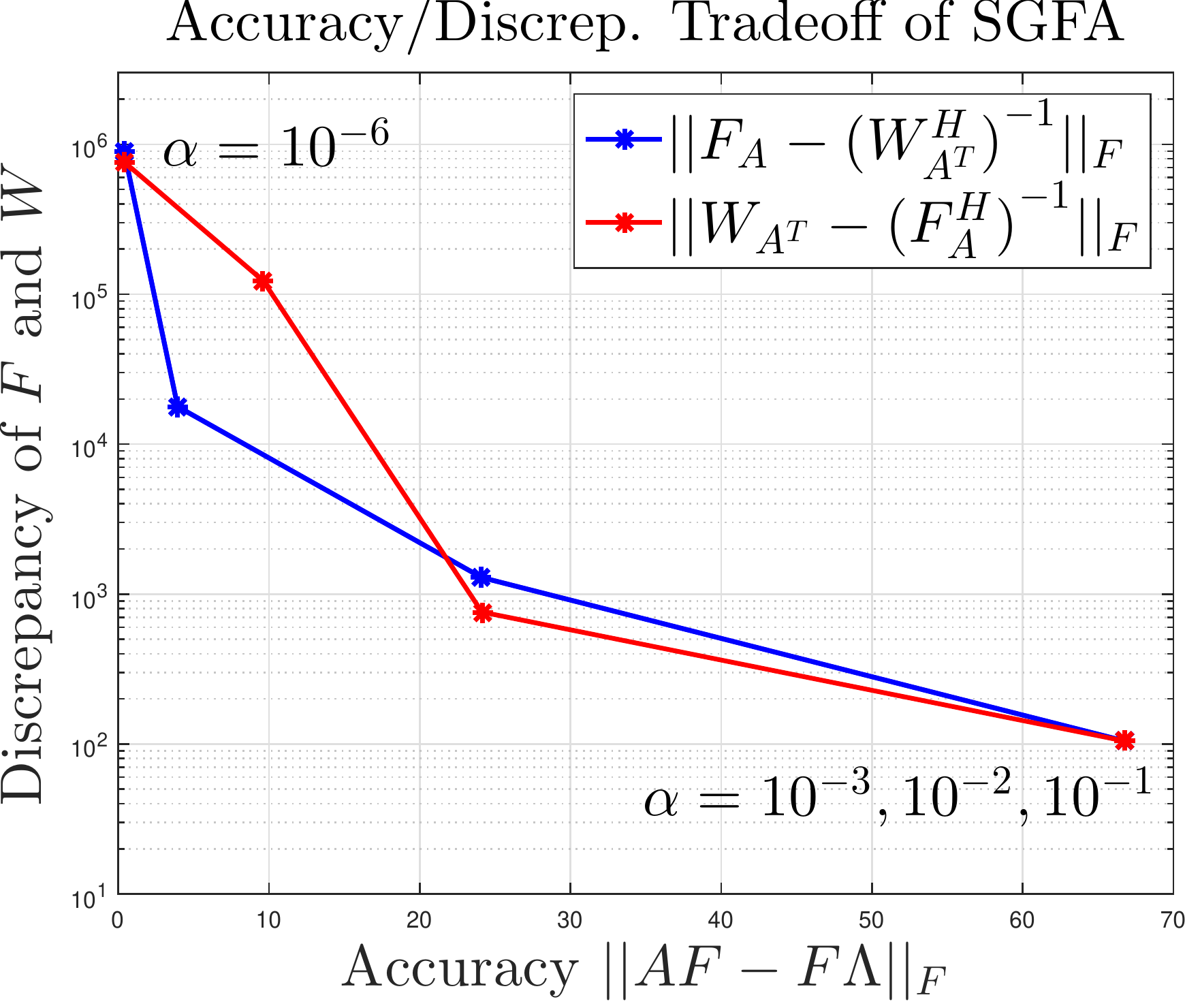}
  	\caption{Manhattan road network: Approximating left  and right eigenvectors, $F_{A}$ and $W_{A^T}$ (respectively), using SGFA, algorithm~\ref{alg:SGFA_alg}
  		with $\beta=0.43$ and considering different $\alpha$ levels.
  	}
  	\label{fig:Left_Right_approx_roadnetwork}
  \end{figure}

Figure~\ref{fig:Left_Right_approx_roadnetwork} plots the discrepancy versus accuracy now for the Manhattan road network graph shift. For~$\alpha=10^{-1},10^{-2},10^{-3}$, SGFA, algorithm~\ref{alg:SGFA_alg}, ran for a single iteration with both~$A$ and~$A^T$ (this explains why there are only four points in the figure). Our conclusions replicate those for Figure~\ref{fig:Left_Right_approx_political_blogs} for the political blogs network: increasing~$\alpha$ decreases accuracy but improves discrepancy. For~$\alpha=10^{-4}$, SGFA, algorithm~\ref{alg:SGFA_alg}, achieves mean accuracy and mean discrepancy of, respectively,
\begin{align}
\label{eqn:meanaccuracy}
  \frac{\left \|AF-F\Lambda \right\|_\mathcal{F}+\left \|A^TW-W\Lambda \right\|_\mathcal{F}}{2} {}&\approx 24\\
  \label{eqn:meandiscrepancy}
  \frac{\left \|F_A-\left(  W_{A^T}^H\right)^{-1} \right\|_\mathcal{F}+\left \|W_{A^T}-\left(  F_{A}^H\right)^{-1} \right\|_\mathcal{F}}{2} {}&\approx 0.19n.
  \end{align}
\section{Directed Erd\H os-R\'enyi graphs}
\label{section:random_graph_experiments}
%
\textcolor{black}{We consider synthetic random graphs; these allow graph properties to be better controlled and to investigate the impact of varying its properties. In concrete, we consider \textit{directed} Erd\H os-R\'enyi graphs (\textcolor{black}{with possible self loops}) of fixed dimension $n=100$ and varying probability of connection~$p$. As seen in figure~\ref{fig:tails_erdos_renyi} (\textcolor{black}{blue curve}) , these random models will have an unstable Fourier basis~$F$ for small and large values of~$p$, i.e., for either very sparse or very dense graph shifts~$A$.}
\subsubsection{Accuracy}
\label{subsec:erdosrenyiaccuracy}
\begin{figure}[htbp]
	\centering
	\begin{subfigure}{0.5\textwidth}
		\centering
		\includegraphics[width=6.9cm,height=4cm]{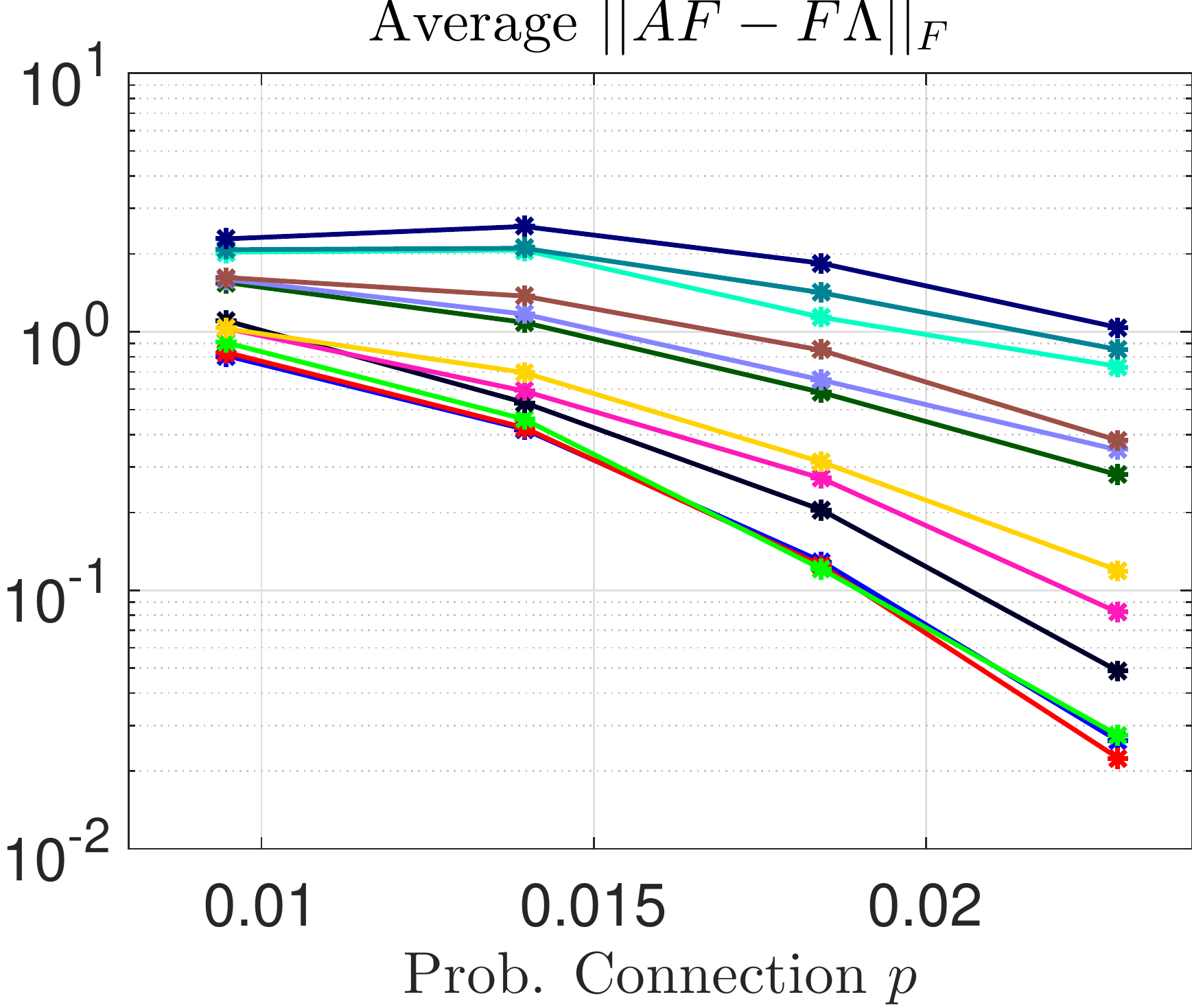}
		\caption{}
		\label{fig:accuracy_erdos_low_p}
	\end{subfigure}
	\begin{subfigure}{0.5\textwidth}
		\centering
		\includegraphics[width=6.9cm,height=4cm]{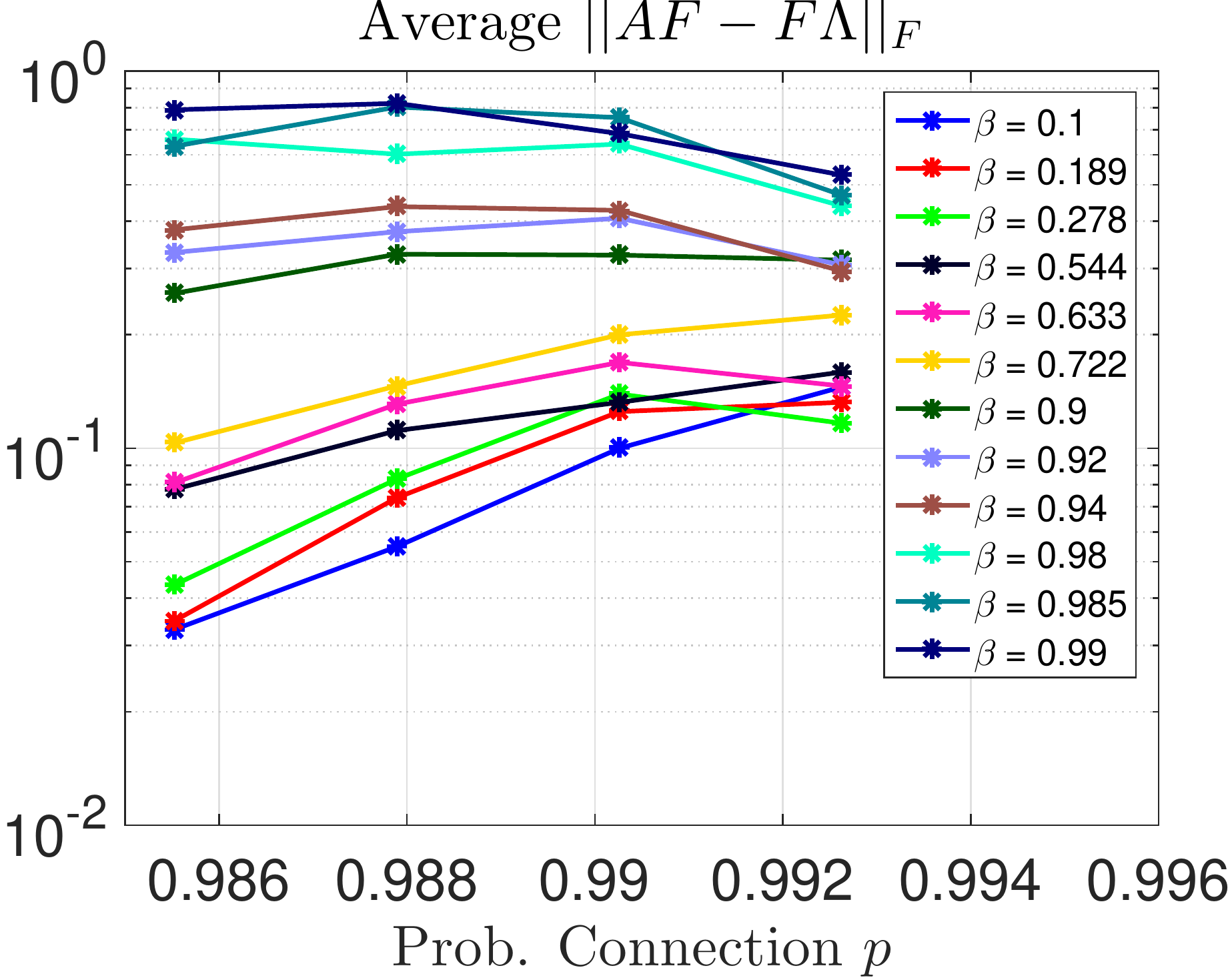}
		\caption{}
		\label{fig:accuracy_erdos_high_p}
	\end{subfigure}
	\caption{Accuracy of SGFA for directed Erd\H os-R\'enyi graphs with probability of connection~$p$. In~\textbf{(a)} low values of~$p$ and in~\textbf{(b)} high values of~$p$. Mean for $100$ Monte Carlo trials.}
		\label{fig:accuracy_erdos}
\end{figure}

\textcolor{black}{Figure~\ref{fig:accuracy_erdos} plots the average accuracy of algorithm~\ref{alg:SGFA_alg} for directed Erd\H os-R\'enyi graphs with varying probability of connection~$p$. The color legend of figure~\ref{fig:accuracy_erdos_high_p} also applies to~\ref{fig:accuracy_erdos_low_p}. For each value of~$p$, the best accuracy results are achieved by considering the smallest values of~$\beta$, i.e., $\beta\in\{0.1,0.189,0.278\}$. This behavior is consistent with that observed in both previous sections where a high value of~$\beta$ was discouraged.}
\textcolor{black}{For sparse models, figure~\ref{fig:accuracy_erdos_low_p} shows that accuracy tends to increase with~$p$ regardless of the value of~$\beta$. Say for $\beta=0.189$, the average approximation error $\left\|AF-F\Lambda\right\|_\mathcal{F}$ is improved by a factor of $37$ by moving from $p=0.0095$ to $p=0.0229$. Similar observations can be taken for dense directed random graphs, see figure~\ref{fig:accuracy_erdos_high_p}. Like for sparse graphs, the average behavior of dense models depends on the value of~$\beta$ and improves, now as~$p$ is reduced. To be concrete, for $\beta=0.189$, the average approximation error $\left\|AF-F\Lambda\right\|_\mathcal{F}$ is improved by a factor of~$4$ by moving from $p=0.9926$ to $p=0.9855$. For $\beta> 0.722$, the mean accuracy plots of figure~\ref{fig:accuracy_erdos_high_p} are approximately constant.}
\subsubsection{Stability}
\label{subsec:erdosrenyistability}
\textcolor{black}{Figure~\ref{fig:random_numerics_stability} plots stability measures for an average iteration of algorithm~\ref{alg:SGFA_alg} for probability of connection $p=0.229$ (right most point in figure~\ref{fig:accuracy_erdos_low_p}).
\begin{figure}[htbp]
	\centering
	\begin{subfigure}{0.4\textwidth}
		\centering
		\includegraphics[width=5.7cm]{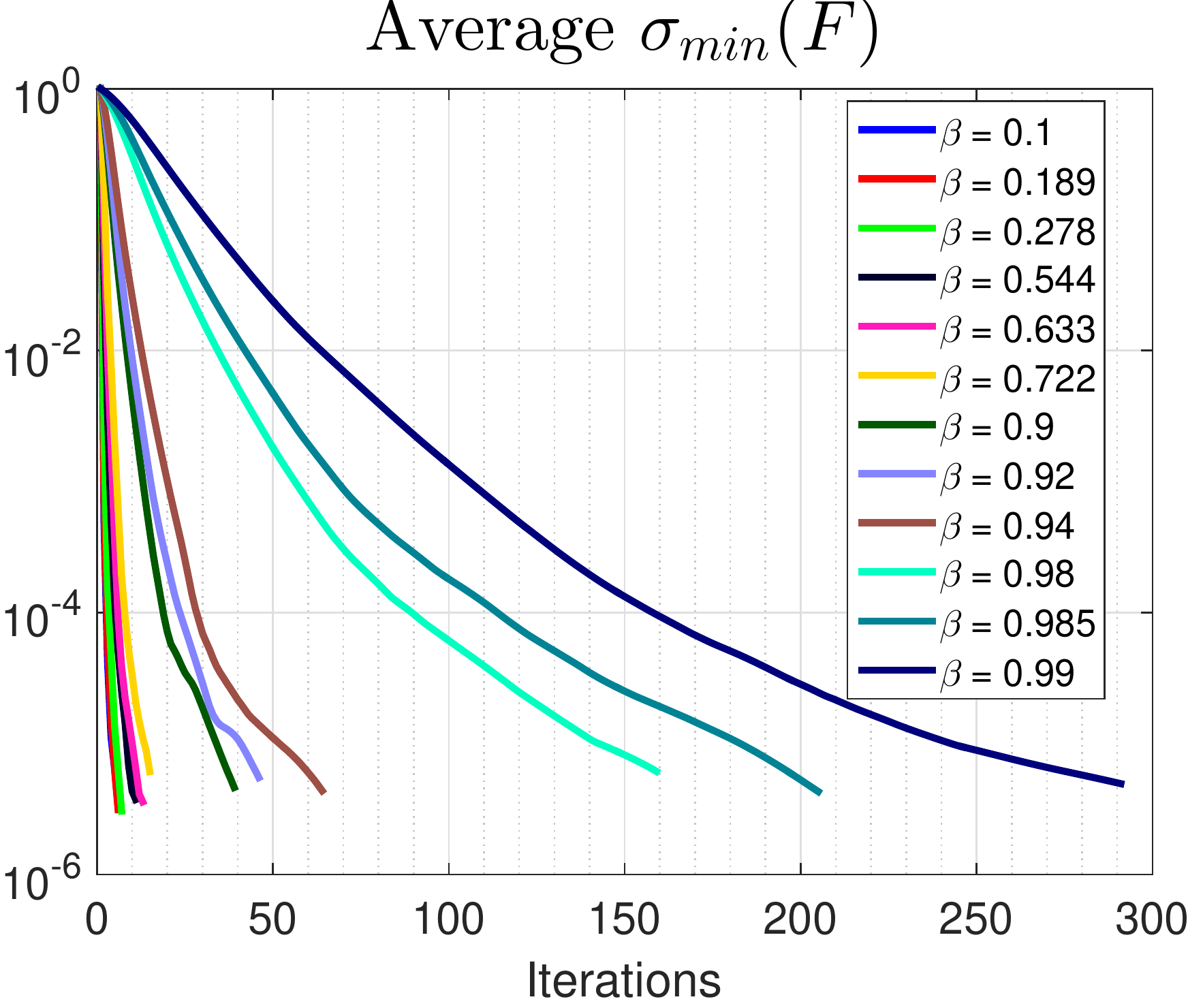}
		\caption{}
		\label{fig:random_numerics_min_svd}
	\end{subfigure}
	\begin{subfigure}{0.4\textwidth}
		\centering
		\includegraphics[width=5.7cm]{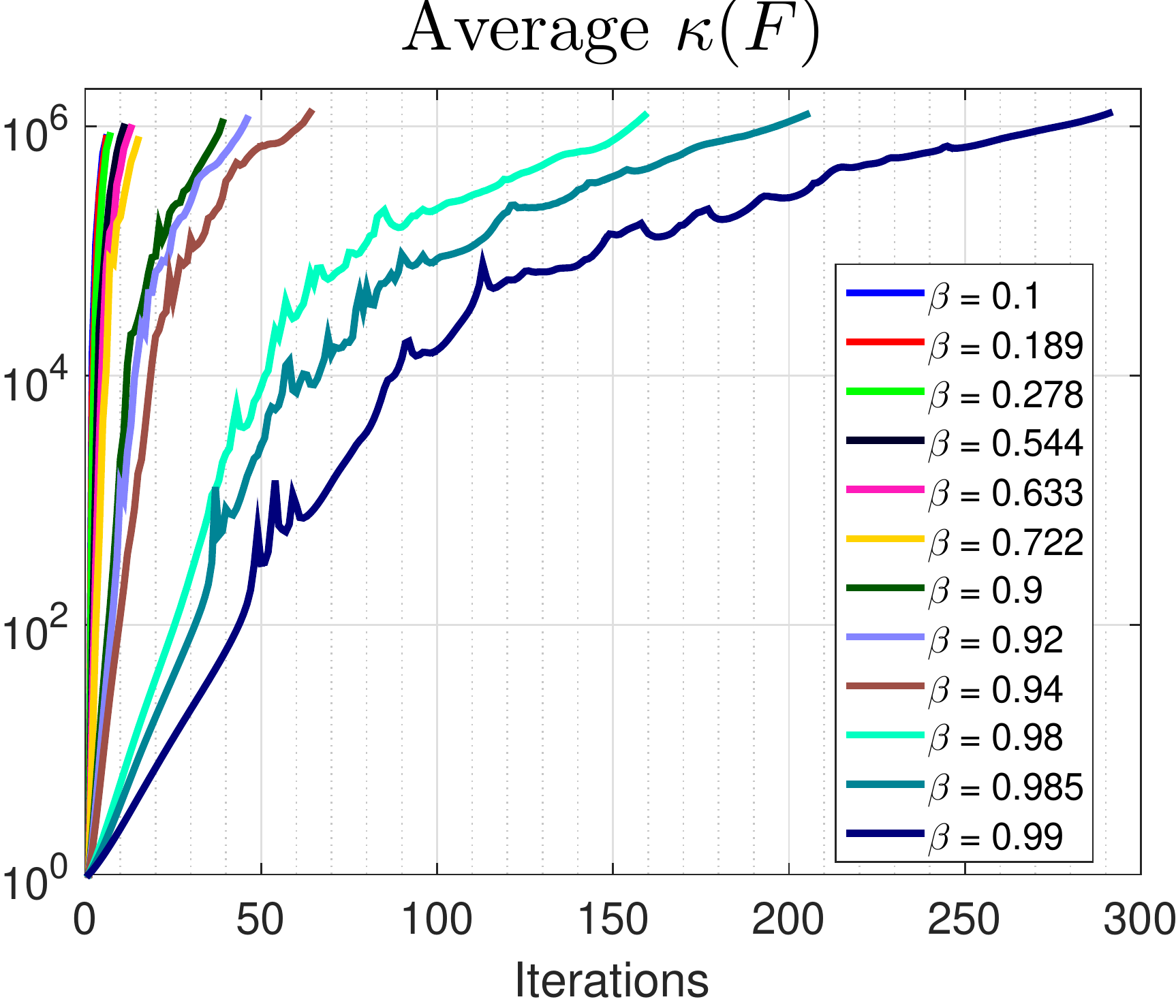}
		\caption{}
		\label{fig:random_numerics_cond}
	\end{subfigure}
	\begin{subfigure}{0.4\textwidth}
		\centering
		\centering
		\includegraphics[width=5.7cm]{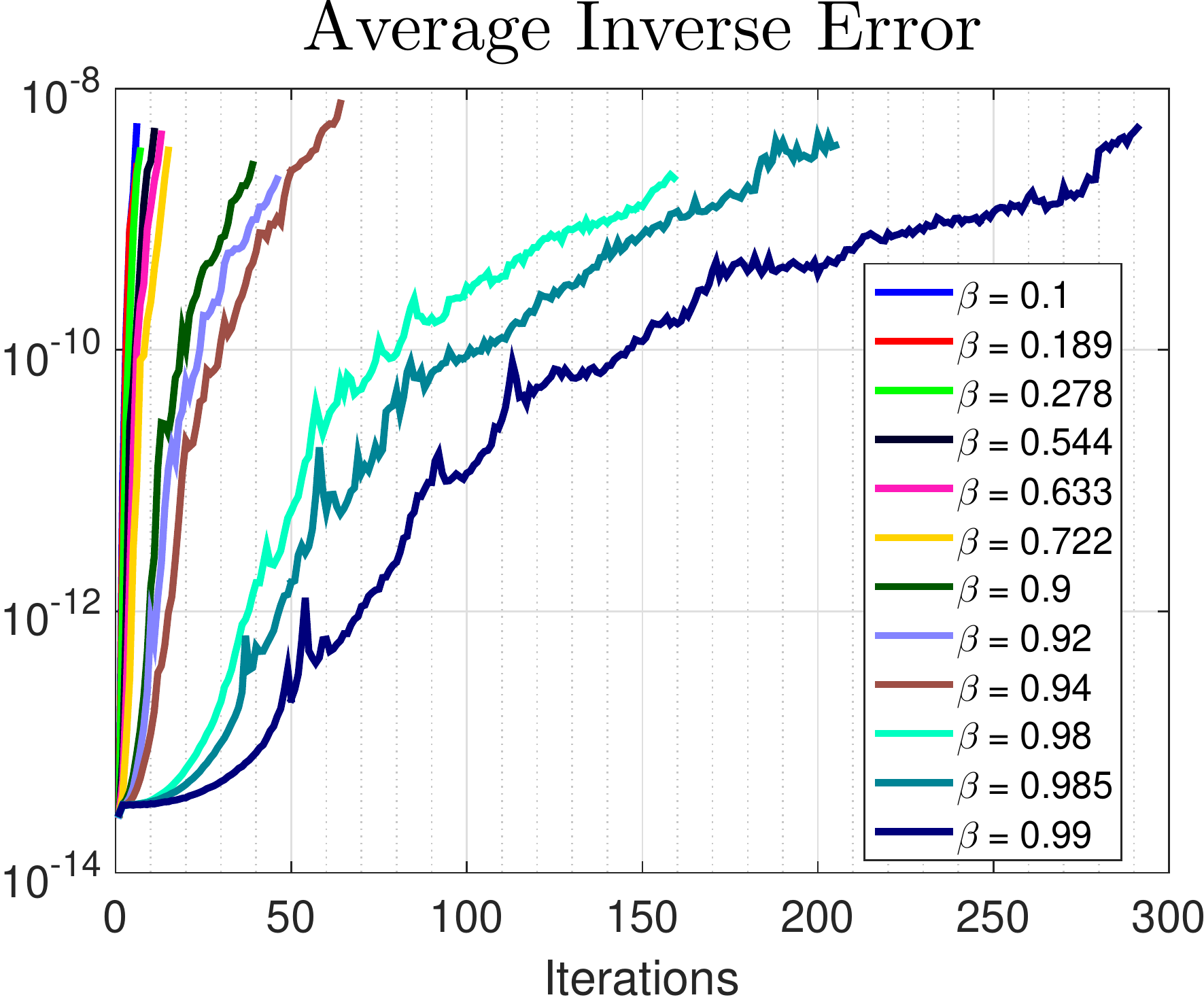}
		\caption{}
		\label{fig:random_numerics_inverse_error}
	\end{subfigure}
	\caption{Stability of SGFA for directed Erd\H os-R\'enyi graphs with $p=0.0184$. Mean for $100$ Monte Carlo trials.}
		\label{fig:random_numerics_stability}
\end{figure}
The number of iterations of SGFA is not fixed, since it depends on the input matrix~$A$ and contraction factor~$\beta$. Hence, we consider $100$ directed Erd\H os-R\'enyi graphs and compute the median number of iterations $\textrm{med}(A,\beta)$ of algorithm~\ref{alg:SGFA_alg}. Since the sample size is $100$ there exists $50$ epochs for which the algorithm ran for more than  $\textrm{med}(A,\beta)$ iterations. Figure~\ref{fig:random_numerics_stability} plots the average among these $50$ epochs for the first  $\textrm{med}(A,\beta)$ iterations, i.e., for each random graph $A$ and factor $\beta$, we average $50$ time series of dimension $\textrm{med}(A,\beta)$. We observe that, on average,  all three stability measures scale linearly with the number of iterations $K$, and this exponential behavior tends to be smoother and faster for low values of $\beta$, say $\beta\leq 0.9$ for figures~\ref{fig:random_numerics_cond} and~\ref{fig:random_numerics_inverse_error}. Again, figure~\ref{fig:random_numerics_cond} suggests that the termination criteria of algorithm~\ref{alg:SGFA_alg} could compare~$\kappa(F_k)$ against $1/\alpha$ instead of $\sigma_{\min}(F_k)$ against~$\alpha$, with $F_k$ the approximated Fourier basis at iteration~$k$.}
\section{Frequency Ordering}
\label{sec:freq_order}
\begin{figure}[htbp]
	\centering
	\begin{subfigure}[t]{0.45\textwidth}
		\centering
		\includegraphics[width=5.5cm]{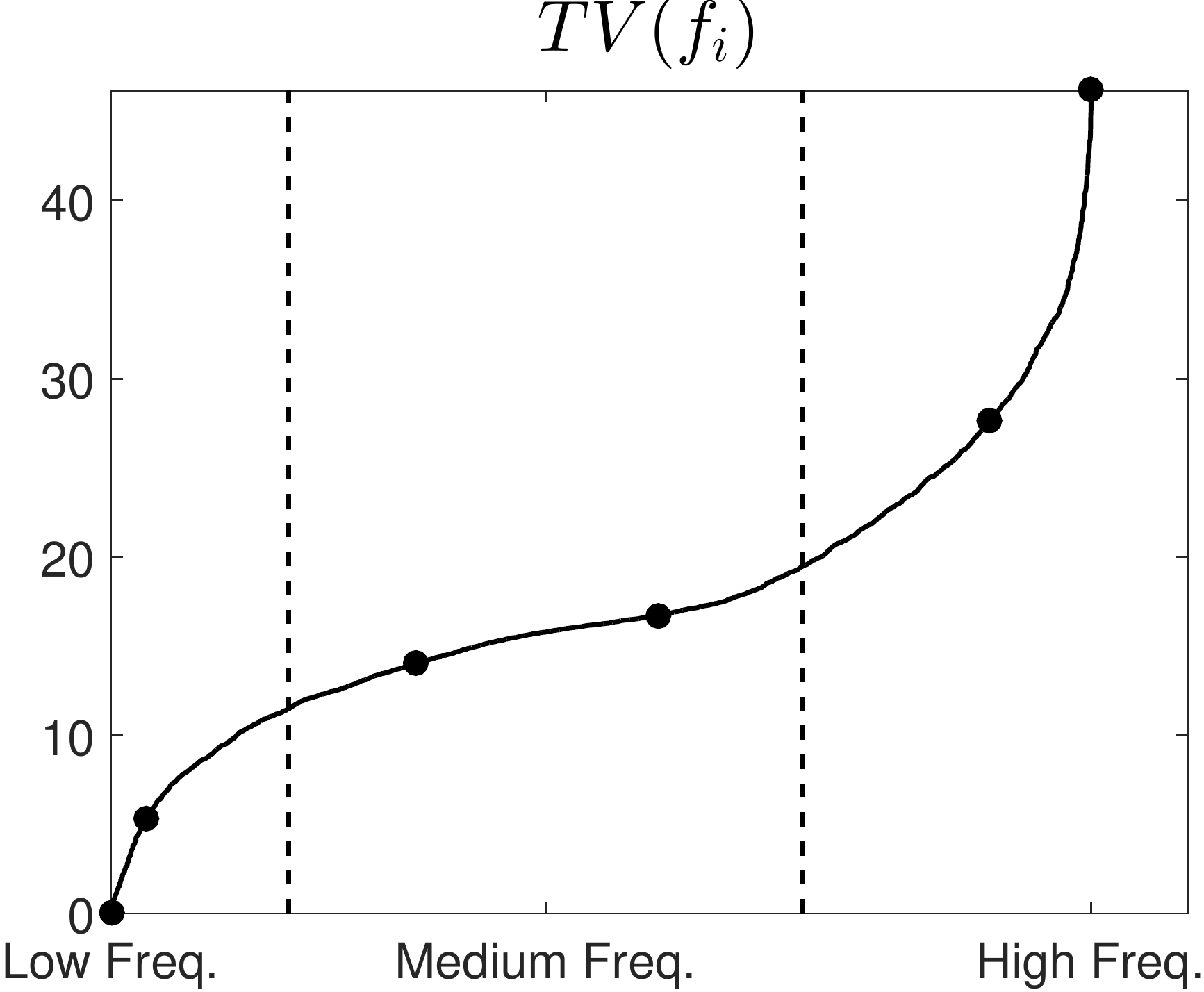}
	\end{subfigure}
	\caption{Manhattan road map: total variation~\eqref{eqn:total_var} of the Fourier basis~$F$ computed in Section~\ref{sec:approx_road_network} with $\beta=0.5$.}
	\label{fig:total_var}
\end{figure}

\begin{figure*}
\begin{tikzpicture}
	\centering
	\draw (0, 5) node[inner sep=0] {\includegraphics[width=5cm]{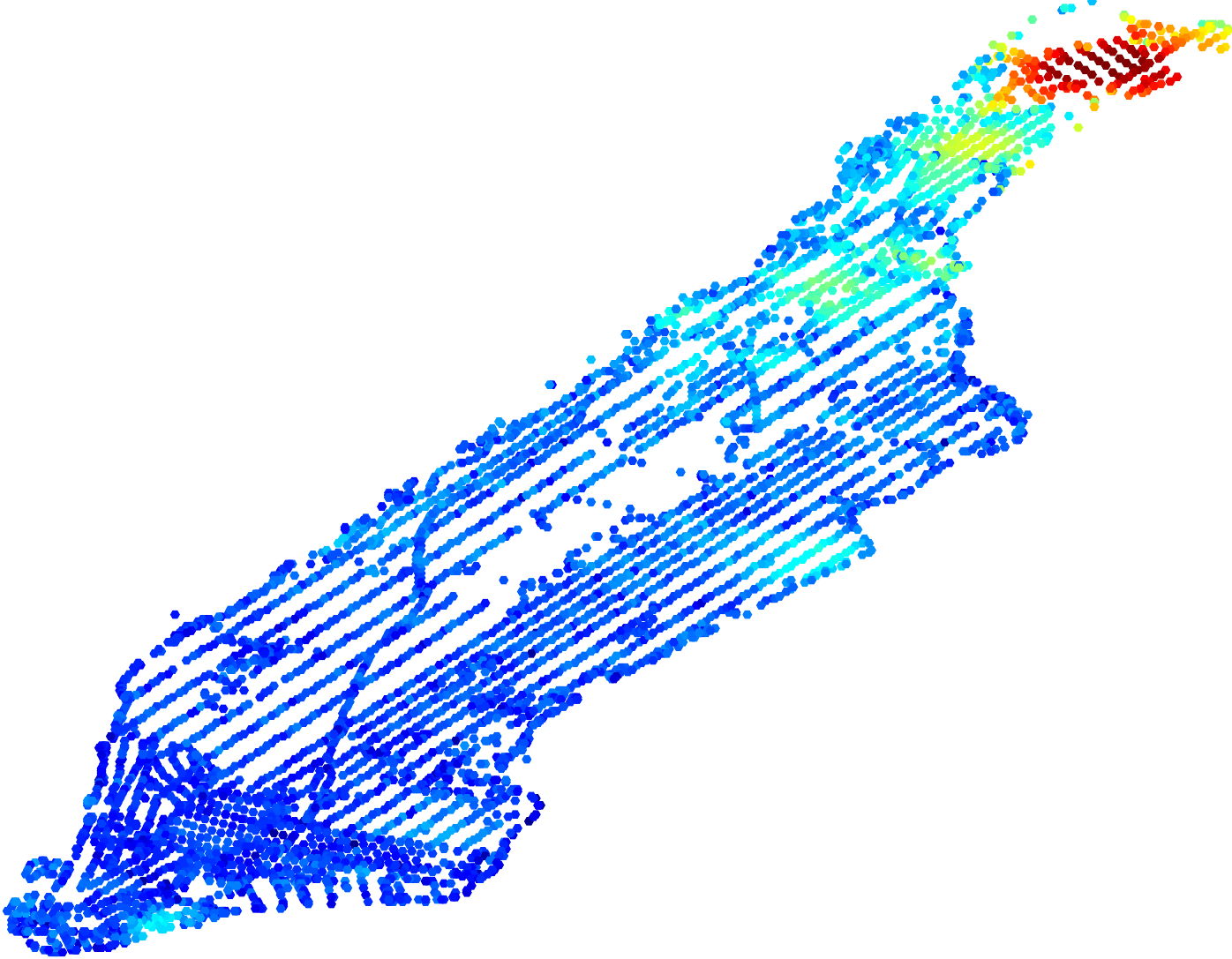}};
	\draw (0, 2.5) node[inner sep=0] {(a) $i=1$.  };
	\draw (6, 5) node[inner sep=0] {\includegraphics[width=5cm]{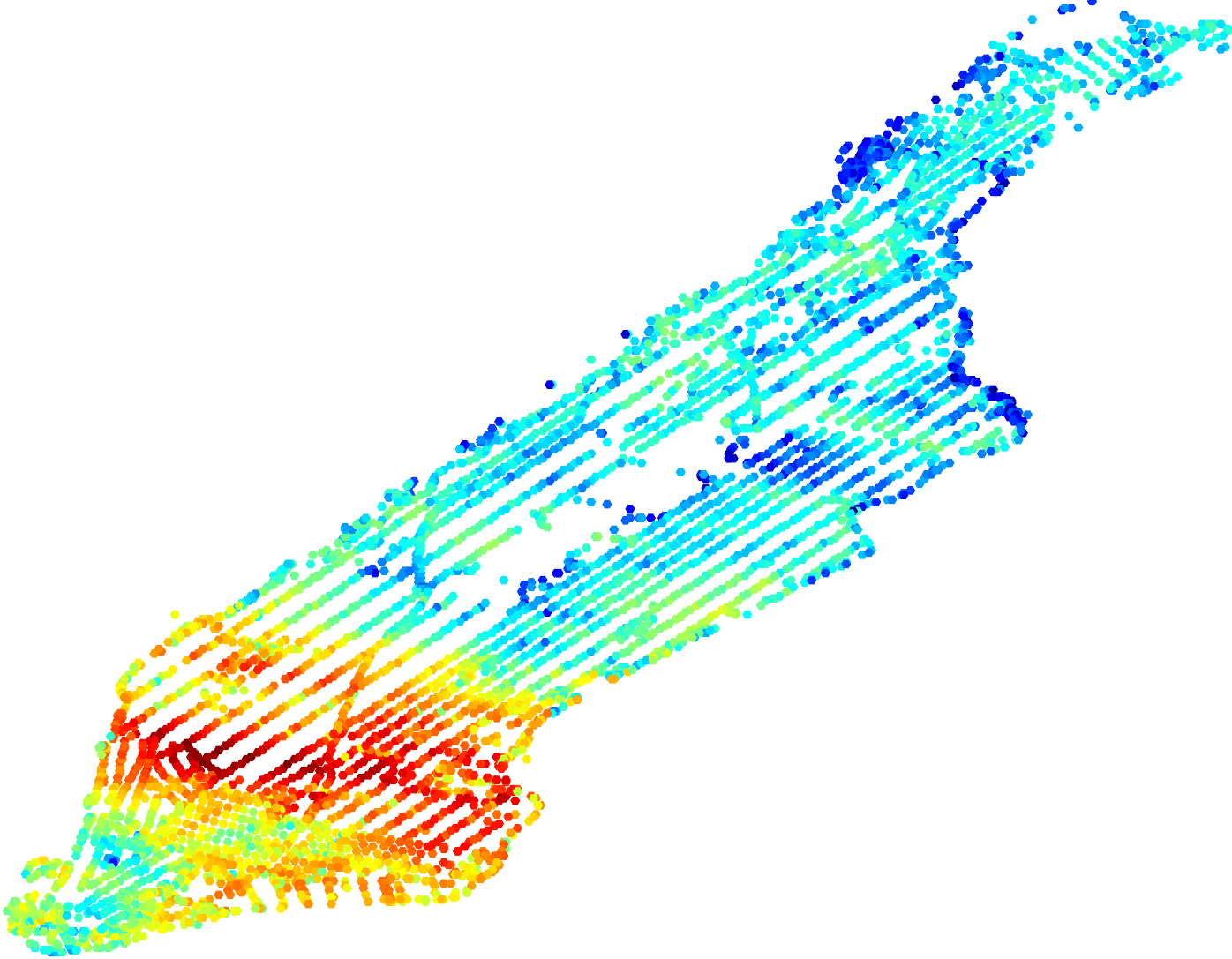}};
	\draw (6, 2.5) node[inner sep=0] {(b) $i=200$.  };
	\centering
	\draw (12, 5) node[inner sep=0] {\includegraphics[width=5cm]{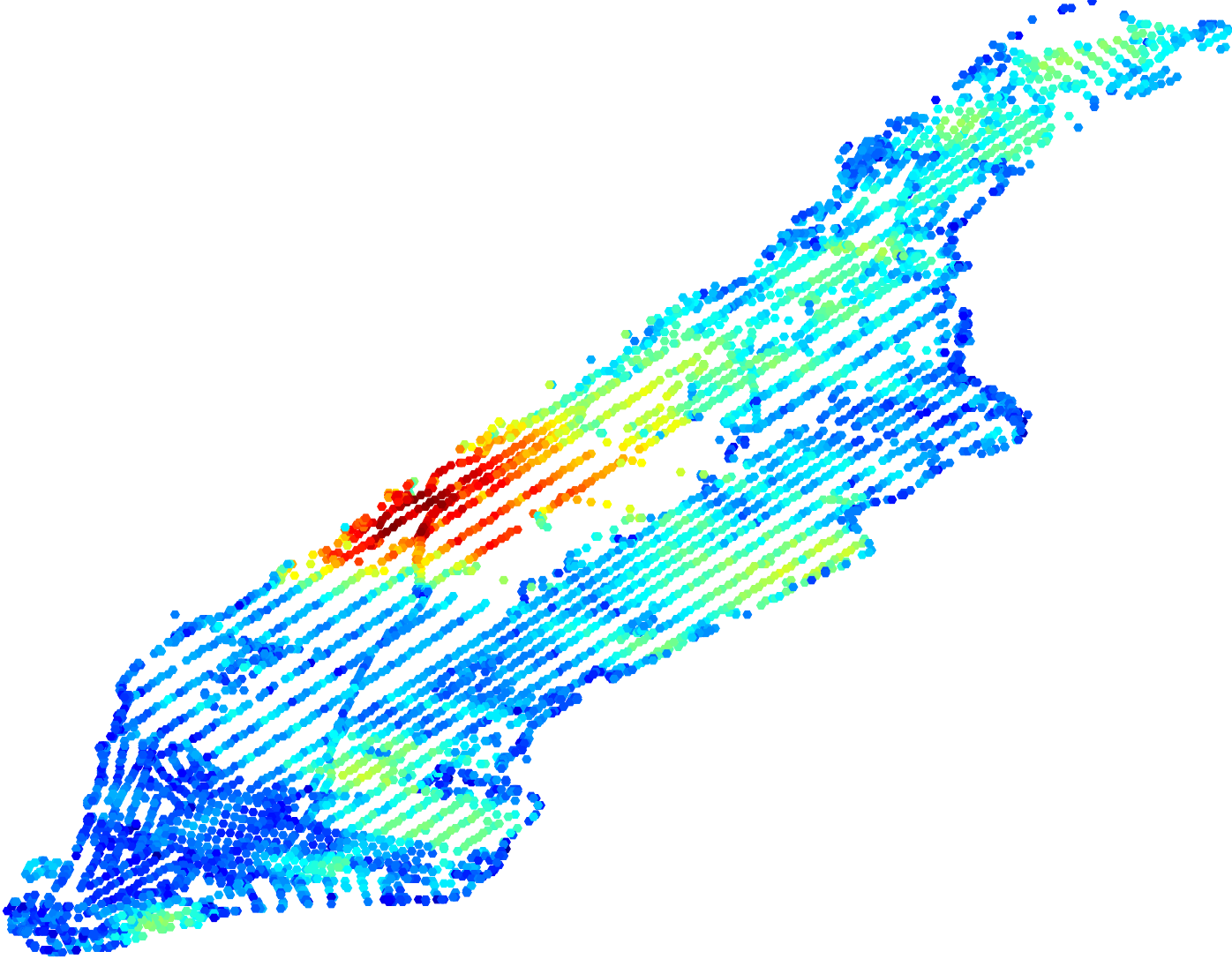}};
	\draw (12, 2.5) node[inner sep=0] {(c) $i=1700$.  };
	\draw (15.5, 2.45) node[inner sep=0] {\includegraphics[width=0.91cm]{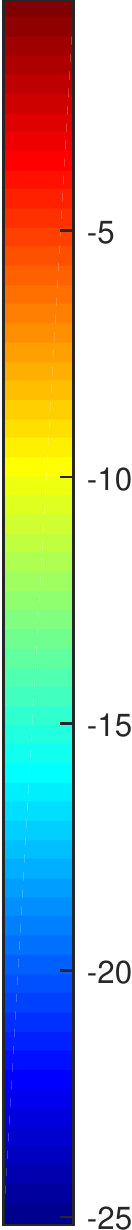}};
	\draw (0, 0) node[inner sep=0] {\includegraphics[width=5cm]{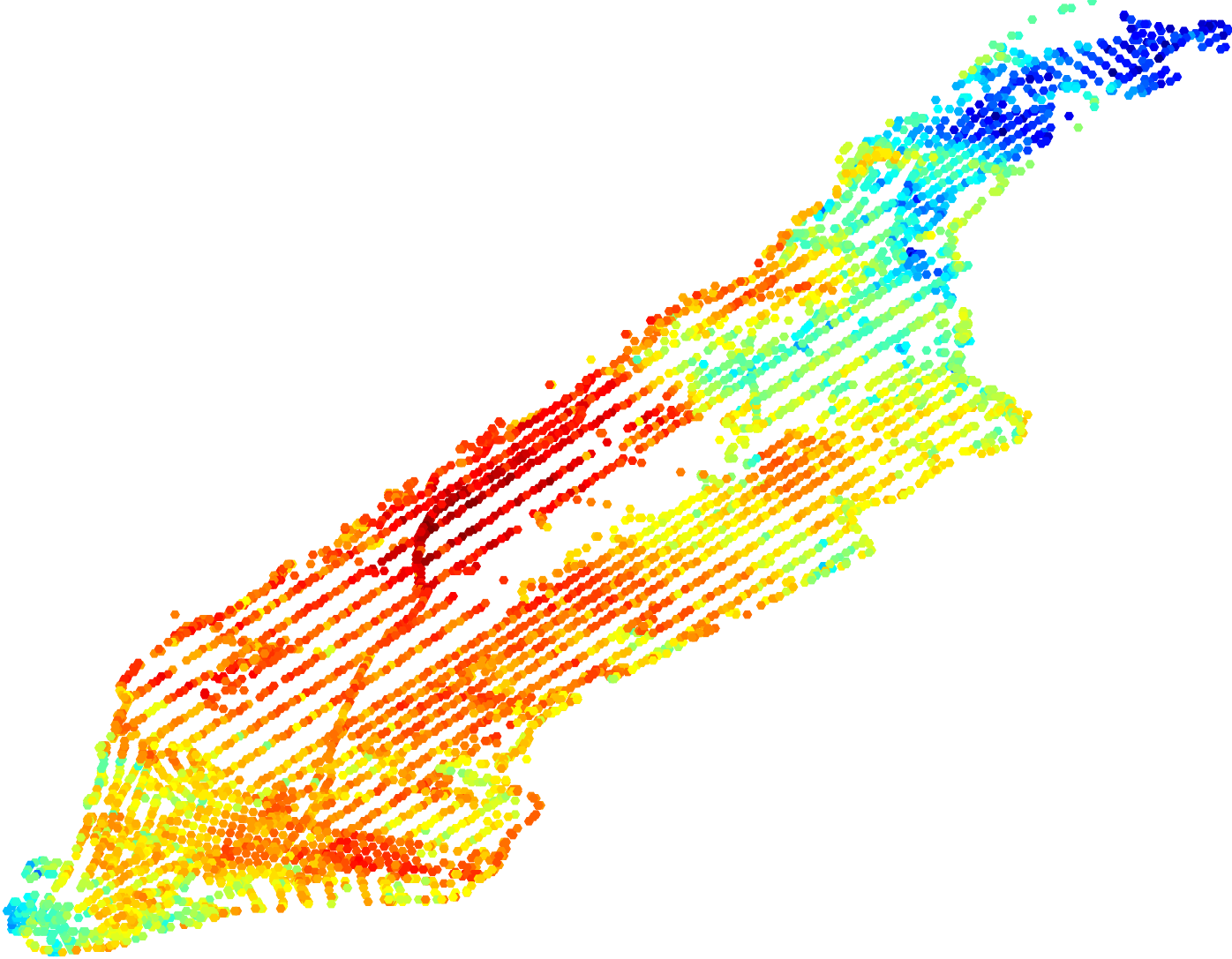}};
	\draw (0, -2.5) node[inner sep=0] {(d) $i=3050$.  };
	\centering
	\draw (6, 0) node[inner sep=0] {
		\includegraphics[width=5cm]{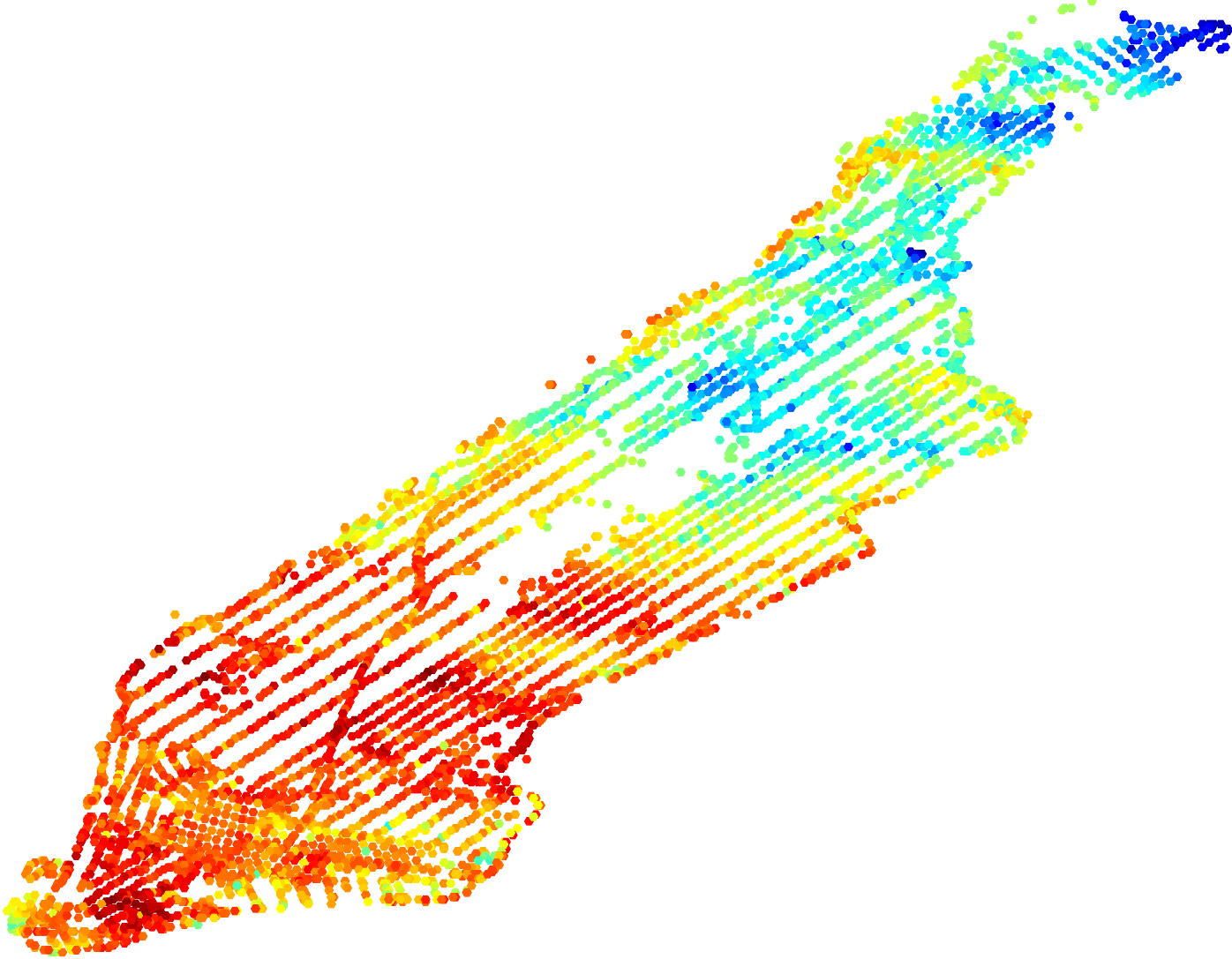} };
	\draw (6, -2.5) node[inner sep=0] {(e) $i=4900$.  };
	\draw (12, 0) node[inner sep=0] {\includegraphics[width=5cm]{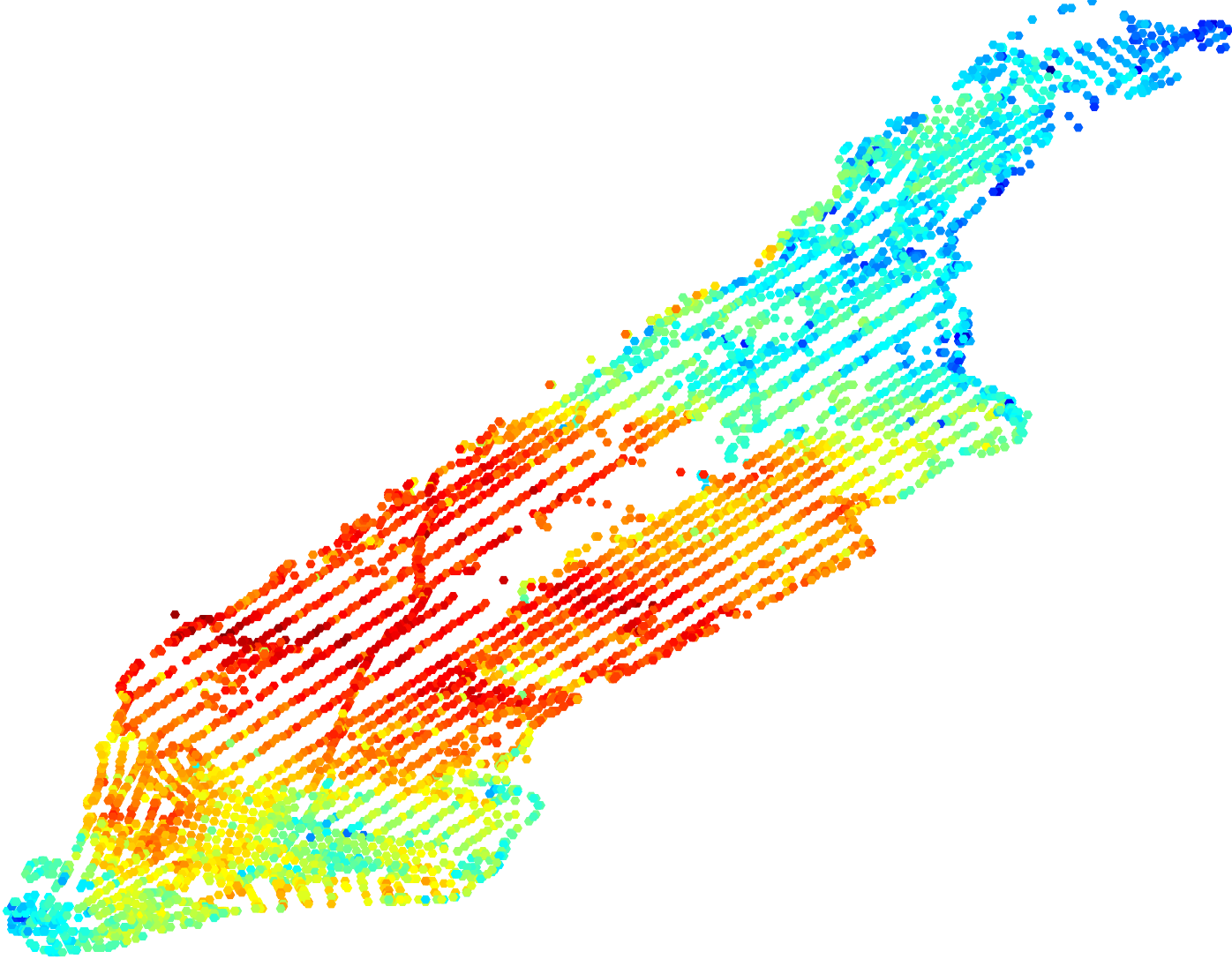}};
	\draw (12, -2.5) node[inner sep=0] {(f) $i=5464$.  };
	\end{tikzpicture}
	\caption{Manhattan road map: Heat Map of the log scale magnitude of several Fourier basis vectors $f_i$. Index~$i$ is ordered according to the total variation in figure~\ref{fig:total_var}\textemdash figures~{(a)-(f)} correspond to the graph frequency components marked by the black dots (from left to right) in figure~\ref{fig:total_var}.
	}
	\label{fig:magnitude_fourier_basis_on_graph}
\end{figure*}

We now illustrate the usefulness of SGFA by analyzing and ordering the graph frequencies and graph spectral components of a (large) real world network, namely, the Manhattan road map in figure~\ref{fig_road_net}. Graph Signal Processing orders the graph frequencies from low to high through the (increasing) total variation $\textrm{TV}$ \cite{GSP_frequencyanalysis,GSP_Big_Data} of the corresponding graph spectral components. The $\textrm{TV}$ of the graph spectral component~$f$ is \cite{GSP_frequencyanalysis,GSP_Big_Data}:
\begin{equation}
\textrm{TV}_{\mathcal{G}}(f)=\left\|f-A_{\scriptsize\textrm{norm}}f\right\|_1,\enspace A_{\scriptsize\textrm{norm}}=\frac{A}{\left|\lambda_{\max}\right|},
\label{eqn:total_var}
\end{equation}
where $\lambda_{\max}$ denotes the eigenvalue of~$A$ with largest magnitude. Figure~\ref{fig:total_var} plots the total variation of the columns of the Fourier basis~$F$ reordered by their $\textrm{TV}$.
Figure~\ref{fig:total_var} labels regions as low, medium, and high frequencies where the boundaries are simply indicative. Figure \ref{fig:magnitude_fourier_basis_on_graph} plots the magnitude of several Fourier basis vectors (columns of~$F$) in the two dimensional plane corresponding to the Manhattan road map. Figures~\ref{fig:magnitude_fourier_basis_on_graph}~(a)-(f) are displayed in order of increasing variation as computed by~\eqref{eqn:total_var}, i.e., corresponding from low to high graph frequencies (dots from left to right in figure~\ref{fig:total_var}).
  The three graph spectral components $f_i$, $i=1,200,1700$, shown in figures~\ref{fig:magnitude_fourier_basis_on_graph}~(a)-(c), tend to be sparse vectors with slowly varying magnitude.  In contrast, the three spectral graph components $f_i$, $i=3050,4900,5646$, displayed in figures~\ref{fig:magnitude_fourier_basis_on_graph}~(d)-(f), are much less smooth and exhibit (visually) much pronounced variation.
\vspace{-1cm}
\section{Conclusion}
\label{sec:conclusion}
This paper addresses an open problem in Graph Signal Processing, namely, it presents SGFA, an approach to compute the graph spectral components for graph signals supported by generic (directed) sparse graphs. With generic graphs, the corresponding graph shift~$A$ may have complex graph frequencies (eigenvalues), non orthogonal graph spectral components (eigenvectors), and, as commonly observed in practical real world applications, possibly repeated eigenvalues. Having an efficient procedure to \textcolor{black}{approximately} diagonalize these shifts~$A$ enables pursuing the graph spectral analysis of graph signals. We formulated the problem of computing accurate stable approximations of the Fourier basis~$F$ (matrix of the graph spectral components) of a generic directed graph shift~$A$ as a constrained optimization\textemdash optimize accuracy, as evaluated by the degree of invariance of the columns of~$F$, while maintaining numerical stability, as measured by the minimum singular value of~$F$ (and corresponding condition number, since the largest singular value of~$F$ is empirically verified not to be very large). This optimization is non-convex, so, we propose an efficient algorithm\textemdash SGFA\textemdash  that decreases, at least exponentially, the value of the objective per each iteration. SGFA attempts to diagonalize a triangular decomposition of~$A$, while guaranteeing the numerical stability of the resulting local solution through a threshold~$\alpha$ that controls the minimum singular value of our approximation. We applied SGFA to two real-world graphs, namely, the $2004$  U.S.~Presidential Election blogs network~\cite{political_blog_data} and the Manhattan road map~\cite{dataset_road_2}, \textcolor{black}{and also to directed Erd\H os-R\'enyi graphs with either very small or very large probability of connection}. The paper shows that SGFA generates efficiently good accurate approximations of~$F$, while insuring its stability, as demonstrated experimentally with respect to several metrics of accuracy and stability (not just the optimized ones). Finally, we illustrate the application of SGFA by computing, ordering, and displaying the graph frequencies (eigenvalues) and graph spectral components (eigenvectors) for the Manhattan road map. \textcolor{black}{Future work includes further analysis of structured random models that may better approximate real world graphs, for example, further analysis of Erd\H os-R\' enyi graphs with community structure.}
\bibliographystyle{IEEEtran}

\bibliography{IEEEabrv,M335}
\end{document}